\documentclass[12pt,preprint]{emulateapj}      
\usepackage{epsfig}

\slugcomment{The Astrophysical Journal, 613: \#\#\#--\#\#\#, 2004 October 1}

\newcommand{\gray }{$\gamma$-ray\ }
\newcommand{\grays}{$\gamma$-rays\ }
\newcommand{\piodecay}{$\pi^0$-decay\ }
\newcommand{\htwo}{H$_2$}
\newcommand{\hi}{H {\sc i}}

\newcommand{\EB}{EGRB}
\newcommand{\Xco}{$X_{\rm CO}$}
\newcommand{\Xcounits}{$\times10^{20}$ molecules cm$^{-2}$/(K km s$^{-1}$)\ }

\newcommand{\icrc}{Int.\ Cosmic Ray Conf.\ }

\newcommand{\plb}{Phys.\ Lett.\ B}

\newcommand{\pubjournal}[5]{#4, #1, #2, #3}
\newcommand{\pubjournala}[5]{#4, #1, #3}

\shorttitle{Diffuse Galactic Continuum Gamma Rays}
\shortauthors{Strong, Moskalenko, \& Reimer}
\begin{document}
\title{Diffuse Galactic Continuum Gamma Rays.\\
A  model compatible with EGRET data and cosmic-ray measurements}

\author{Andrew W.~Strong}
\affil{Max-Planck-Institut f\"ur extraterrestrische Physik,
   Postfach 1603, D-85740 Garching, Germany; aws@mpe.mpg.de}

\author{Igor V.~Moskalenko\altaffilmark{1}}
\affil{NASA/Goddard Space Flight Center, Code 661, Greenbelt, MD 20771; 
moskalenko@gsfc.nasa.gov}
\altaffiltext{1}{Joint Center for Astrophysics, University of Maryland, 
   Baltimore County, Baltimore, MD 21250}

\and

\author{Olaf Reimer}
\affil{Ruhr-Universit\"at Bochum, D-44780 Bochum, Germany; olr@tp4.ruhr-uni-bochum.de}


\begin{abstract}

We present a study of the compatibility of some current models of the
diffuse Galactic continuum \grays with EGRET data. A set of regions
sampling the whole sky is chosen to provide a comprehensive range of
tests. The range of EGRET data used is extended to 100 GeV. The models
are computed with our GALPROP cosmic-ray propagation and gamma-ray
production code. We confirm that the ``conventional model'' based on
the locally observed electron and nucleon spectra is inadequate, for
all sky regions. A conventional model plus hard sources in the inner
Galaxy is also inadequate, since this cannot explain the GeV excess
away from the Galactic plane. Models with a hard electron injection
spectrum are inconsistent with the local spectrum even considering the
expected fluctuations; they are also inconsistent with the EGRET data
above 10 GeV.

We present a new model which fits the spectrum in all sky regions
adequately. Secondary antiproton data were used to fix the Galactic
average proton spectrum, while the electron spectrum is adjusted using
the spectrum of diffuse emission itself. The derived electron and
proton spectra are compatible with those measured locally considering
fluctuations due to energy losses, propagation, or possibly details of
Galactic structure.  This model requires a much less dramatic
variation in the electron spectrum than models with a hard electron
injection spectrum, and moreover it fits the \gray spectrum better and
to the highest EGRET energies. It gives a good representation of the
latitude distribution of the \gray emission from the plane to the
poles, and of the longitude distribution. We show that secondary
positrons and electrons make an essential contribution to Galactic
diffuse \gray emission.
 
\end{abstract}



\section{Introduction}

Diffuse continuum \grays from the interstellar medium are potentially
able to reveal much about the sources and propagation of cosmic rays
(CR), but in practice the exploitation of this well-known connection
is problematic. 
While the basic processes governing the CR propagation and production
of diffuse \gray emission seem to be well-established, some puzzles
remain. In particular, the spectrum of \grays\ calculated under the
assumption that the proton and electron spectra in the Galaxy resemble
those measured locally reveals an excess above 1 GeV in the EGRET
spectrum \citep{hunter97}.

The Galactic diffuse continuum \grays are produced in
energetic interactions of nucleons with gas via neutral pion
production, and by electrons via inverse Compton (IC) scattering and
bremsstrahlung. These processes are dominant in different parts of
the spectrum, and therefore if deciphered the \gray spectrum can
provide information about the large-scale spectra of nucleonic and
leptonic components of CR. In turn, having an improved understanding
of the Galactic diffuse \gray emission and the role of CR is essential
for unveiling the spectra of other components of the diffuse emission
and is thus of critical importance for the study of many topics in
\gray astronomy, both Galactic and extragalactic.

The puzzle of the ``GeV excess'' has lead to an attempt to re-evaluate
the reaction of $\pi^0$-production in $pp$-interactions. However, a
calculation made using modern Monte Carlo event generators to simulate
\emph{high-energy} $pp$-collisions has shown \citep{mori97} that the
\gray flux agrees rather well with previous calculations. A flatter
Galactic nucleon spectrum has been suggested as a possible solution to
the ``GeV excess'' problem \citep{gralewicz97,mori97}. Explaining the
excess requires an ambient proton spectrum power-law index of about
--2.4--2.5, compared to --2.75 measured locally \citep[for a summary
of recent data see][]{M02}.  Such large variations in the proton
spectrum are, however, improbable. A sensitive test of the
large-scale-average proton spectrum has been proposed by
\citet*{MSR98} based on the fact that secondary antiprotons and \grays
are produced in the same $pp$-interactions. The secondary antiprotons
sample the proton spectrum in a large region of the Galaxy, and a
flatter nucleon spectrum in distant regions would lead to
overproduction of secondary antiprotons and positrons. The ``hard
nucleon spectrum'' hypothesis has effectively been excluded by recent
measurements of $\bar p/p$ ratio at high energies
\citep[e.g.,][]{Beach2001}. In addition, new accurate measurements of
the local proton and Helium spectrum allow less freedom for deviations
in the \piodecay component.

A ``hard electron spectrum'' hypothesis has been investigated by
\citet{porter97}, \citet{pohl98}, and \citet{aharonian00}. 
An essential idea of this
approach is that the locally-measured CR spectrum of electrons is not
a good constraint because of the spatial fluctuations due to energy
losses and the stochastic nature of the sources in space and time; the
average interstellar electron spectrum responsible for \grays via IC
emission (and bremsstrahlung) can therefore be quite different from
that measured locally. An extensive study of this hypothesis has been
made by \citet*{SMR00}; in this model a less dramatic but essential
modification of the proton and Helium spectrum (for the \piodecay
component) was also required. The latter was still consistent with the
locally-observed proton spectrum, as it should be since the proton
fluctuations are expected to be small \citep{SM01a} as the result of
their negligible energy losses. The ``hard electron spectrum''
hypothesis suffers however from the following problems:
\begin{itemize}

\item It is hardly compatible with the local electron spectrum
\emph{even} considering the fluctuations due to stochastic sources
and energy losses, as shown by a 3D time-dependent study \citep{SM01};

\item The fit to the shape of the \gray spectrum is still poor above 1
GeV \citep{SMR00};

\item It cannot reproduce the spectrum in the inner and outer Galaxy
and intermediate/high latitudes simultaneously \citep*{SMR03a};

\end{itemize}

These problems were already evident before the present study, but now
we show in addition that
\begin{itemize}

\item it predicts significantly higher intensities than the EGRET data
above 10 GeV.

\end{itemize}

Another suggestion which has been made \citep{volk} is that the \gray
spectrum contains a contribution from accelerated particles confined
in SNR. The SNR proton and electron spectra, being much harder than
the interstellar CR spectra, produce \piodecay and IC \grays adding to
the apparently diffuse $\gamma$-rays, while the SNRs themselves are
too distant to be resolved into individual sources.

An alternative model involving spatial variation of the CR propagation
conditions has been proposed by \citet{ew02a,ew02b}.

A shortcoming of previous analyses was that the comparison with EGRET
data was limited to particular regions, and the rich EGRET data have
remained not fully exploited. \citet{hunter97} made an extensive
comparison of the spectra near the Galactic plane
$|b|<10^\circ$. Other analyses have concentrated on particular
molecular clouds: Ophiuchus \citep{hunter94}, Orion \citep{digel99},
Cepheus and Perseus \citep{digel96}, Monoceros \citep{digel01}, high
latitudes \citep{sreekumar98}. The previous analysis by \citet{SMR00}
was limited to the inner Galaxy at low latitudes, and profiles
integrated over large regions of longitude or latitude. In that
study, we compared a range of models, based on our CR propagation
code GALPROP, with data from the Compton Gamma Ray Observatory.
Relative to the work of \citet{hunter97} we emphasized the connection
with CR propagation theory and the importance of IC emission, and less
to fitting to structural details of the Galactic plane. The study
confirmed that it is rather easy to get agreement within a factor
$\sim$2 from a few MeV to 10 GeV with a ``conventional'' set of
parameters, however, the data quality warrant considerably better fits.

In the present paper we attempt to exploit the fact that the models
predict quite specific behaviour for different sky regions and this
provides a critical test: the ``correct'' model should be consistent
with the data in \emph{all} directions.  We show that a new model,
with less dramatic changes of electron and nucleon spectra relative to
the ``conventional'' model, can well reproduce the \gray data. The
changes consist in renormalization of the \emph{intensities} of the
electron and proton spectra, and a relatively small modification of
the proton spectrum at \emph{low} energies. The model is compatible
with locally observed particle spectra considering the expected level
of spatial fluctuations in the Galaxy. We extend the \gray data
comparisons over the entire sky and to 100 GeV in energy. We also
exploit the recent improved measurements of the local proton, Helium,
as well as antiproton, and positron spectra which are used as
constraints on the proton spectrum in distant regions.

Our approach differs from that of \citet{hunter97} in that it is based
on a model of CR propagation while \citeauthor{hunter97} use CR-gas
coupling and a relatively small IC component. It is also different
from \citet{strongmattox96} in that it is based on a physical model,
while that work was based on model-fitting to gas surveys to
determine the \gray emissivity spectrum as a function of
Galactocentric radius.
The current study concentrates on spectral aspects of the
\gray emission; the question of the CR source gradient
and the distribution of molecular hydrogen is addressed in 
\citep*{SMR04a}.

The selection of a good model for the diffuse Galactic emission is
critical to another topic, the extragalactic \gray background (\EB).
We have argued in \citet{SMR00} and \citet{MS00} that IC from a large
halo can make up a substantial fraction of the high-latitude emission
and hence reduce the residual \EB\ \citep[and modify its spectrum
relative to][]{sreekumar98}. In a companion paper \citep*{SMR04} we
present a comprehensive discussion of the \EB, with a new estimate
which is used in the present paper.

\section{Models}
\subsection{GALPROP code}

\begin{deluxetable*}{lcccccccc}
\tablecolumns{9}
\tablewidth{18cm}
\tabletypesize{\footnotesize}
\tablecaption{Particle injection spectra and normalizations.
\label{model_parameters1}}
\tablehead{
\colhead{} &
\colhead{} & 
\multicolumn{3}{c}{Proton spectrum} &
\colhead{} &
\multicolumn{3}{c}{Electron spectrum} \\ 
\cline{3-5} \cline{7-9}
\colhead{} &
\colhead{} &
\colhead{Injection} &
\colhead{Break} &
\colhead{Normalization\tablenotemark{b}} &
\colhead{} &
\colhead{Injection} &
\colhead{Break} &
\colhead{Normalization\tablenotemark{b}} \\
\colhead{Model} &
\colhead{ID} &
\colhead{index\tablenotemark{a}} &
\colhead{rigidity, GV} &
\colhead{@ 100 GeV} &
\colhead{} &
\colhead{index\tablenotemark{a}} &
\colhead{rigidity, GV} &
\colhead{@ 32.6 GeV}
}
\startdata
Conventional & 44\_500180 & 
1.98/2.42 & 9 & $5.0\times10^{-2}$ & &
1.60/2.54 & 4 & $4.86\times10^{-3}$ \\

Hard electron    & 44\_500181 & 
1.98/2.42 & 9 & $5.0\times10^{-2}$ & &
1.90      & \nodata & $1.23\times10^{-2}$ \\

Optimized       & 44\_500190 &
1.50/2.42 & 10 & $9.0\times10^{-2}$ & &
1.50/2.42 & 20 & $2.39\times10^{-2}$ \\

\enddata
\tablecomments{The GALPROP model IDs are given for future reference; 
the corresponding parameter files contain a complete specification of the models.}
\tablenotetext{a}{Below/above the break rigidity.}
\tablenotetext{b}{Normalization of the local spectrum (propagated).
Units are m$^{-2}$ sr$^{-1}$ s$^{-1}$ GeV$^{-1}$.}
\end{deluxetable*}

The principles of the GALPROP code for CR propagation and \gray
emission have been described in \citet{SM98} and \citet{SMR00}. Since
then the code\footnote{As usual the code and documentation is
available at http://www.mpe.mpg.de/$\sim$aws/aws.html} has been
entirely re-written in C++ \citep[][and references therein]{M02}
using the experience gained from the original (fortran) version, with
improvements in particular in the generation of \gray skymaps. Both
2D (radially symmetric) and full 3D options are available, the latter
allowing also explicit time-dependence with stochastic SNR source
events \citep{SM01}. For this paper the 2D option is sufficient since
we need only kpc-scale averaged CR spectra (even if these differ from
local CR measurements).

An important point to note is that even in the 2D case, the symmetry
applies only to the CR distribution; for the gas-related components
(\piodecay and bremsstrahlung) of the \gray skymaps we use 21-cm line
survey data for \hi\ and CO ($J=1 \rightarrow 0$) survey data for
\htwo, in the form of column densities for Galactocentric ``rings,''
using velocity information and a rotation curve (see Appendix
for details). In this way details of Galactic
structure are included at least for the gas, at a sufficient level for
the present limited state of knowledge on e.g. the relation of
cosmic rays to spiral structure. The longitude range
$350^\circ<l<10^\circ$ is not included in the \hi\ and CO survey data
due to lack of kinematic information; for the analysis interpolated
values are used, and this is found to be fully consistent will the
\gray data. The interstellar radiation field (ISRF) for computing IC
emission and electron energy losses is the same as that described and
used in \citet{SMR00}; pending a new calculation (an ambitious
project) this is the best we have available. Although the uncertainty
in the ISRF is a shortcoming, note that since we fit to the \gray
data by adjusting the electron spectrum, inaccuracies in the ISRF
spectrum will tend to be compensated.

The radial distribution of CR sources used is the same as in
\citet{SMR00}, since we find this empirically-derived form still gives
a good reproduction of the \gray longitude distribution\footnote{
For earlier work on the CR distribution see \citet{stecker77},
\citet{harding85}, \citet{bloemen86}, \citet{strong88}, 
\citet{strongmattox96}.}. Although
flatter than the SNR distribution \citep[e.g.,][]{CaseBhattacharya98},
this may be compensated by the gradient in the CO-to-\htwo\ conversion
factor whose metallicity and temperature dependences have the net
effect of causing the factor to increase with $R$
\citep*{papadopoulos02,israel97}.  We use a uniform value of \Xco =
1.9\Xcounits as in \citet{SMR00} and \citet{strongmattox96}; this is
consistent with the value $(1.8\pm0.3)$\Xcounits from a recent
(non-$\gamma$-ray) CO survey analysis by \citet{dame01}.

\placetable{model_parameters1}

\begin{figure}[!thb]
\centering
\includegraphics[width=9cm]{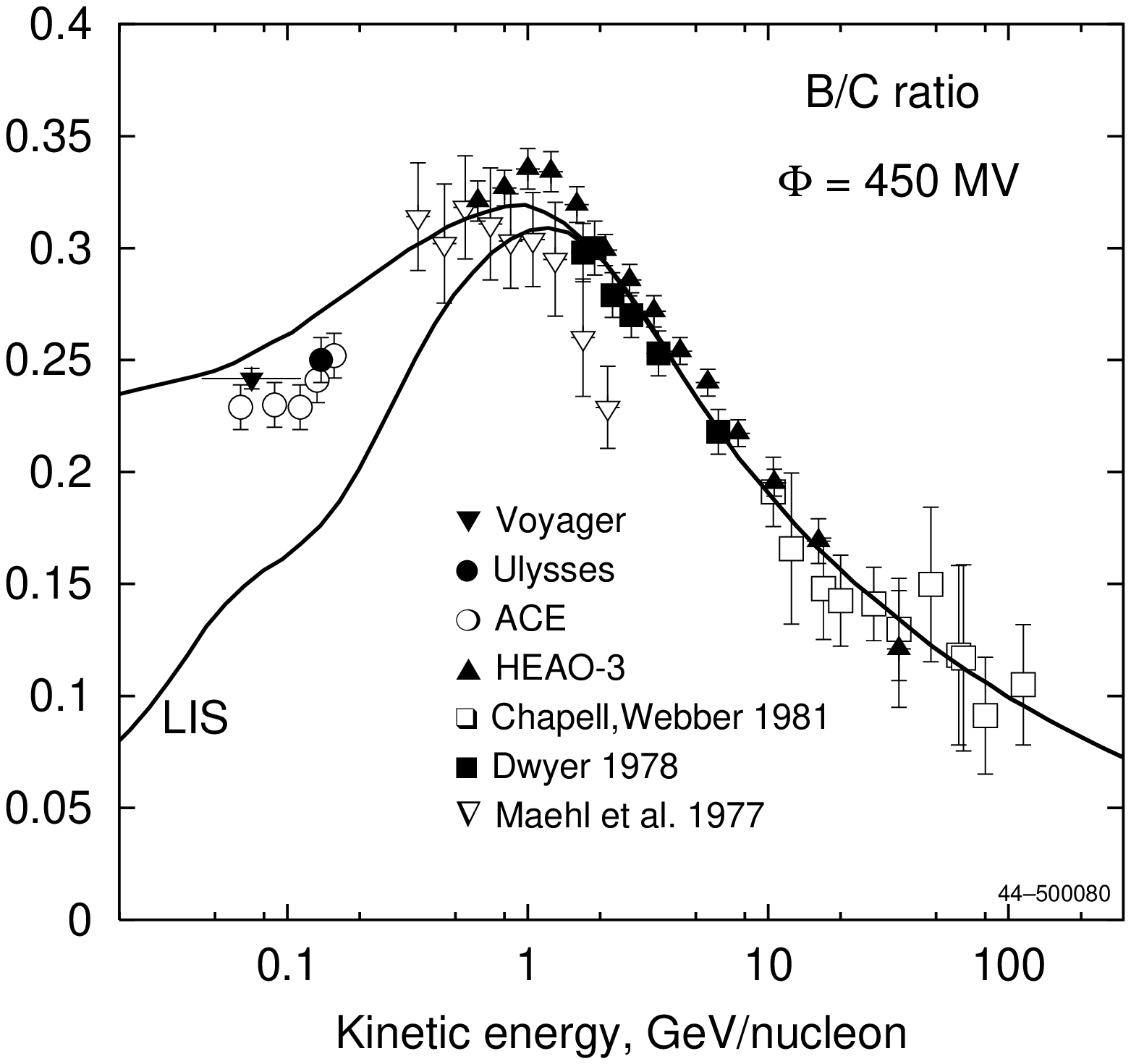}
 \caption{B/C ratio as calculated in reacceleration model.
Lower curve -- LIS, upper -- modulated ($\Phi=450$ MV).
Data below 200 MeV/nucleon: ACE \citep{davis}, Ulysses
\citep*{ulysses_bc}, Voyager \citep*{voyager}; high energy data:
HEAO-3 \citep{Engelmann90}, for other references see
\citet{StephensStreitmatter98}.
\label{fig:bc}}
\end{figure}

\placefigure{fig:bc}

\placetable{sky_regions}

The parameters of the models are summarized in Table
\ref{model_parameters1}. The models differ only in the injection
spectra of protons (and He) and electrons, while the injection spectra
of heavier nuclei are assumed to have the same power-law in rigidity,
for all models. For propagation, we use essentially the same diffusion
reacceleration model, model DR, as described in \citet{M02}. The
propagation parameters have been tuned to fit the B/C ratio (Fig.\
\ref{fig:bc}) using improved cross-sections \citep{MM03}. The spatial
diffusion coefficient is taken as $\beta D_0(\rho/\rho_0)^\delta$,
where $D_0=5.8\times10^{28}$ cm s$^{-1}$ at $\rho_0=4$ GV, and
$\delta=1/3$ (Kolmogorov spectrum). The Alfv\'en speed is $v_A=30$ km
s$^{-1}$. The halo height is taken as $z_h=4$ kpc as in
\citet{SMR00}, in accordance with our analysis of CR
secondary/primary ratios \citep*[][and references
therein]{MMS01,M02}. However, values $z_h$ differing by 50\% (the
estimated error) with corresponding adjustment of $D_0$ would provide
essentially similar results  since the IC contribution scales mainly
with the electron spectrum which is here treated as a free parameter.

The spectra are compared in the regions summarized in
Table~\ref{sky_regions}.  Region A corresponds to the ``inner
radian'', region B is the Galactic plane excluding the inner radian,
region C is the ``outer Galaxy'', regions D and E cover higher
latitudes at all longitudes, region F is ``Galactic poles''. Region H
is the same as in \citet{hunter97} and is used for comparison with
results of \citeauthor{hunter97} In addition to spectra, profiles in
longitude and latitude are an essential diagnostic; our latitude
profiles are plotted logarithmically because of the large dynamic
range from the Galactic plane to the poles.

The \EB\ used here is based on the new determination by \citet{SMR04}.
Since this was derived for the EGRET energy bands, it is interpolated
in order to produce a continuous spectrum for combining with the model
Galactic components. The present analysis is however not sensitive to
the details of the \EB. Since our COMPTEL data do not contain the \EB\
(see Section \ref{sec:comptel}), we do not extrapolate the \EB\ beyond
the EGRET energy range when comparing with data.


\begin{deluxetable}{crrl}
\tablecolumns{4}
\tablewidth{0mm}
\tabletypesize{\footnotesize}
\tablecaption{Sky regions used for comparison of models with data.
\label{sky_regions}}
\tablehead{
\colhead{Region} &
\colhead{$l$, deg} & 
\colhead{$|b|$, deg} &
\colhead{Description}
}
\startdata
H & 300--60 & 0--10  &\citet{hunter97} region        \\
A & 330--30 & 0--5   &inner Galaxy \\
B & 30--330 & 0--5   &Galactic plane avoiding inner Galaxy \\
C & 90--270 & 0--10  &outer Galaxy                   \\
D & 0--360  & 10--20 &intermediate latitudes 1       \\
E & 0--360  & 20--60 &intermediate latitudes 2       \\
F & 0--360  & 60--90 &Galactic poles                 \\
\enddata
\bigskip
\end{deluxetable}

\subsection{Presentation of results}

The output of the GALPROP runs is in the form of FITS files; the
visualization\footnote{An additional program GALPLOT has been
developed for this purpose, with flexible plotting options and
convolution; this will be made available with future versions of
GALPROP.} in the form of spectra and profiles, and comparison of the
results with data involves integrations over sky regions and energy as
well as convolution. The predicted model skymaps are convolved with
the EGRET point-spread function as described in \citet{SMR00}. For the
profiles the convolved model is directly compared with the observed
intensities. For the spectra, the procedure is slightly different: the
predicted (unconvolved) intensities are compared with intensities
corrected for the effect of convolution as given by the model under
study. This procedure has the advantage that the spectra are spatially
\emph{deconvolved}, allowing for more direct interpretation and also
the combination of data with other experiments, such as COMPTEL, with
different instrument response functions. The effect of this procedure
on the spectra is only significant below 500 MeV.

\subsection{Statistical test}

The choice of model in this work is mainly subjective, based on visual
inspection of spectra and profiles. In order to give also an objective
criterion, a $\chi^2$ statistic has been computed over the full sky,
for each EGRET energy range between 30 MeV  and 50 GeV
(Table~\ref{chisq}); The binning for this test is the same as used in
\citet{SMR04}: raster scanned bins in longitude for latitude width
$2^\circ$, giving 78 sky bins. As in \citet{SMR04}, a lower limit of
10 counts per bin were accepted, so at high energies the number of
bins is reduced.  The error is computed as the sum of the statistical
error and the systematic error as described in section~\ref{sec:egret}.


\begin{deluxetable}{ccccc}
\tablecolumns{5}
\tablewidth{0mm}
\tabletypesize{\footnotesize}
\tablecaption{Comparison of models using $\chi^2$ for full sky.
\label{chisq}}
\tablehead{
\colhead{Energy range} &
\colhead{44\_500180} &
\colhead{44\_500181} &
\colhead{44\_500190} &
\colhead{Number of}\\
\colhead{MeV} &
\colhead{conventional} &
\colhead{hard electron}&
\colhead{optimized} &
\colhead{data points} 
}
\startdata
   30--50       &   90   &   38 &   34   & 78\\
   50--70       &   19   &   31 &   26   & 78\\
   70--100       &   18   &   49 &   30   & 78\\
  100--150       &   38   &   89 &   47   & 78\\
  150--300       &   33   &   82 &   41   & 78\\
  300--500       &   43   &   24 &   24   & 78\\
  500--1000       &   140  &   59 &   32   & 78\\
 1000--2000       &   382  &  216 &   61   & 78\\
 2000--4000       &   441  &  243 &   93   & 76\\
 4000--10000       &   247 &    53 &   49   & 66\\
10000--20000       &   56   &   54 &   21   & 23\\
20000--50000       &   22   &   35 &    4   &  7\\
\\
   30--50000       & 1528   &  974 &  462   &796\\
\enddata
\end{deluxetable}

\section{\gray and cosmic ray measurements}
\subsection{EGRET data}\label{sec:egret}
We use the co-added and point-source removed EGRET counts and exposure
maps in Galactic coordinates with 0.5$^\circ$ $\times$ 0.5$^\circ$
binsize at energies between 30 MeV and 10 GeV, as described in
\citet{SMR00}. Apart from the most intense sources, the removal of
sources has little influence on the comparison with models. For the
spectra, the statistical errors on the EGRET data points are very
small since the regions chosen have large solid angle; the systematic
error dominates and we have conservatively adopted a range $\pm$15\%
in plotting the observed spectra \citep{sreekumar98,esposito}. For
longitude and latitude profiles only the statistical errors are plotted. In
addition we use EGRET data in the energy ranges 10--20, 20--50 and
50--120 GeV. Because the instrumental response of EGRET determined at
energies above 10 GeV is less certain compared to energies below 10
GeV, it is necessary to account for additional uncertainties. In
particular the EGRET effective area can only be deduced by
extrapolation from the calibrated effective area at lower energies
\citep{Thompson1993a}. We accordingly adopt values of 0.9, 0.8, and
0.7 times the 4--10 GeV effective area, respectively. On top of the
statistical and systematic uncertainties as described above we account
for the uncertainties due to the uncalibrated effective area of the
EGRET telescope above 10 GeV with an additional systematic error of
$\pm$5\%. However, the actual number of photons $>$10 GeV is small:
1091, 362 and 53 events respectively, and concentrated mainly in the
inner Galaxy; hence the comparison with models above 10 GeV can only
be made in this region.

At low energies the EGRET effective area includes the so-called
``Kniffen factor'' \citep{Thompson1993b} derived by fitting the Crab
spectrum; this additional uncertainty (factor = 2--3.4 for 30--50
MeV and 1.2--1.6 for 50--70 MeV) should be borne in mind when
comparing models with EGRET data.

\subsection{COMPTEL data}\label{sec:comptel}
The intensities are based on COMPTEL maximum entropy all-sky maps in
the energy ranges 1--3, 3--10 and 10--30 MeV, as published in
\citet{strong99}. The intensities are averaged over the appropriate
sky regions, with high latitudes being used to define the zero
level. COMPTEL data is only used for the inner Galaxy spectra, since
the skymaps do not show significant diffuse emission elsewhere. For
this reason, the COMPTEL data shown in the figures does not include
the \EB.

\subsection{Cosmic rays}\label{sec:cr}

\begin{figure}[!tb]
\centering
\includegraphics[width=9cm]{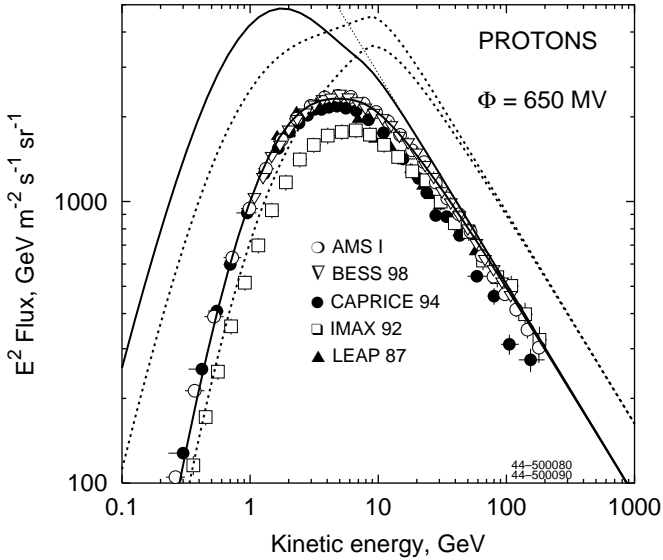}
 \caption{Proton spectra as calculated in conventional (solid lines) 
and optimized (dots) models 
compared with the data (upper curve -- LIS, lower -- modulated to 650 MV).
Thin dotted line shows the LIS spectrum best fitted to the data above 20 GeV 
\citep{M02}. 
Data: AMS \citep{p_ams}, BESS 98 \citep{sanuki00}, CAPRICE 94 \citep{Boez99},
IMAX 92 \citep{Menn00}, LEAP 87 \citep{p_leap}.
\label{fig:protons}}
\end{figure}

\begin{figure}[!tb]
\centering
\includegraphics[width=9cm]{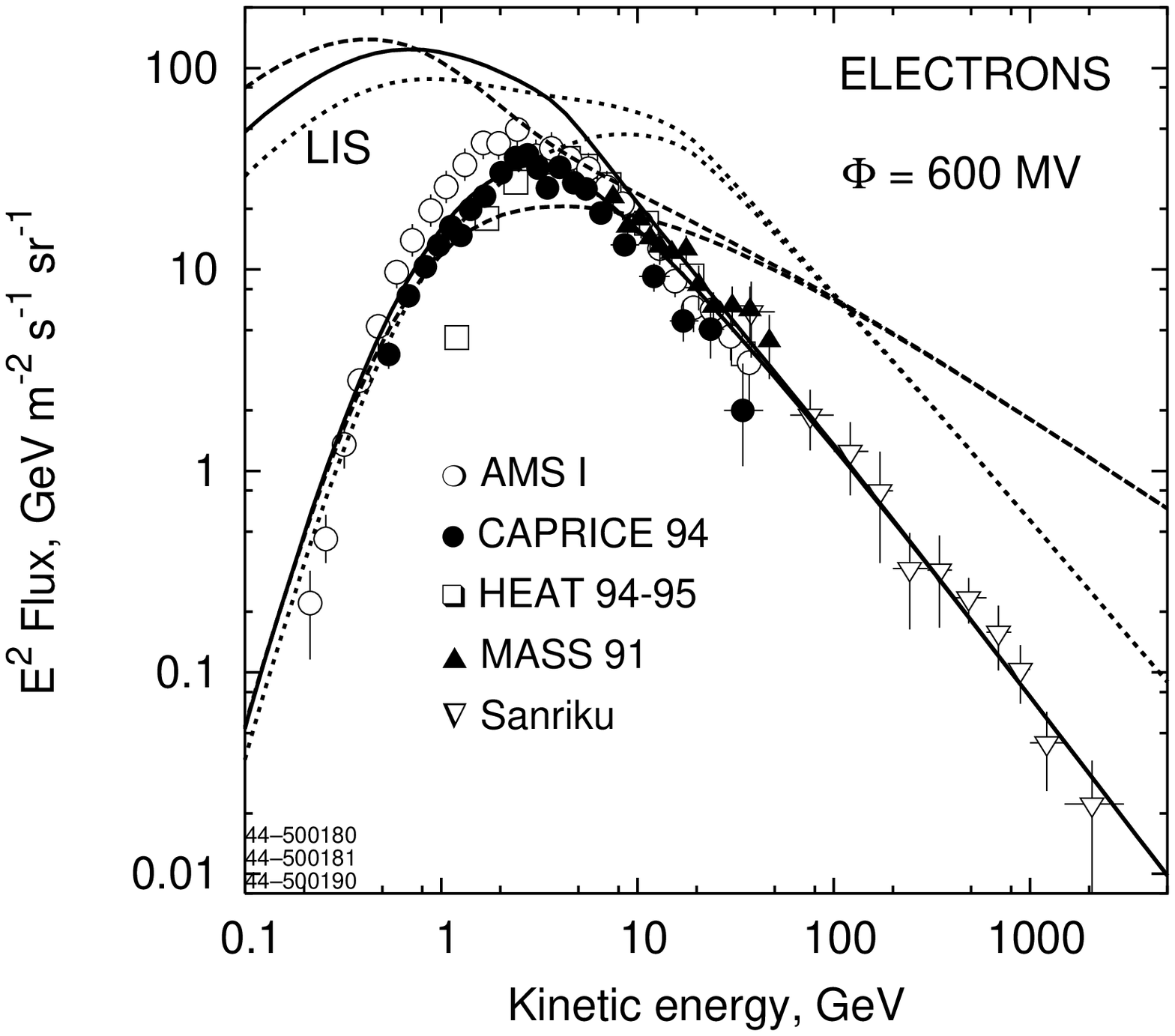}
 \caption{Electron spectra for conventional (solid), hard electron (dashes), 
and optimized models (dots), compared with data 
(upper curve -- LIS, lower -- modulated to 600 MV).
Data: AMS \citep{leptons_ams}, CAPRICE 94 \citep{Boez00}, HEAT 94-95 \citep{duvernois01}
MASS 91 \citep{grimani02}, Sanriku \citep{kobayashi99}.
\label{fig:electrons}}
\end{figure}

In our {\it conventional model} we use the locally-observed proton,
He, and electron spectra (solid lines in Figs.\ \ref{fig:protons},
\ref{fig:electrons}). The nucleon data are now more precise than
those which were available for \citet{SMR00}. The proton (and Helium)
injection spectra and the propagation parameters are chosen to
reproduce the most recent measurements of primary and secondary
nuclei, as described in detail in \citet{M02}. The error on the
dominant proton spectrum in the critical (for $\pi^0$-decay) 10--100 
GeV range is now only $\sim5\%$ for BESS \citep{sanuki00}.
Relative to protons, the contribution of He in CR to the \gray\ flux
is about 17\%, and the CNO nuclei in CR contribute about 3\%. The He
nuclei in the ISM contribute about 25\% relative to hydrogen for the
given ratio He/H = 0.11 by number. The total contribution of nuclei
with $Z>1$ is about 50\% relative to protons.

\placefigure{fig:protons} 

\placefigure{fig:electrons}

In our {\it optimized model} we use the proton and He high-energy
spectral shape derived from the local data (dotted lines in Figs.\
\ref{fig:protons}, \ref{fig:electrons}). We allow however for some
deviations in the normalization. The antiproton
\citep{Orito00,Beach2001} and positron data provide an important
constraint \citep{MSR98,SMR00} on the proton spectrum on a large
scale. Since the low-energy protons and nuclei are undetectable in the ISM, we
allow more freedom in the proton and He spectrum below 10 GeV. We
introduce a break at 10 GeV which enables us to fit the \gray spectrum
while still remaining within the constraints provided by the
locally-observed antiproton and positron spectra. The deviations from
the local measurements at low energies can be caused by the effect of
energy losses and spatial fluctuations in the Galaxy.
The modification of the low-energy proton spectrum may also be partly a
compensation for errors in the models of neutral pion production at
\emph{low energies} \citep[e.g.,][]{stecker70}, which rely on the
data of 1960's \citep[see][and references therein]{dermer86} and do
not provide the required accuracy now. Besides, the low-energy
protons are strongly affected by solar modulation; while the effect of
solar modulation is not fully understood, it is essential below 10
GeV. We refer a reader to Section \ref{sec:discussion} where various
aspects of the uncertainties are discussed in more detail. For
electrons, the injection index near $\sim$1.8 at $\sim$1 GeV is
consistent \citep[see][]{SMR00} with observations of the synchrotron
index $\beta = 2.40-2.55$ for 22--408 MHz \citep{roger99} and $\beta =
2.57\pm0.03$ for 10--100 MHz \citep*{webber80}.

Secondary and tertiary antiprotons are calculated as described in
\citet{M02}. Secondary positron and electron production is computed
using the formalism described in \citet{MS98}, which includes a
reevaluation of the secondary $\pi^\pm$- and $K^\pm$-meson decay
calculations. Antiprotons, positrons, and electrons including
secondary electrons are propagated in the same model as other CR
species.

\section{Conventional model}

\begin{figure*}[!thb]
\centering 
\includegraphics[height=58mm, width=58mm]{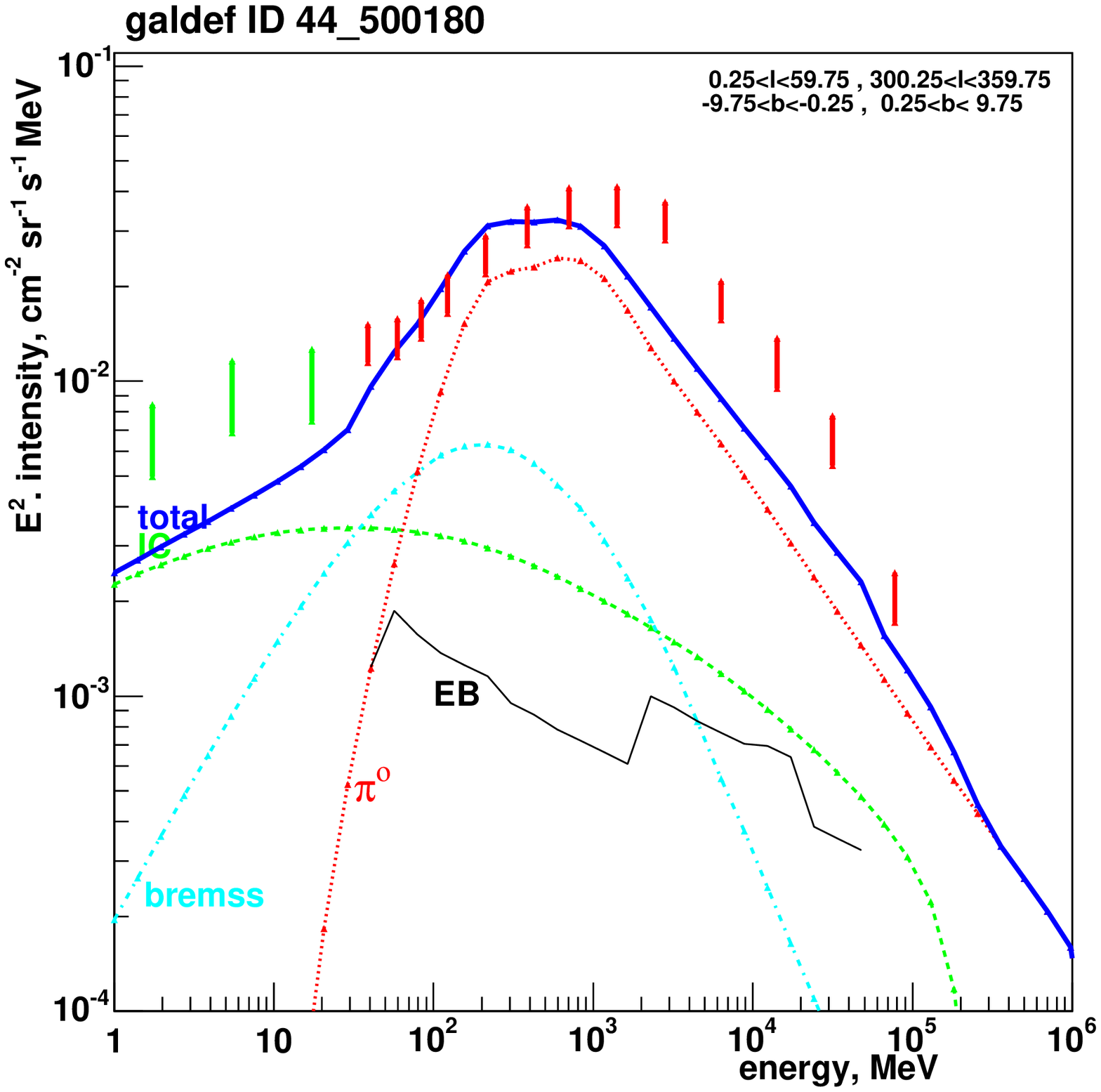}
\includegraphics[height=58mm, width=58mm]{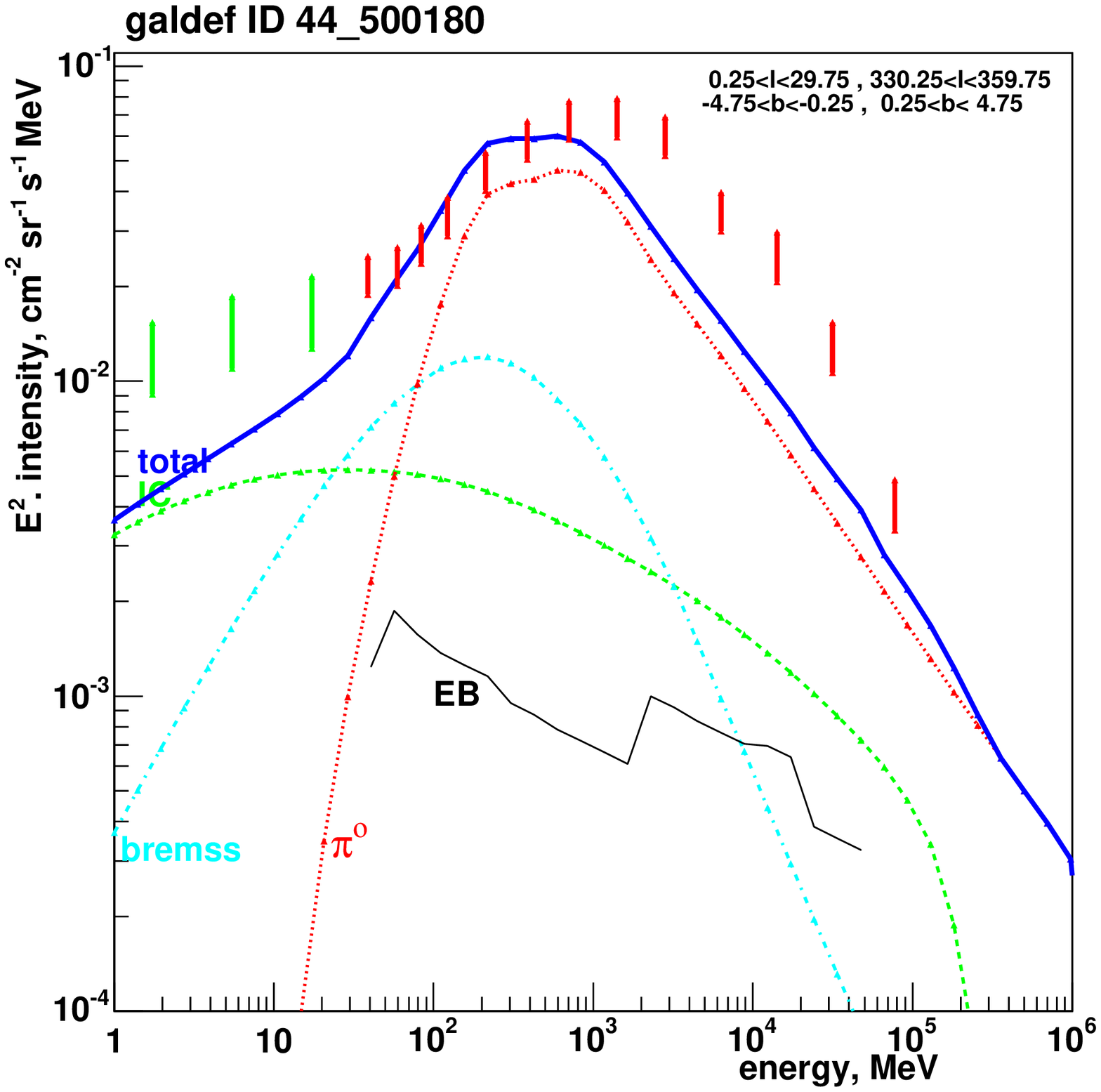}
\includegraphics[width=58mm]{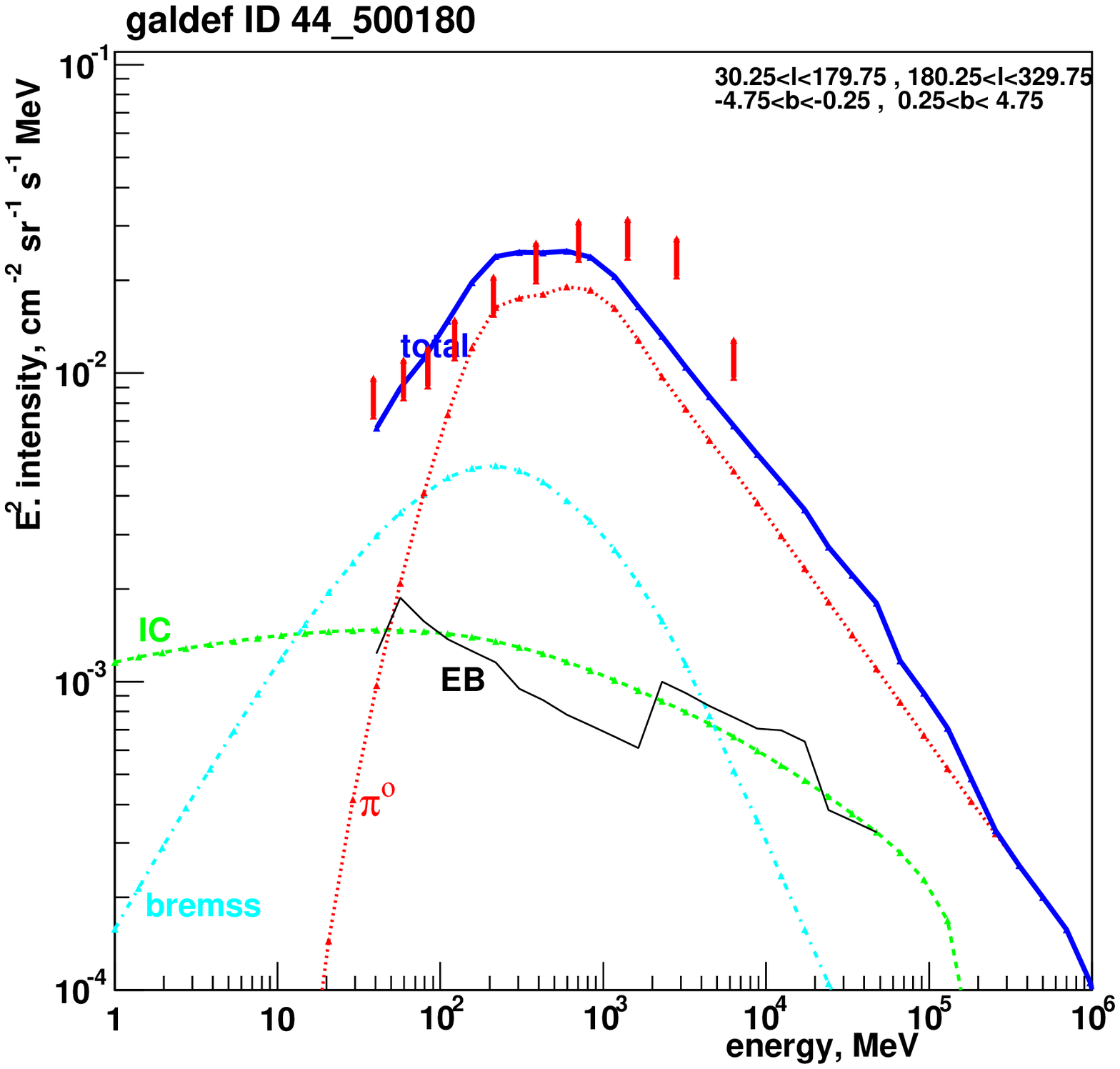}
\includegraphics[width=58mm]{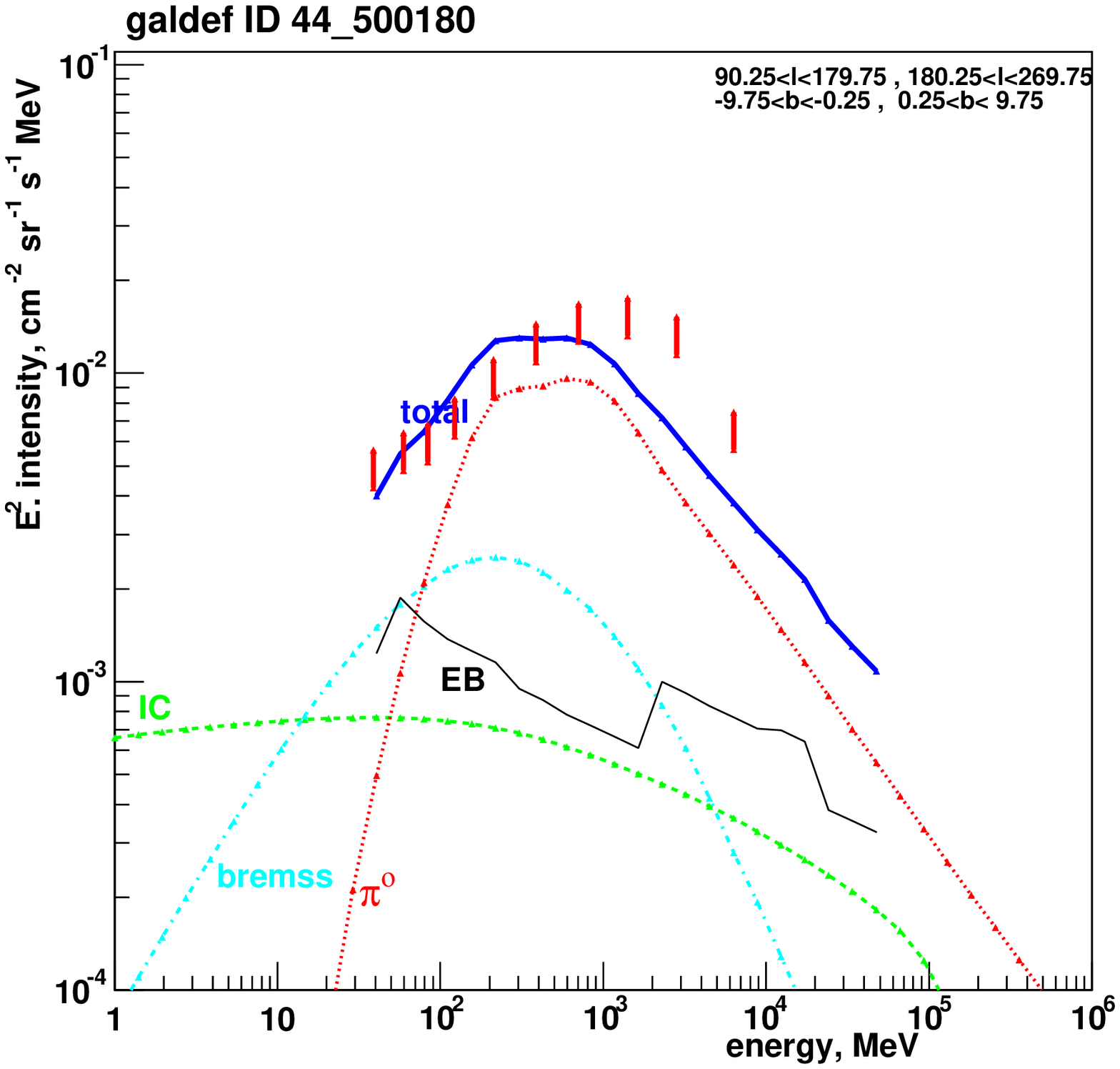}
\includegraphics[width=58mm]{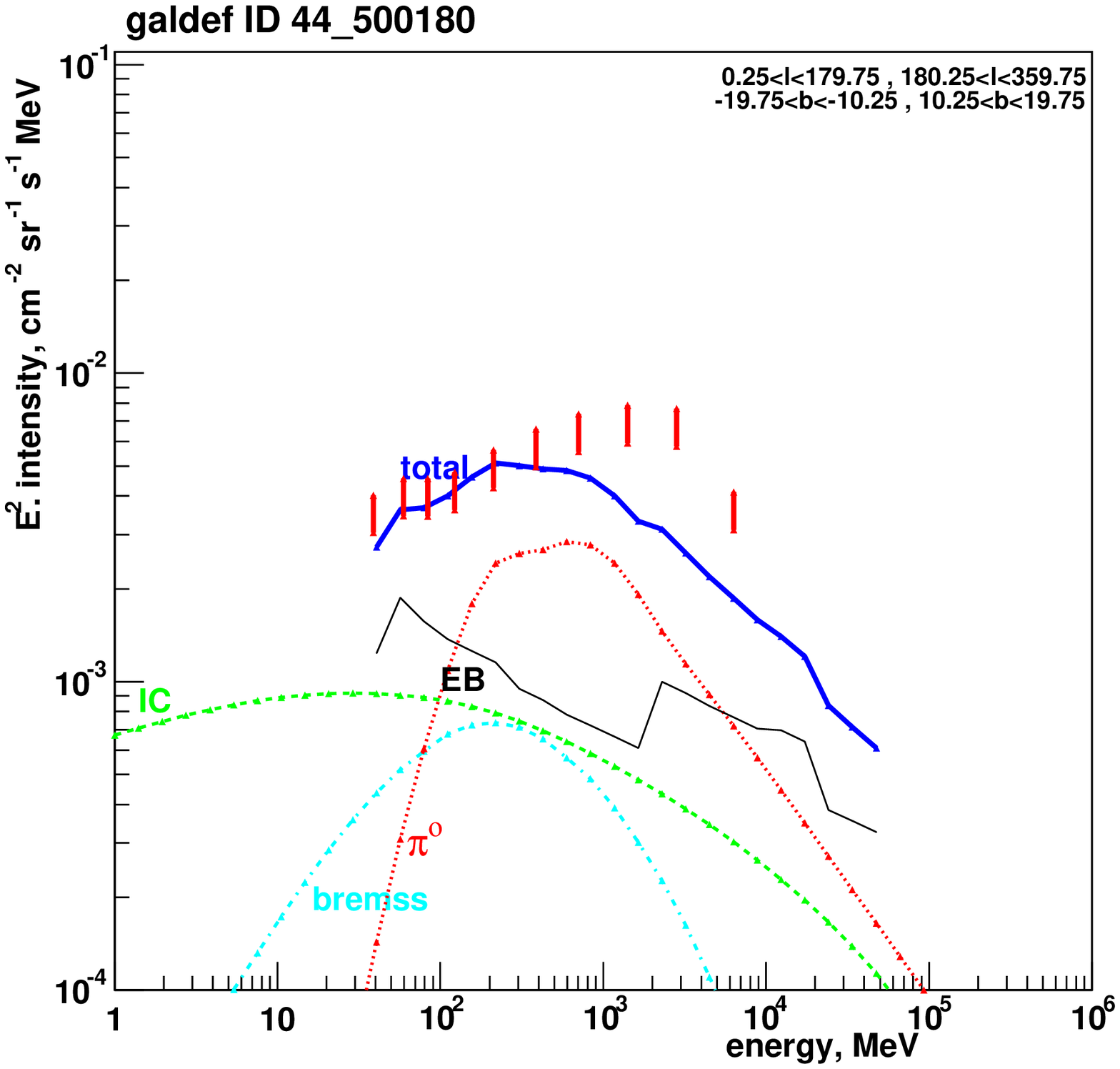}
\includegraphics[width=58mm]{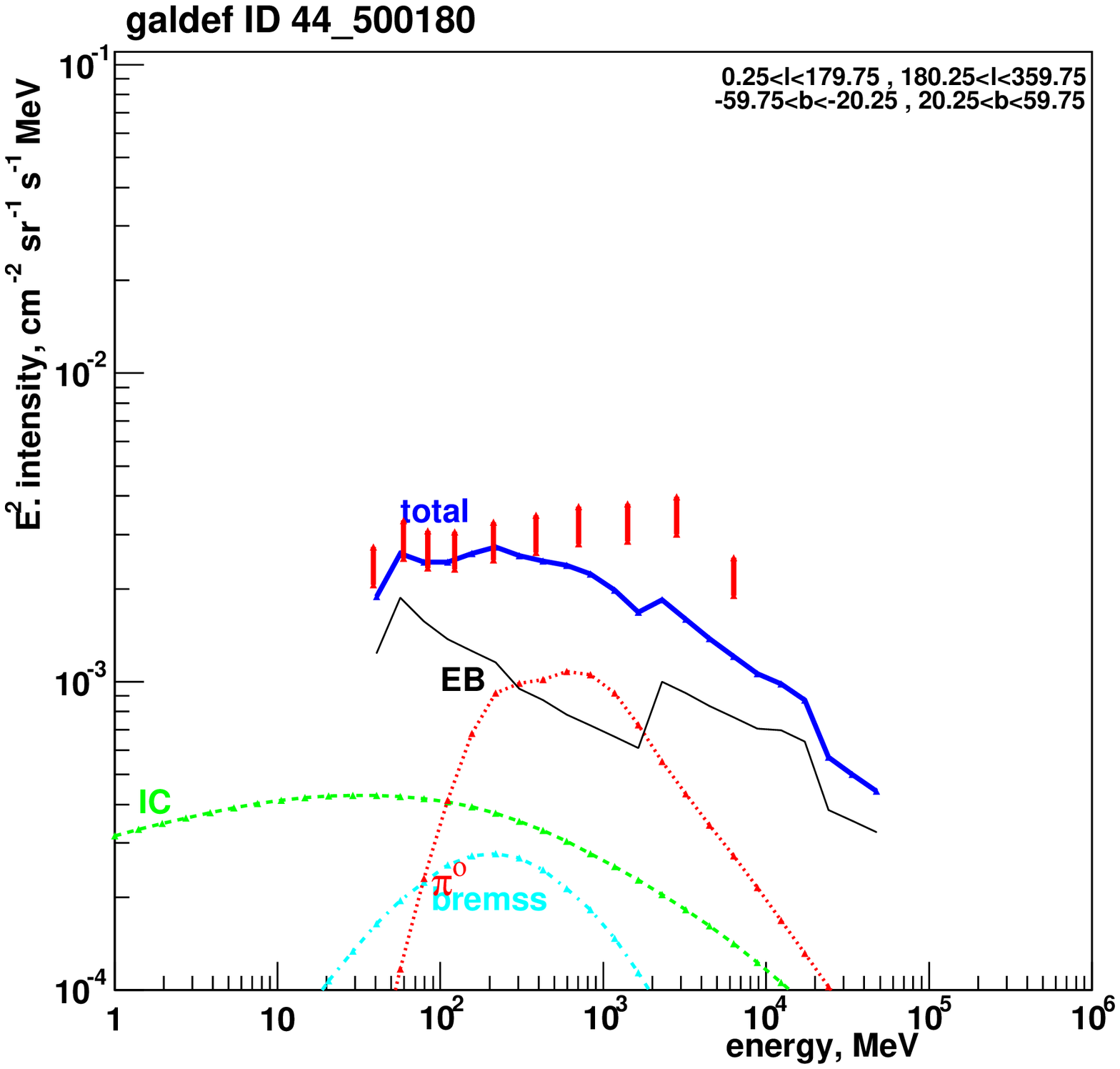}
\includegraphics[width=58mm]{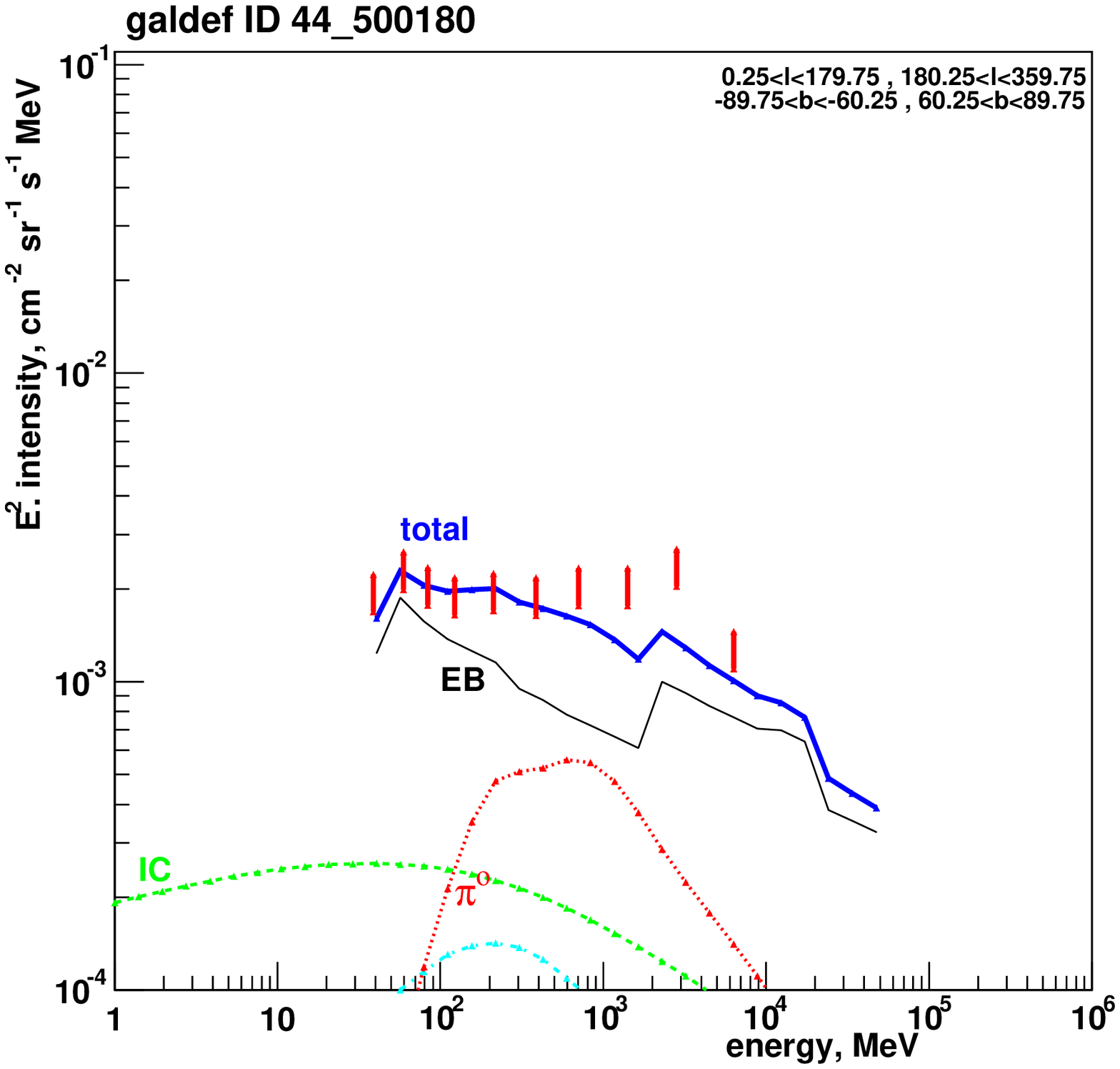}

\caption{\gray spectrum of conventional model (44-500180)
for the sky regions described in Table \ref{sky_regions}:
top row H--A--B, middle row C--D--E, bottom F.
The model components are: \piodecay (dots, red), IC (dashes,
green), bremsstrahlung (dash-dot, cyan), \EB\ (thin solid, black),
total (thick solid, blue).
EGRET data: red vertical bars.  COMPTEL data: green vertical bars.
NB \EB\ is added to the total prediction for the EGRET energy range only.
\label{fig:spectrum_conventional} }
\end{figure*}

We start by repeating the test of the ``conventional'' model; the
\gray spectra in the 7 test regions are shown in  Fig.\
\ref{fig:spectrum_conventional}. As required by the ``conventional''
tag, the proton and electron spectra are consistent with the locally
observed spectra (Figs.\ \ref{fig:protons}, \ref{fig:electrons}).
This is the same ``conventional'' model as in \citet{SMR00}, with
updated nucleon spectra, but because we compare with a more complete
set of EGRET data than in \citet{SMR00}, the discrepancies become more
explicit, and we can check whether they arise only in particular sky
regions. Note that IC plays only a minor role in this type of model.
As found in previous work, the GeV energy range shows an excess
relative to that predicted; what is now evident is that this excess
appears {\it in all latitudes/longitude ranges}. This is consistent
with the results of \citet{hunter97} and \citet{digel01}. It already
shows that the GeV excess is not a feature restricted to the Galactic
ridge or the gas-related emission. Further it is clear that a simple
upward rescaling of the \piodecay component will not improve the fit
in any region, since the observed peak is at higher energies than the
\piodecay peak. In other words, since the spectrum is very different
from \piodecay even at intermediate latitudes, a substantial IC
component is required.
The $\chi^2$ values (Table~\ref{chisq}) confirm the visual conclusion
that this model is unacceptable.

\placefigure{fig:spectrum_conventional}

Note that this version of the ``conventional'' spectrum is
nevertheless in rather {\it better} agreement with EGRET data than in
\citet{SMR00}, due to inclusion of secondary positrons/electrons,
general improvements in the model (e.g., $\pi^0$-decay, improved gas
data) and the EGRET data treatment \citep[direct use of the count and
exposure data instead of the model-fitting analysis
of][]{strongmattox96}. The improvement is especially evident in the
30--100 MeV range, where secondary positrons/electrons make a
substantial contribution (see Section \ref{sec:secondary}).

\begin{figure}[!tb]
\centering
\includegraphics[width=9cm]{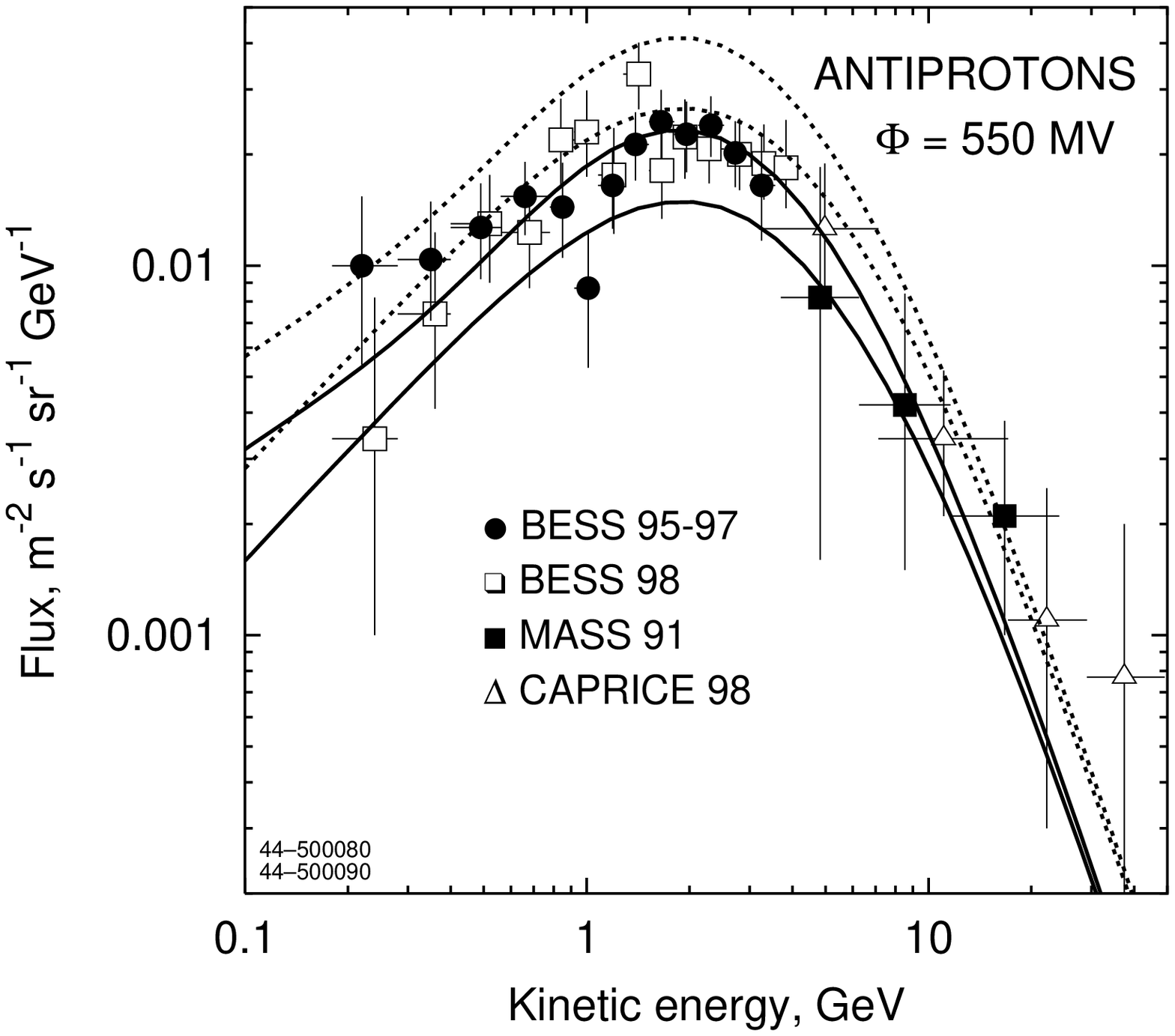}
 \caption{Antiproton flux as calculated in conventional
and optimized models 
compared with the data (upper curve -- LIS, lower -- modulated to 550 MV).
The lines are coded as in Fig.~\ref{fig:protons}.
Data: BESS 95-97 \citep{Orito00}, BESS 98 \citep{Asaoka02},
MASS 91 \citep{basini99}, CAPRICE 98 \citep{boezio01}.
\label{fig:pbars}}
\end{figure}

A test against antiproton and positron data also shows ``excesses''.
The conventional model with reacceleration is known \citep{M02} to
produce a factor of $\sim$1.5 ($\sim$2.5$\sigma$) less antiprotons at
2 GeV than measured by BESS \citep{Orito00}. The antiproton spectrum
for the conventional model is shown in Fig.\ \ref{fig:pbars}.
Positron data, though scattered, also show some ``excess'' at high
energies (Fig.\ \ref{fig:positrons}). It is thus clear that the
``excesses'' in GeV $\gamma$-rays in all directions, in GeV
antiprotons, and in positrons above several GeV found in the
conventional model indicate that the \emph{average} high-energy
proton flux in the Galaxy should be more intense \emph{or} our
reacceleration model is invalid \emph{or} there is a contribution
from unconventional sources (e.g. dark matter). For more discussion
of antiproton and positron tests see Section \ref{sec:secondary}.

\placefigure{fig:pbars}

\placefigure{fig:positrons}

In the ``SNR source'' scenario of \citet{volk} the \gray spectrum in
the inner Galaxy is attributed to an additional population of
unresolved SNR, but this component cannot explain the excess observed
at high latitudes, and hardly in the outer
Galaxy\footnote{\citet{volk} did not address the question of regions
away from the inner Galaxy.}.
The presence of the GeV excess in all sky regions is also a problem
for the suggestion by \citet{aharonian00} of a hard proton spectrum
in the inner Galaxy.
These explanations are therefore by themselves
insufficient, although they could give a contribution.

\section{Hard electron injection spectrum model}

This model is essentially as in \cite{SMR00}, recomputed with the
current GALPROP code. The main feature is the electron injection index
of 1.9. Comparison of the spectra in the 7 sky areas (Fig.\
\ref{fig:spectrum_hard}) show that this model reproduces the GeV
excess except in the inner Galaxy (region A) where it is still too
low. However the spectral shape is not well reproduced. More
significant, comparing with the new EGRET data above 10 GeV in the
inner Galaxy, the spectrum is much too hard. 
The $\chi^2$ values (Table~\ref{chisq}) confirm the visual conclusion
that this model is only marginally unacceptable.

Fig.\ \ref{fig:electrons}
compares the locally observed electron spectrum with that from the
model; the deviation at high energies is much larger than expected
from the 3D study by \citet{SM01}. As discussed in the Introduction,
there are therefore a number of reasons to lead us to consider this
model as after all untenable.

\section{Optimized model}

Since the conventional model fails to reproduce the observed \gray
spectrum, and the hard electron spectrum model is untenable, we use
the diffuse \grays themselves to obtain an optimized solution. The
\emph{average} interstellar electron spectrum is sufficiently
uncertain that we can look for a ``solution'' involving a less
drastic change in the electron injection spectrum than the hard
electron injection spectrum model. We find that an injection spectrum
of electrons with a steepening from 1.5 to 2.42 at 20 GeV (see Table
\ref{model_parameters1}) produces sufficient curvature in the inverse
Compton spectrum to explain the observed shape of the \gray spectrum,
{\it provided the electron spectrum is suitable normalized upwards by
a factor of about 4 relative to the locally observed spectrum}. The
proton injection spectrum is also normalized upward, by a factor 1.8;
it has the same shape as for the electrons, as a function of rigidity,
but the break energy is 10 GeV instead of 20 GeV. It has exactly the
same slope above 10 GeV as the conventional proton spectrum.
(The proton re-normalization factor 1.8 
is not taken \emph{ad hoc} but is chosen
to reproduce the antiproton data, see Section~\ref{sec:secondary}
for more details).

\begin{figure*}[!thb]
\centering 
\includegraphics[width=58mm]{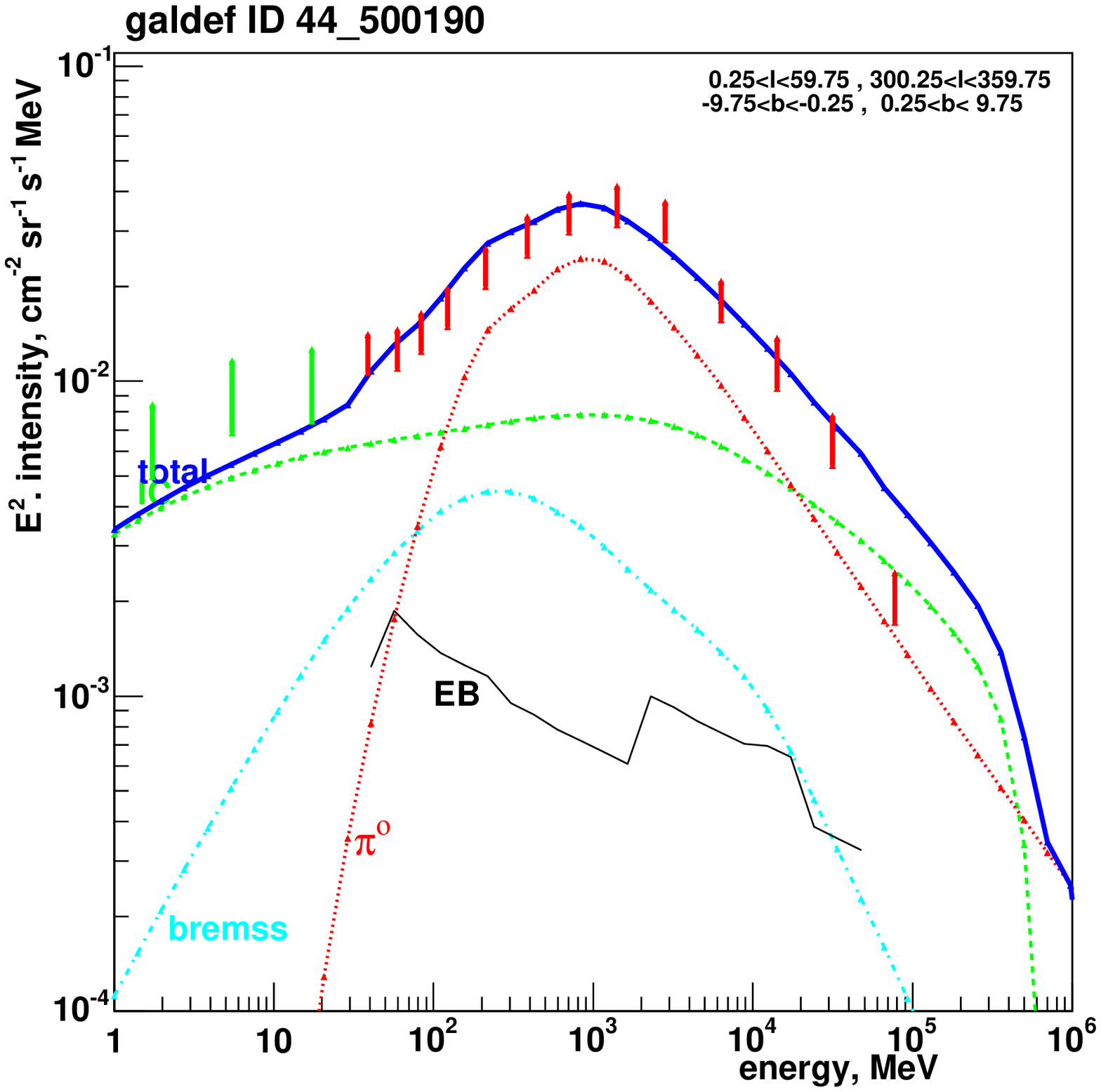}
\includegraphics[width=58mm]{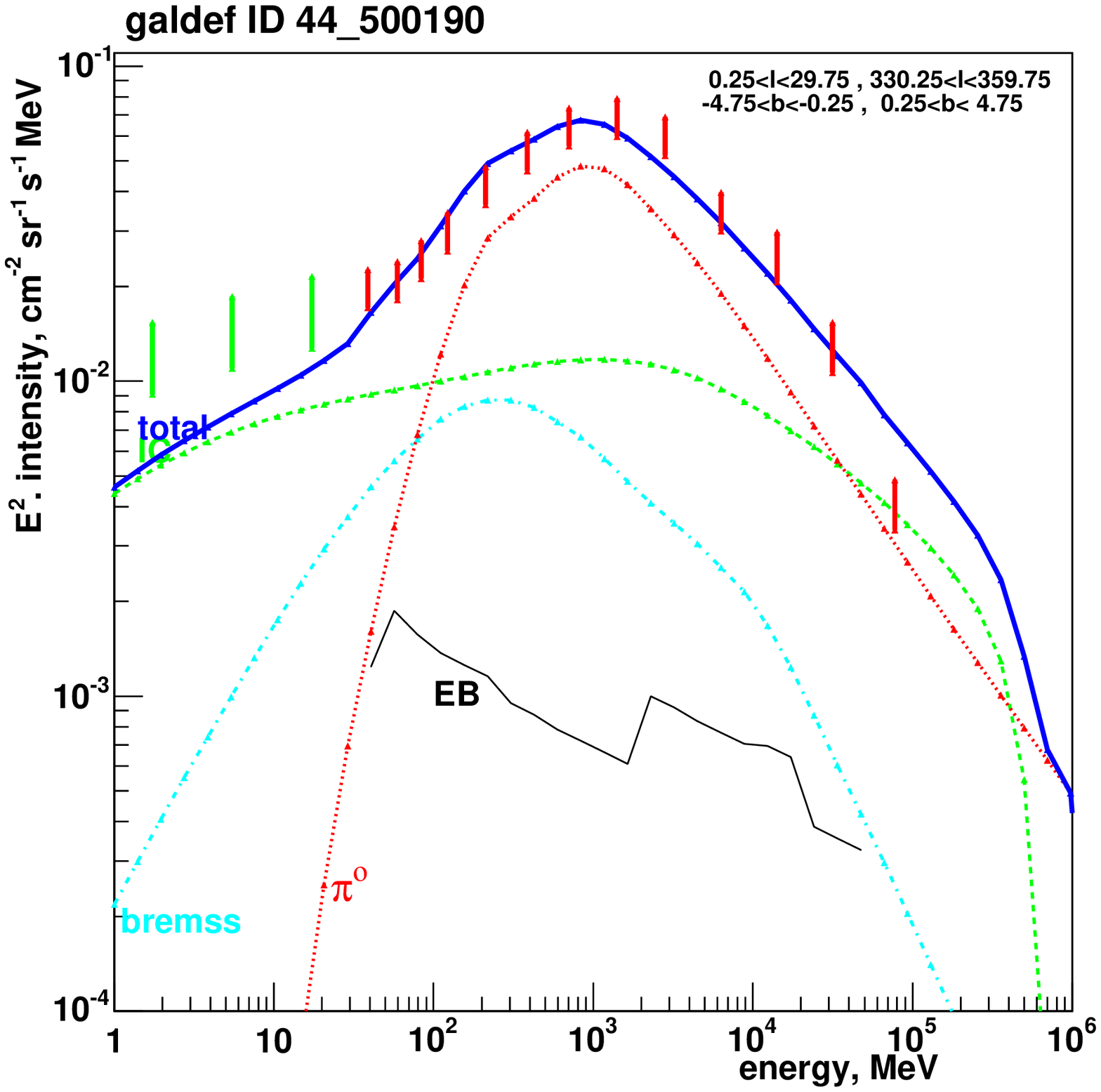}
\includegraphics[width=58mm]{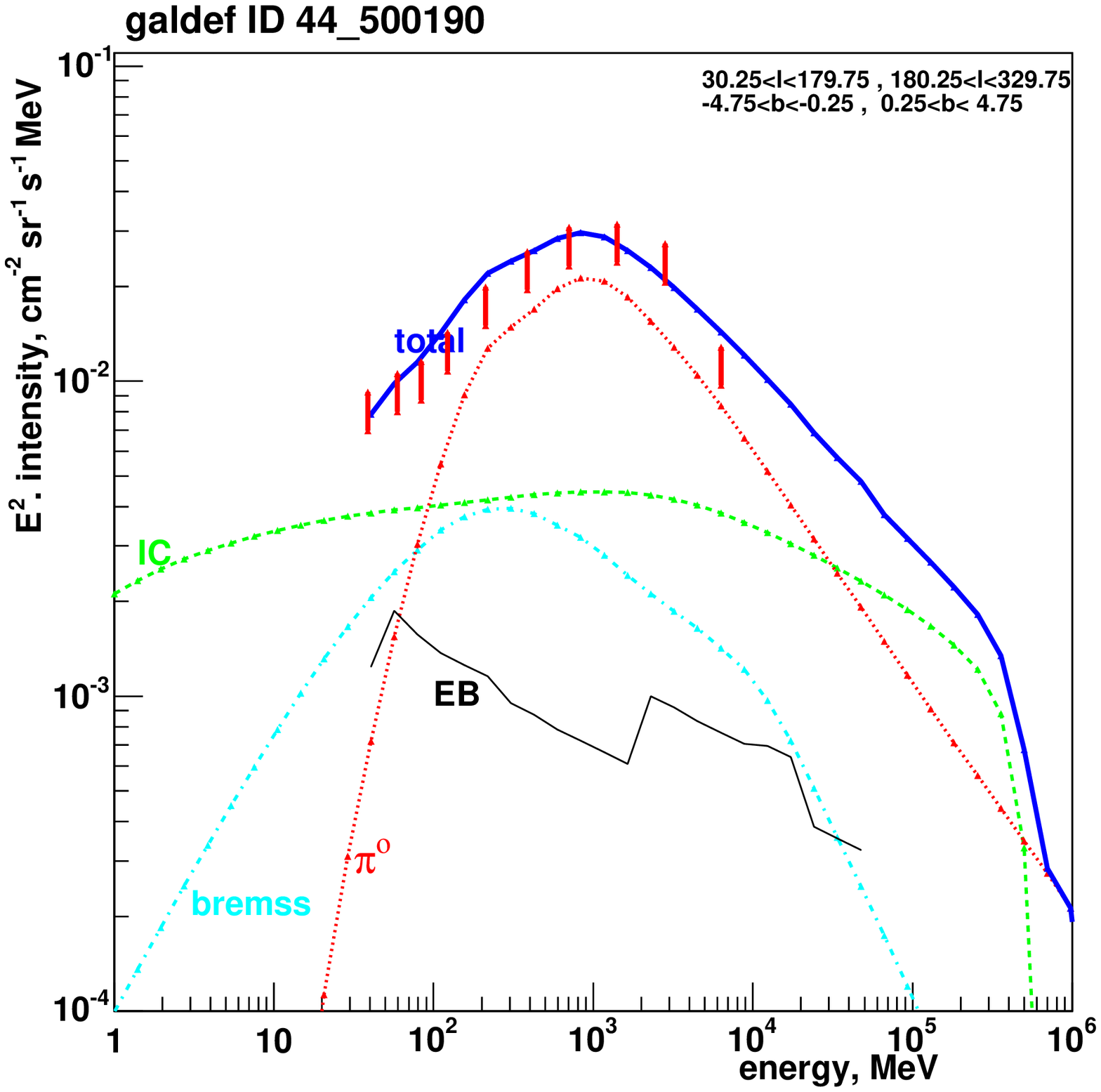}
\includegraphics[width=58mm]{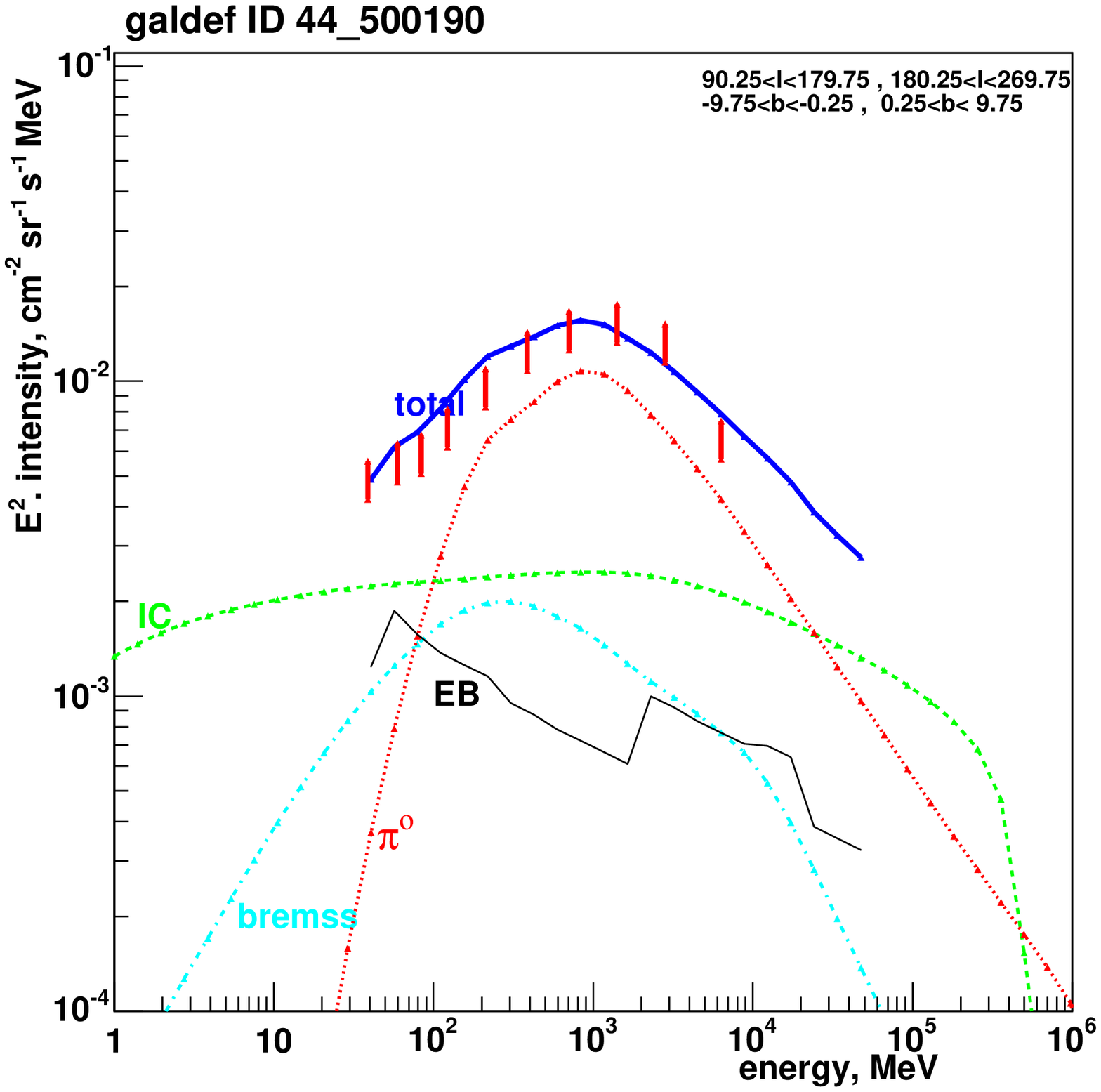}
\includegraphics[width=58mm]{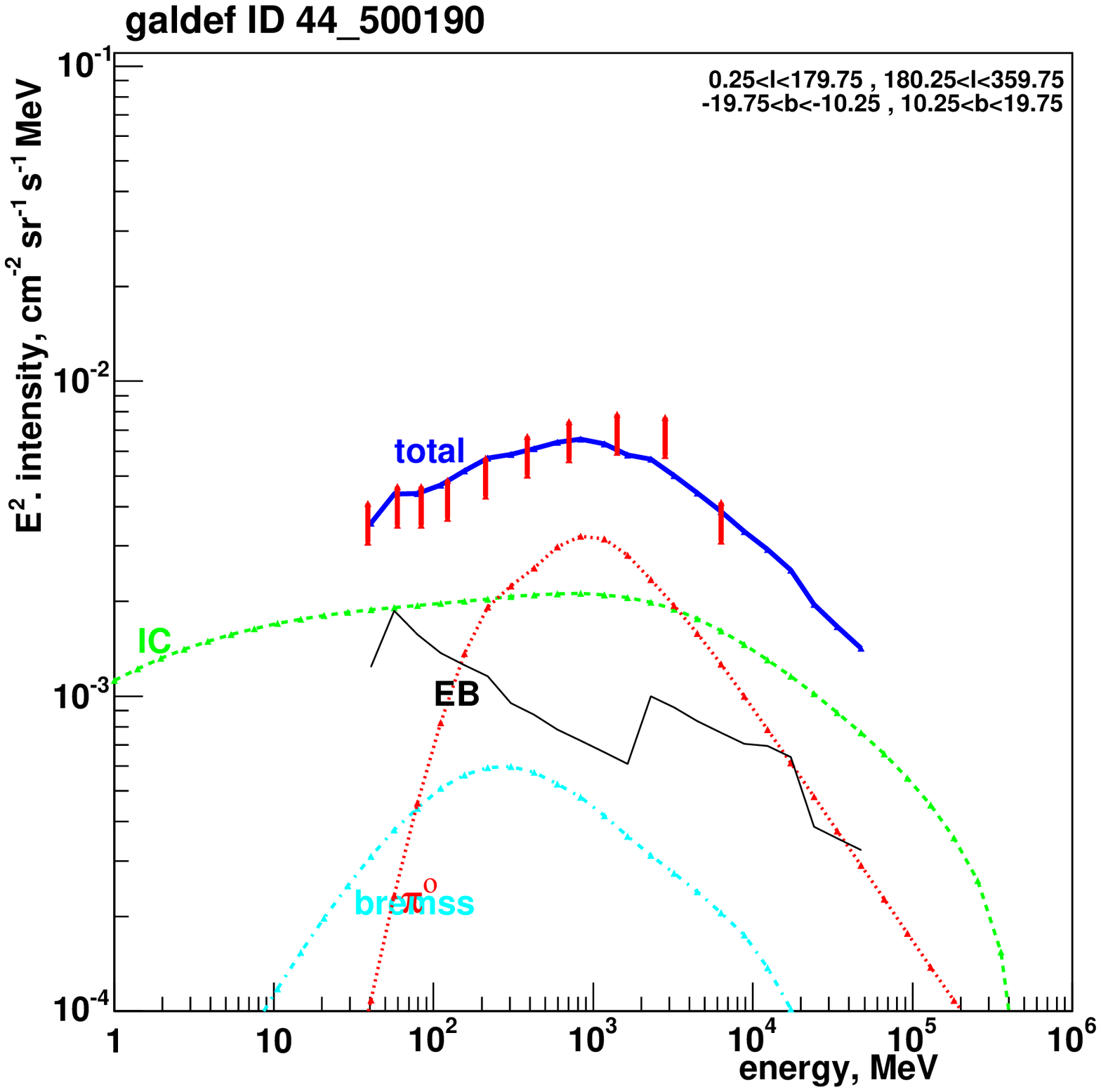}
\includegraphics[width=58mm]{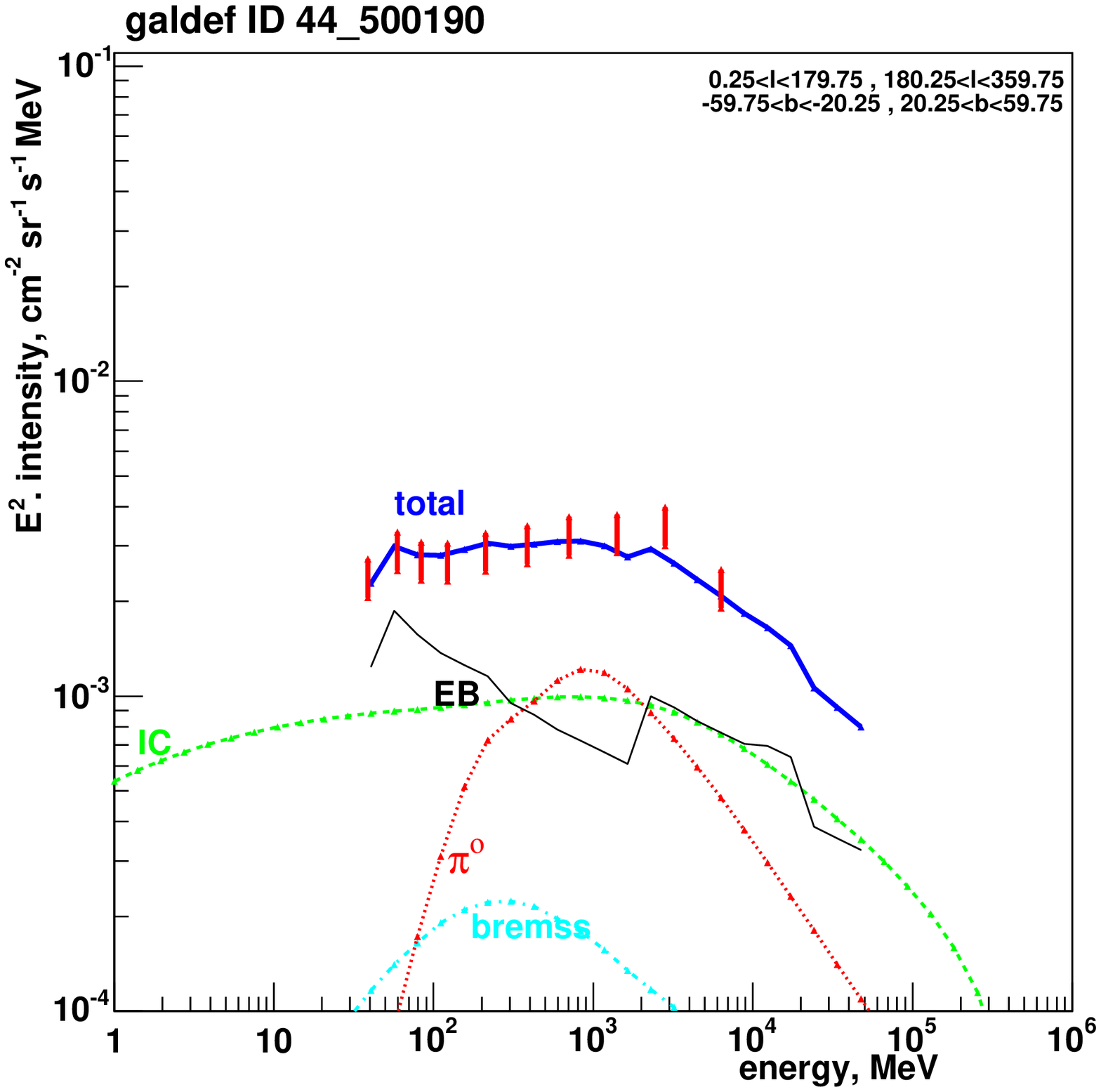}
\includegraphics[width=58mm]{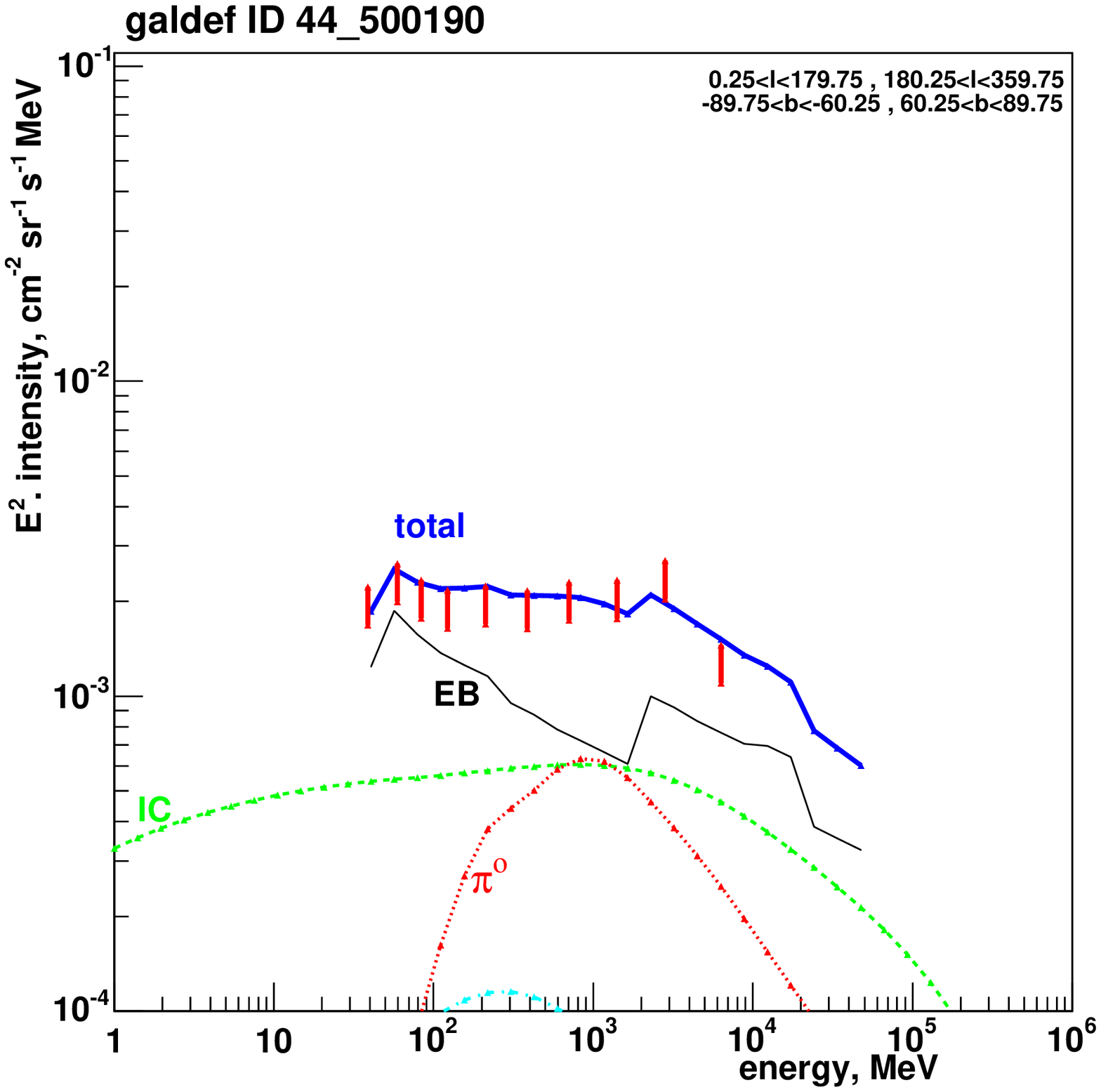}

\caption{\gray spectra of optimized model (44-500190); regions and
coding as for Fig.~\ref{fig:spectrum_conventional}.
\label{fig:spectrum_optimized} }
\end{figure*}

The \gray spectra in the 7 test regions are shown in
Fig.~\ref{fig:spectrum_optimized}. The fits to the observed \gray
spectra are better than for the conventional and hard electron
spectrum models, both in the 1--10 GeV region and above 20 GeV. The
spectra in different regions are satisfactorily reproduced and there
is no longer a significant GeV excess. Hence the spectrum can now be
reproduced from 30 MeV to 100 GeV. The proposed scenario implies a
substantial contribution from IC at all energies, but especially below
100 MeV and above 1 GeV. Also IC dominates at latitudes $|b|>10^\circ$
at all energies.

\begin{figure*}[p]
\centering
\includegraphics[width=58mm]{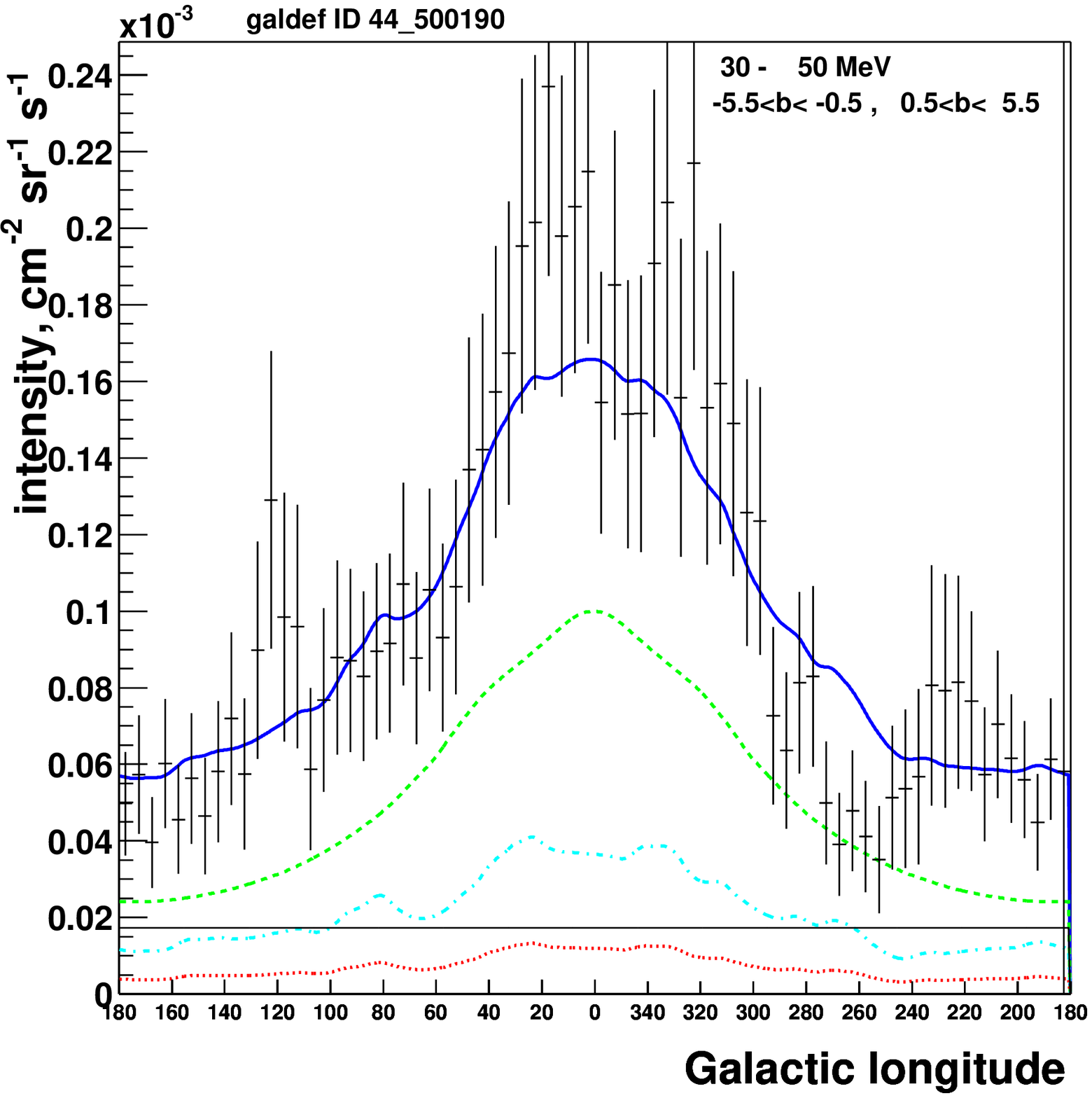}
\includegraphics[width=58mm]{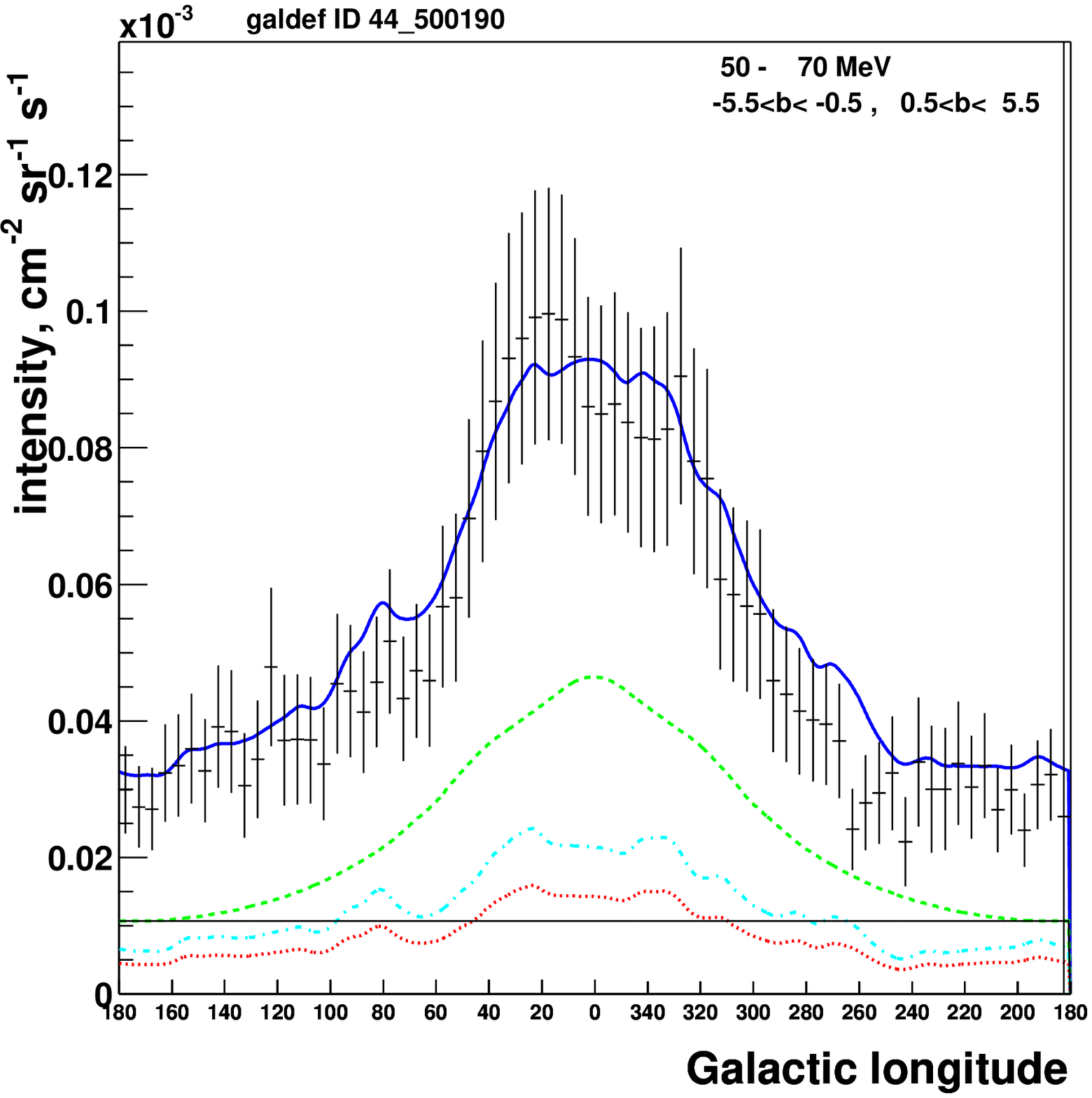}
\includegraphics[width=58mm]{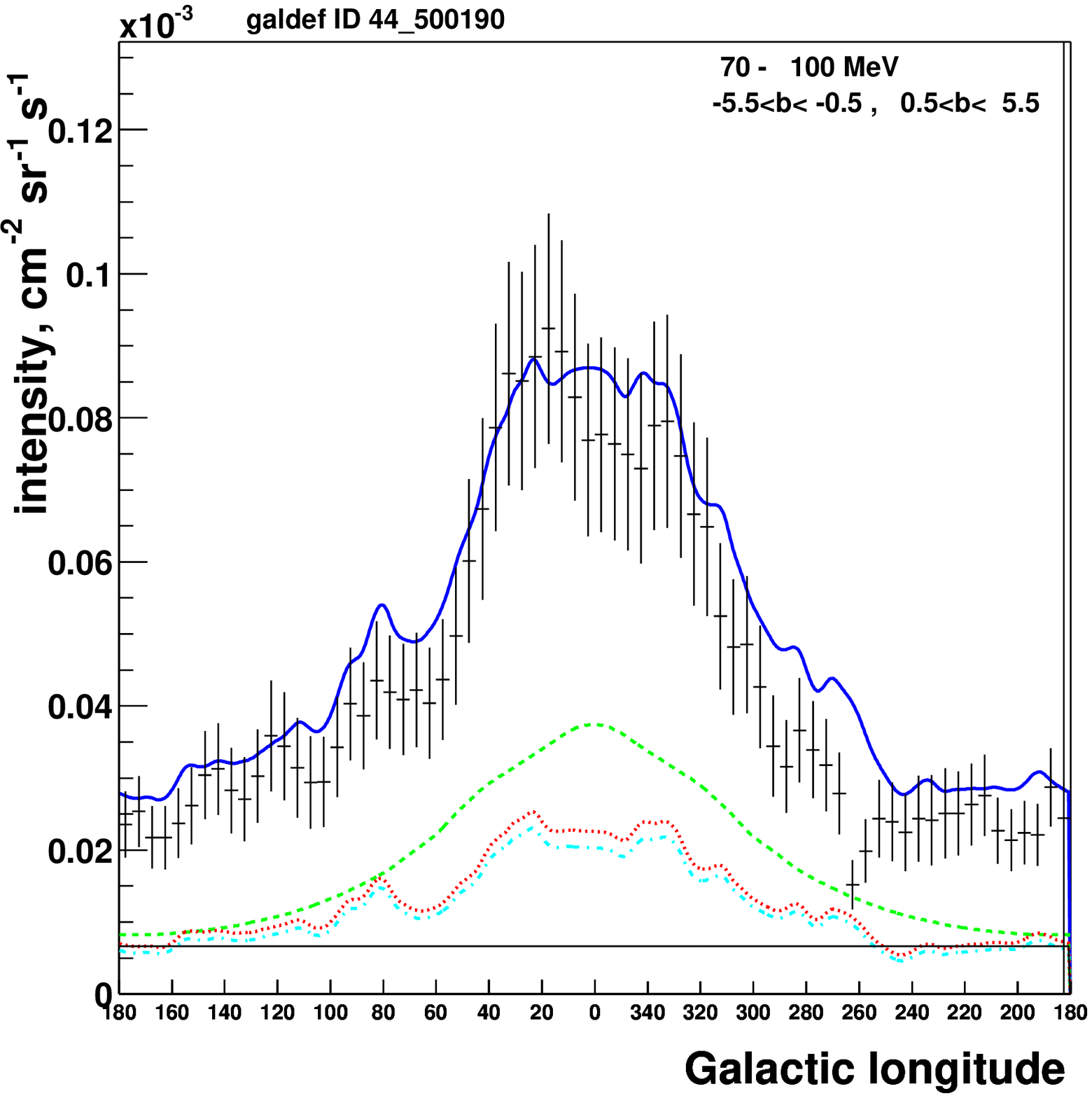}
\includegraphics[width=58mm]{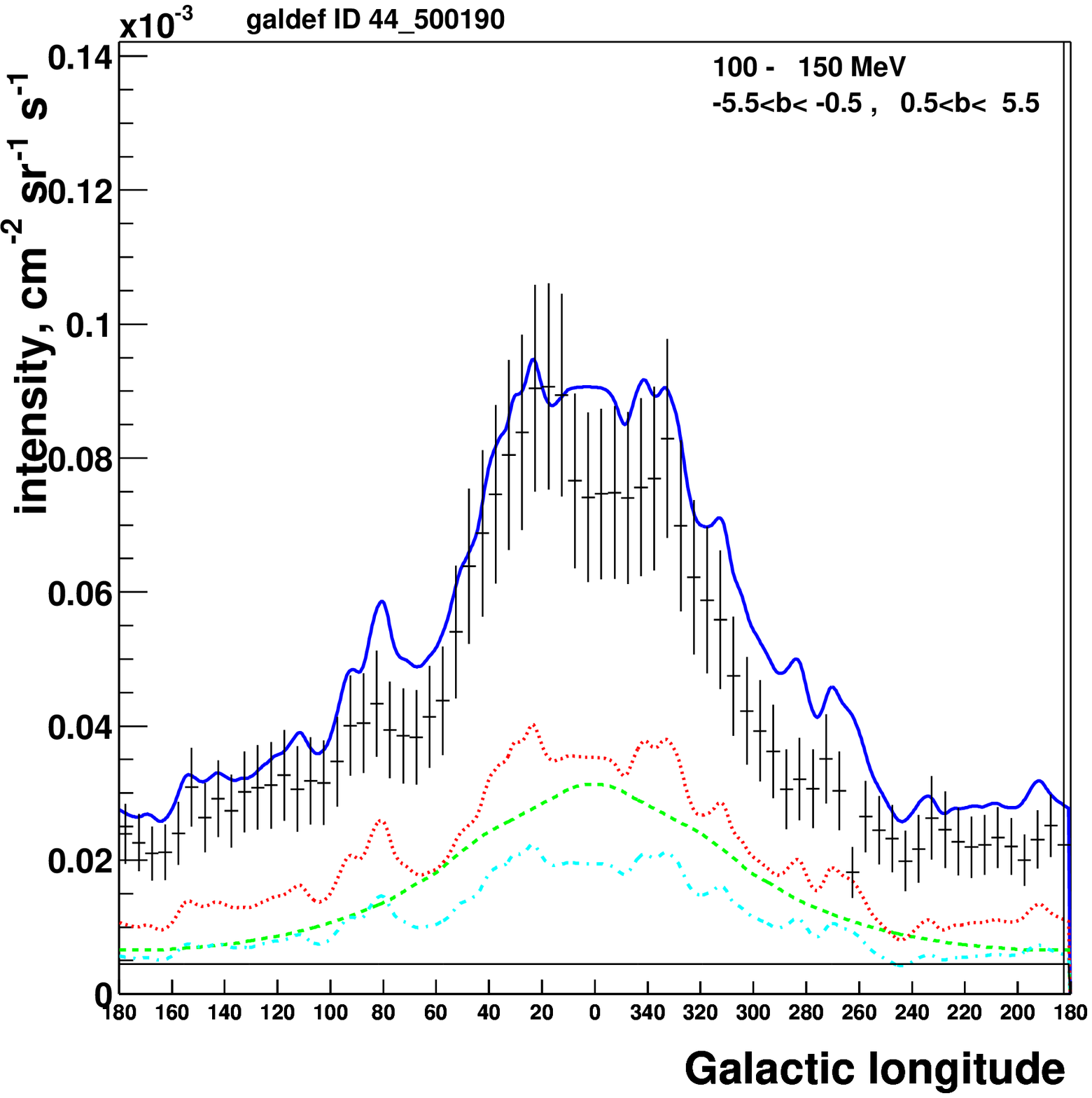}
\includegraphics[width=58mm]{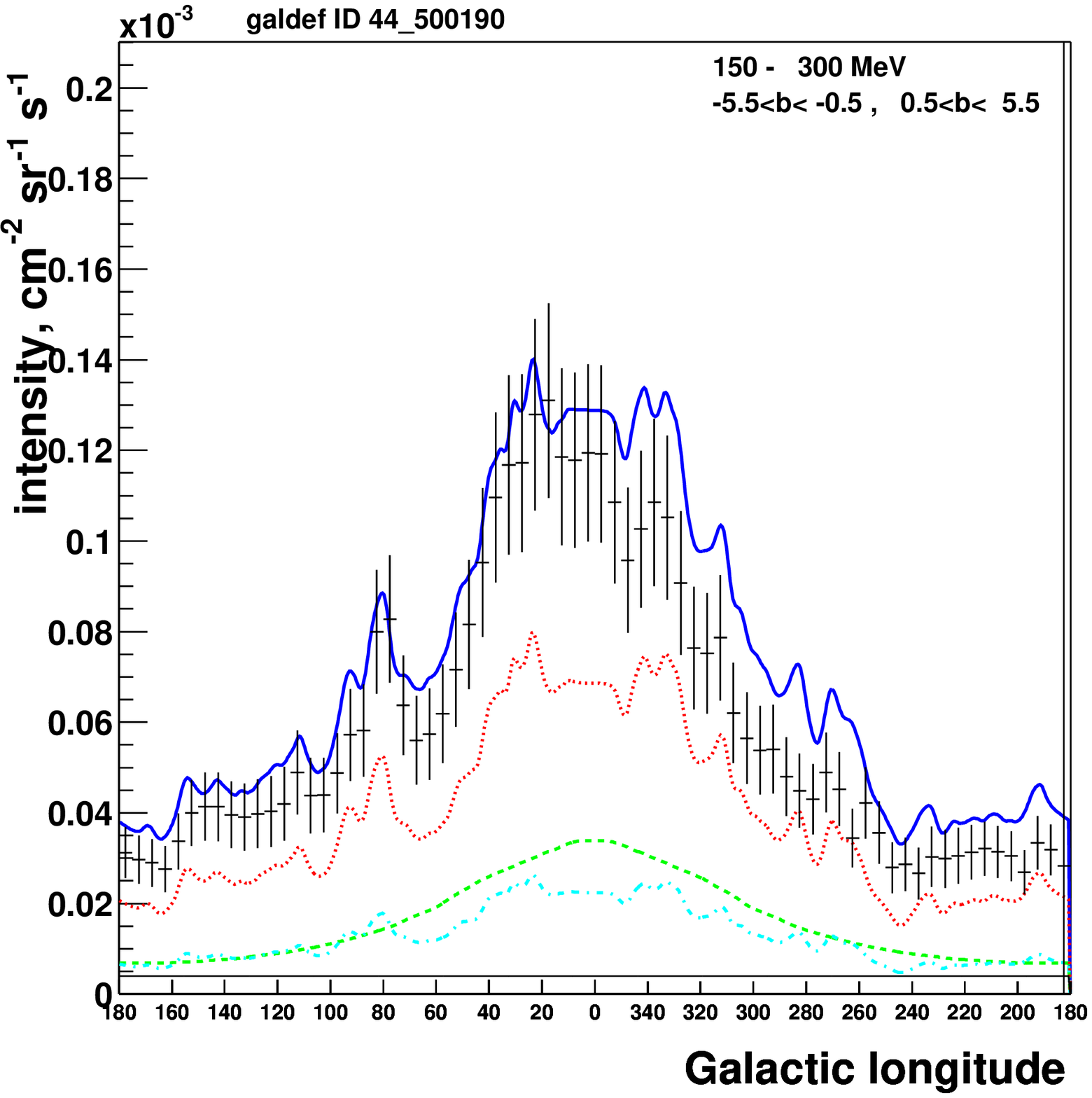}
\includegraphics[width=58mm]{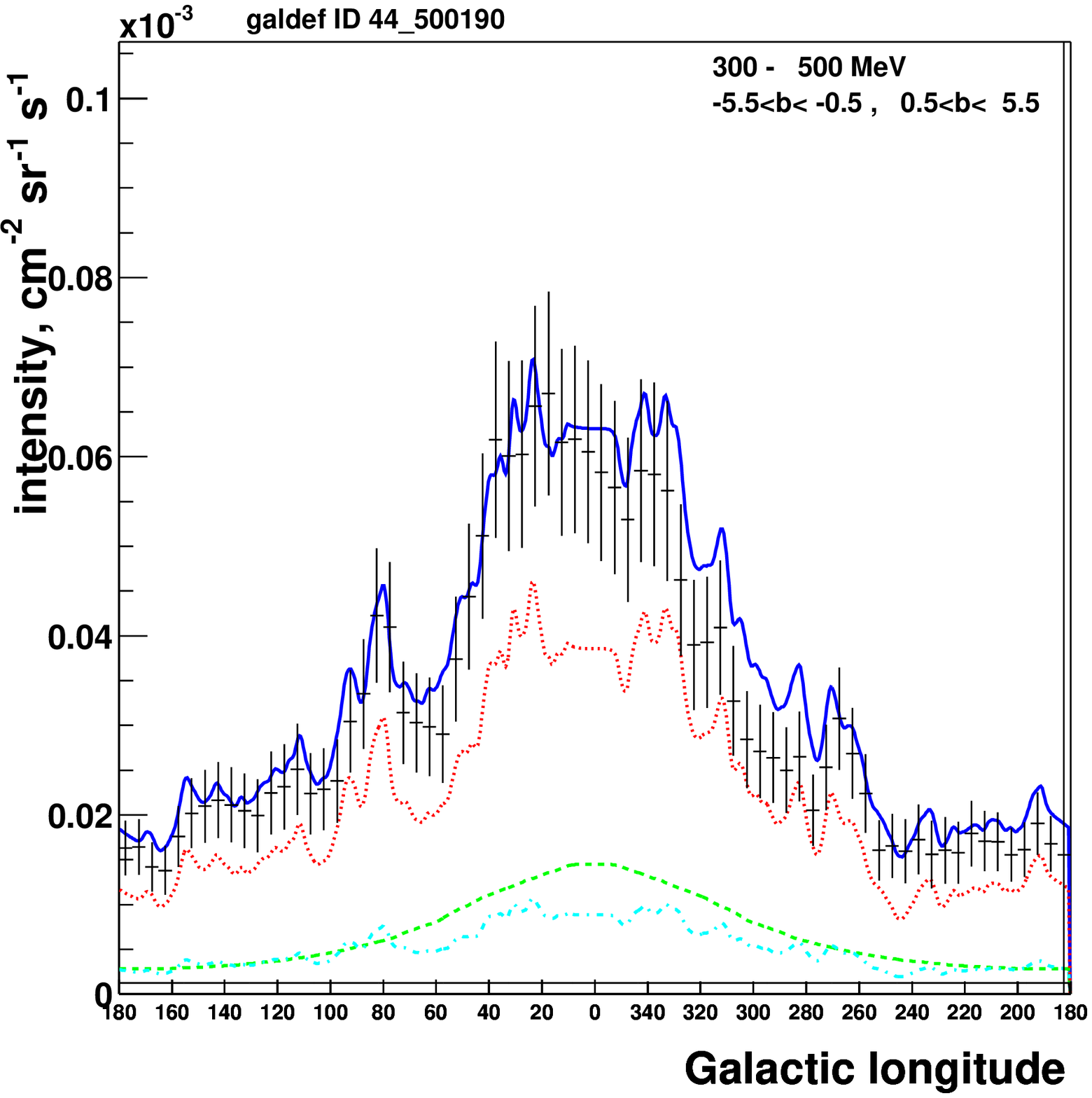}
\includegraphics[width=58mm]{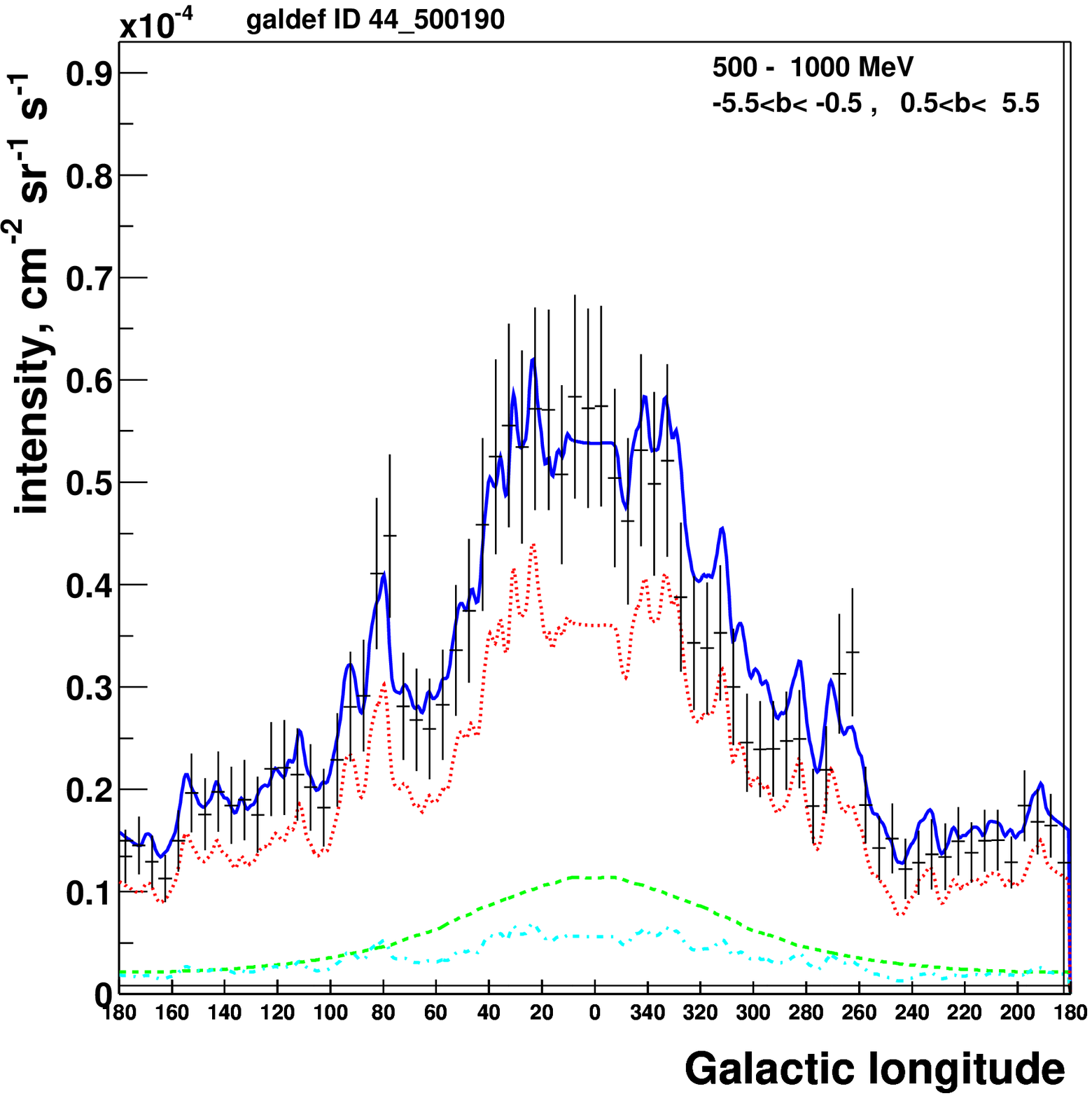}
\includegraphics[width=58mm]{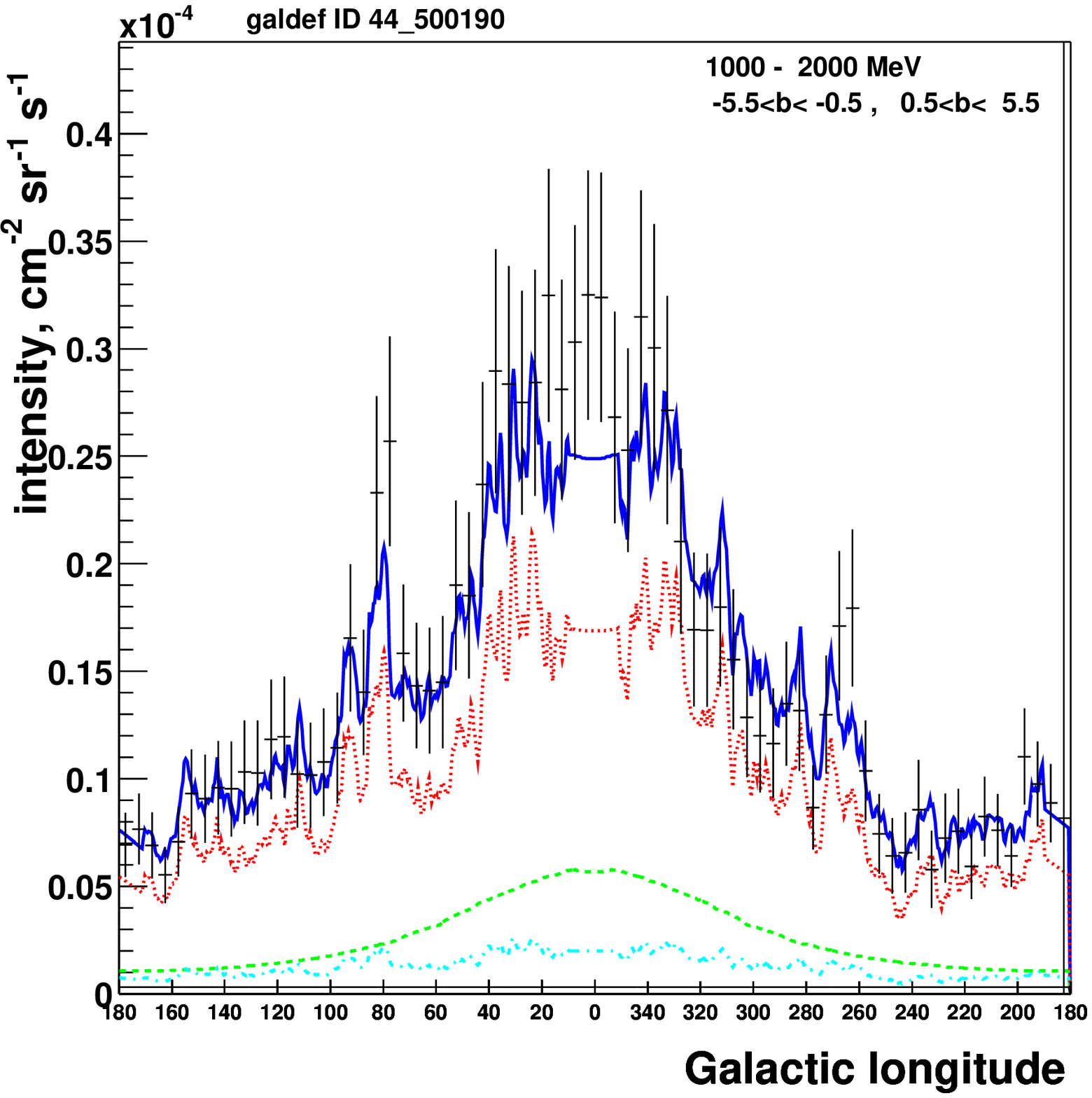}
\includegraphics[width=58mm]{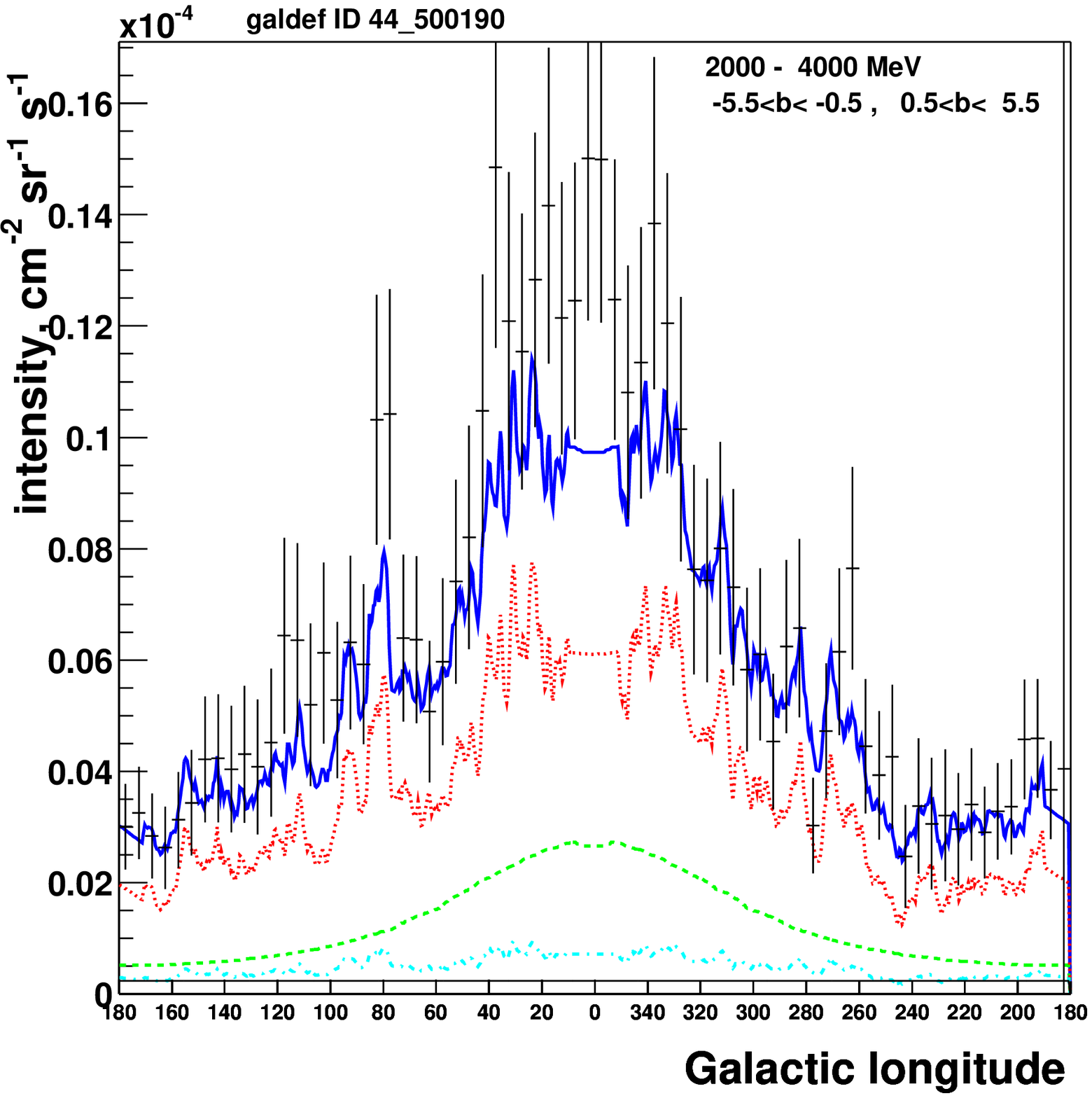}
\includegraphics[width=58mm]{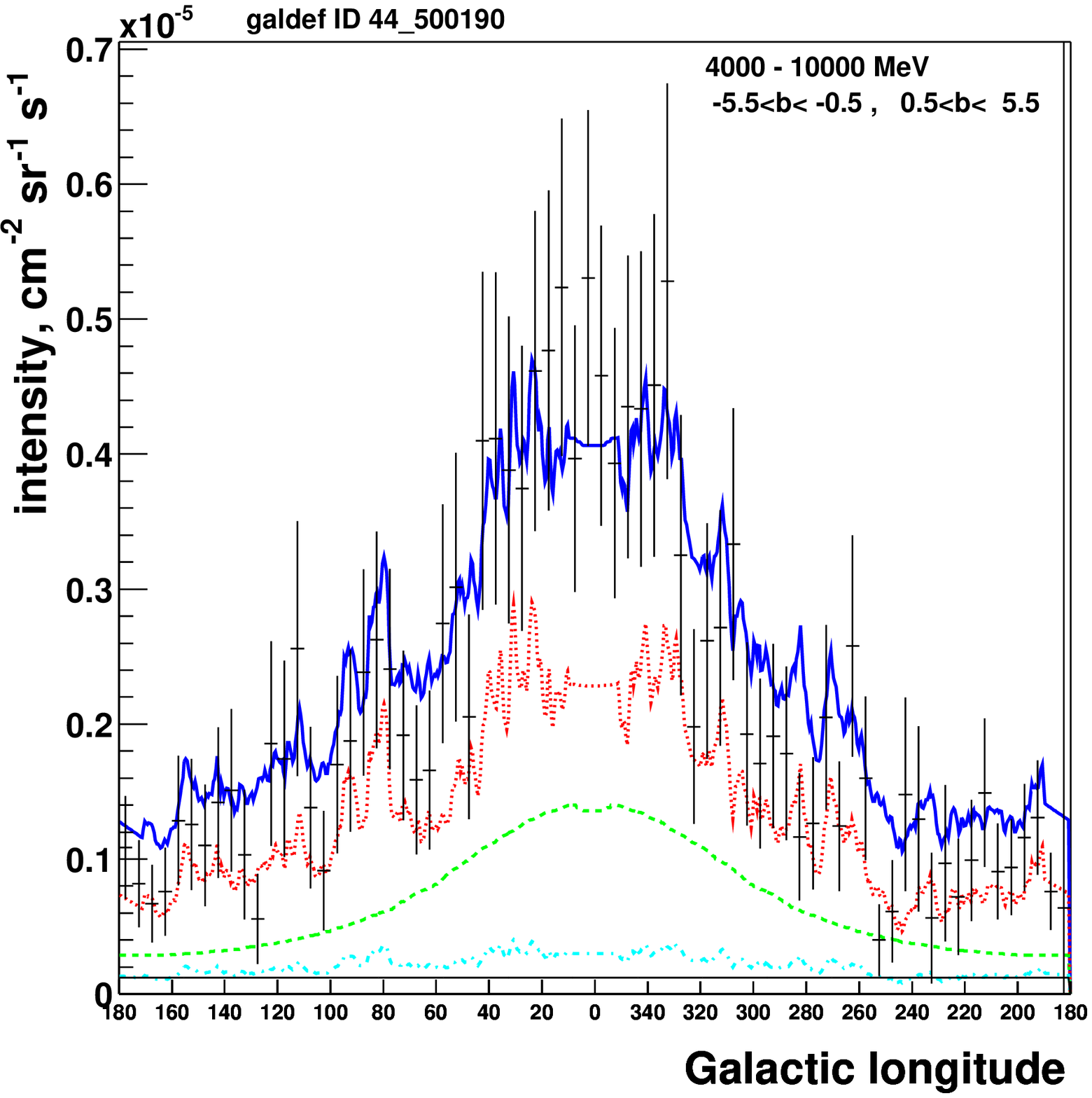}

\caption{Longitude profiles ($|b|<5.5^\circ$) 
for optimized model (500190),
compared with EGRET data in 10 energy ranges, 30 MeV -- 10 GeV.  
Lines are coded as in Fig.~\ref{fig:spectrum_conventional}.}
\label{fig:longitude_profiles_optimized}
\end{figure*}

\begin{figure*}
\centering 
\includegraphics[width=58mm]{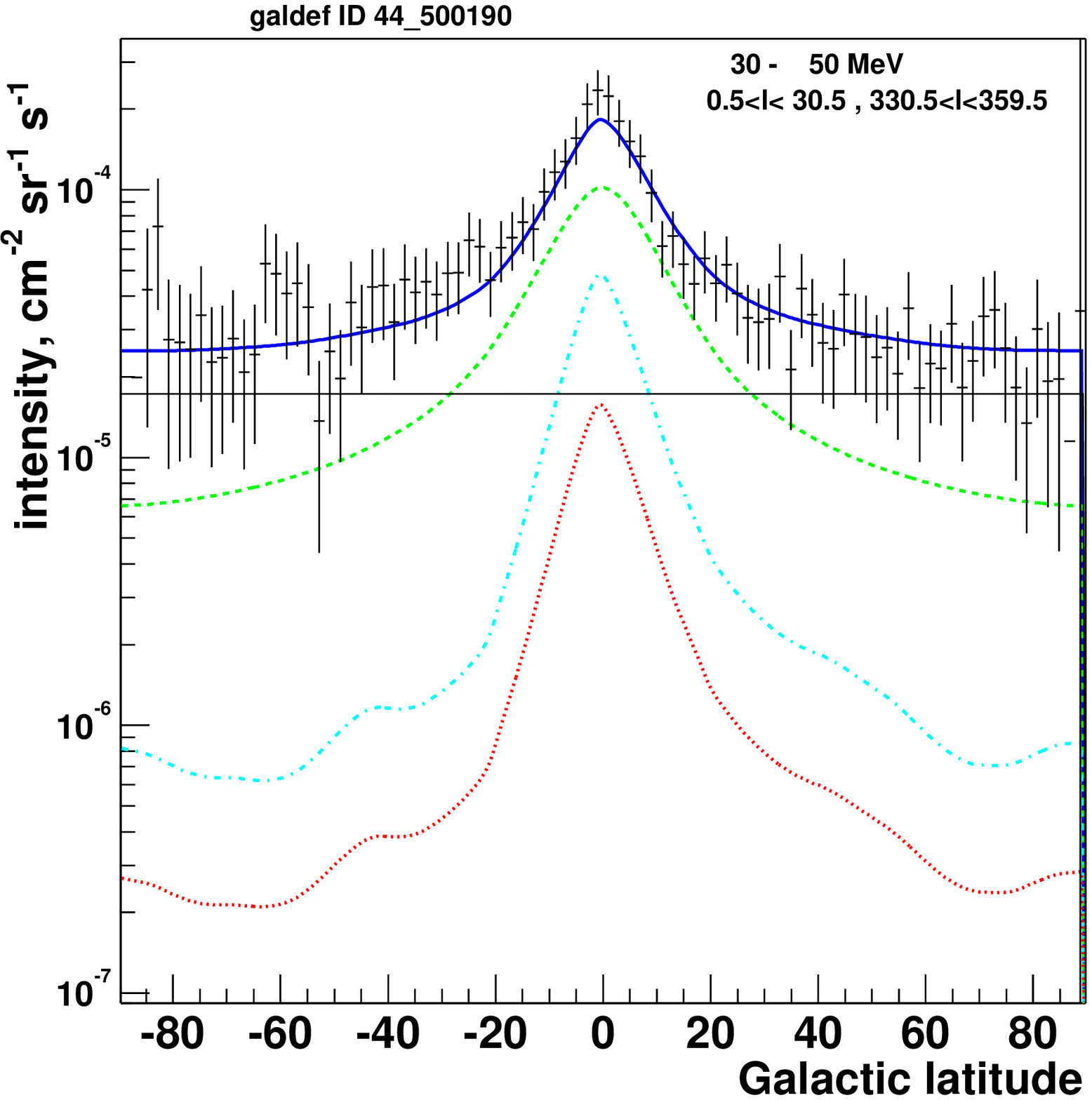}
\includegraphics[width=58mm]{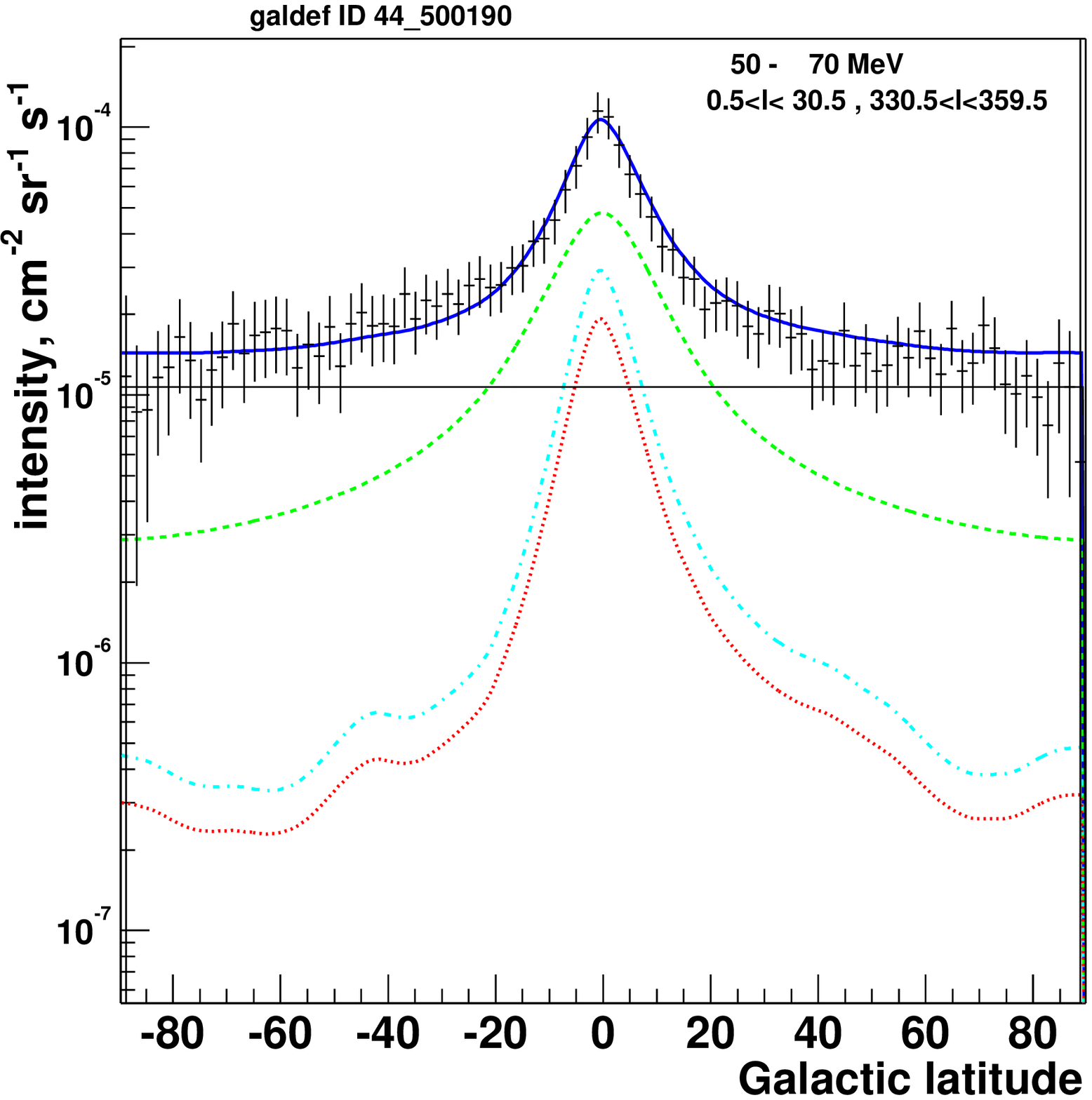}
\includegraphics[width=58mm]{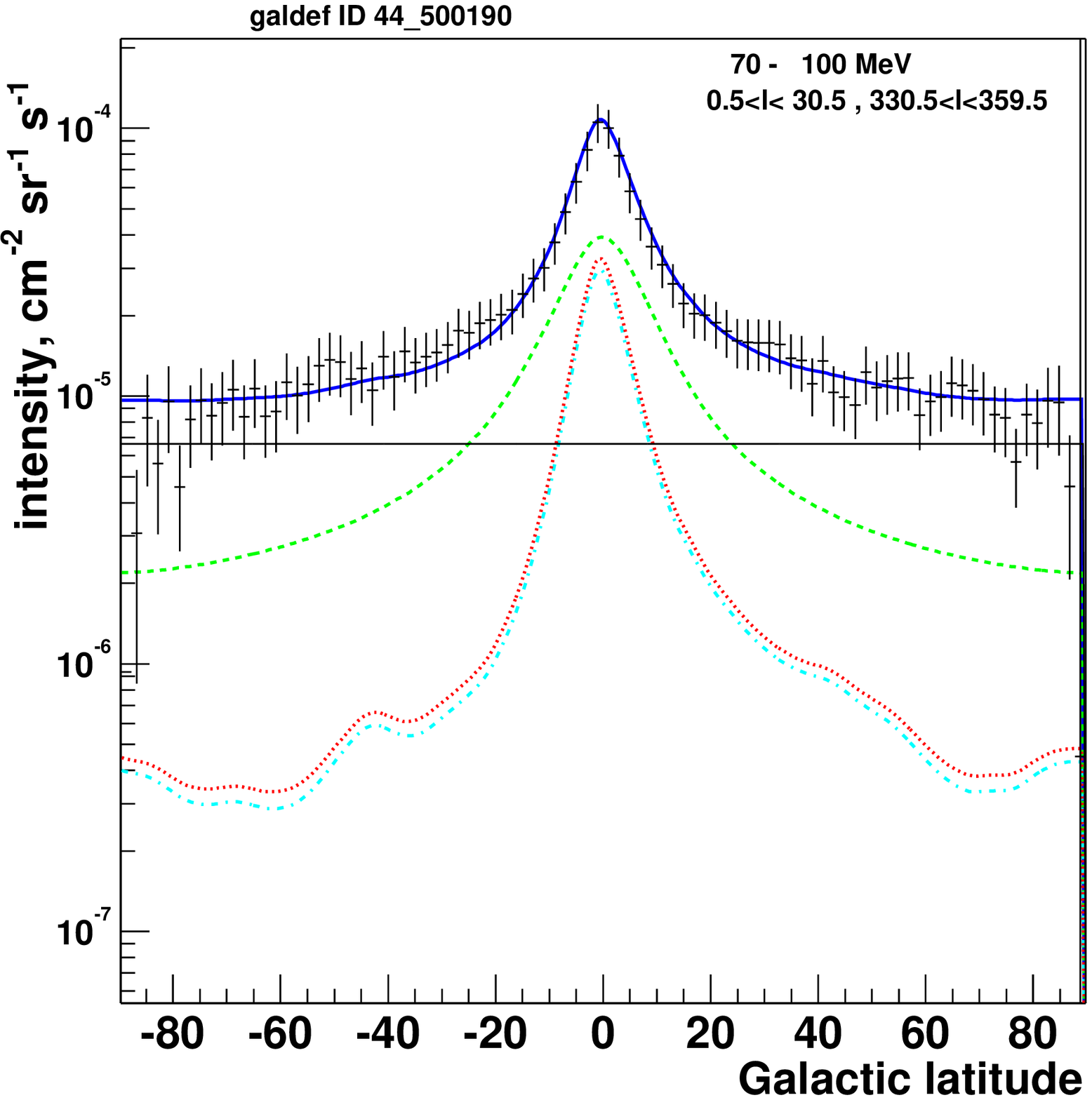}
\includegraphics[width=58mm]{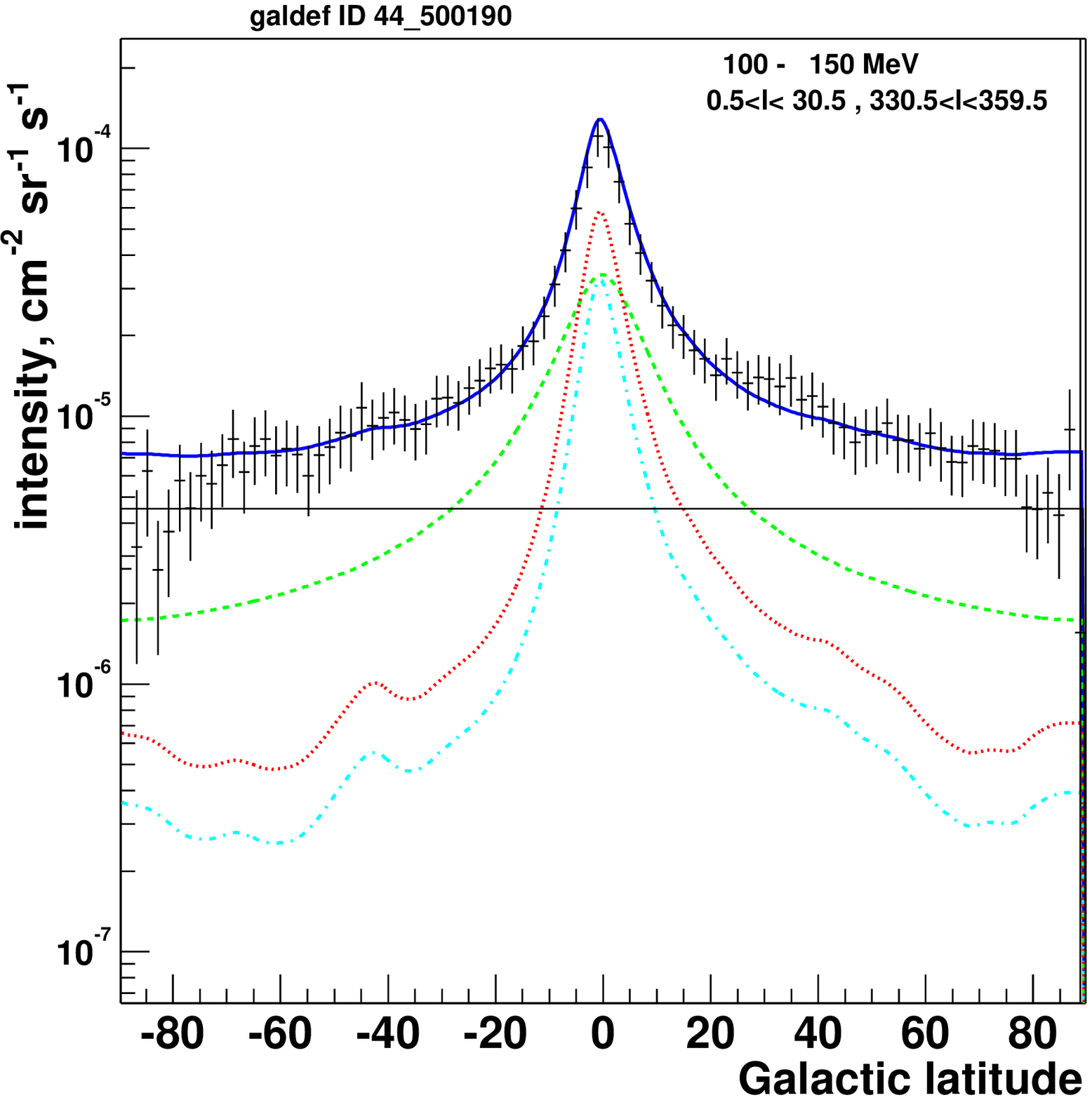}
\includegraphics[width=58mm]{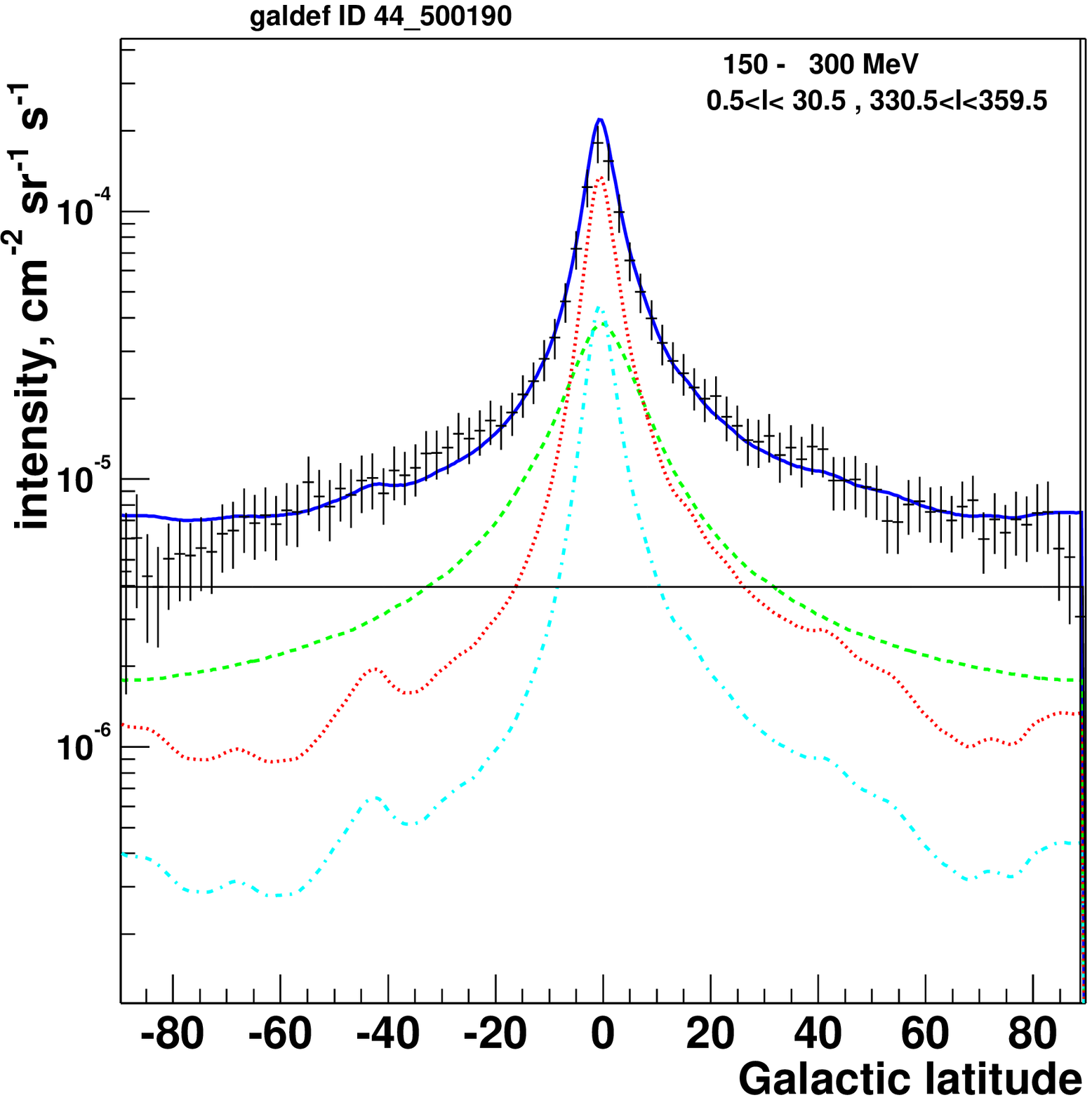}
\includegraphics[width=58mm]{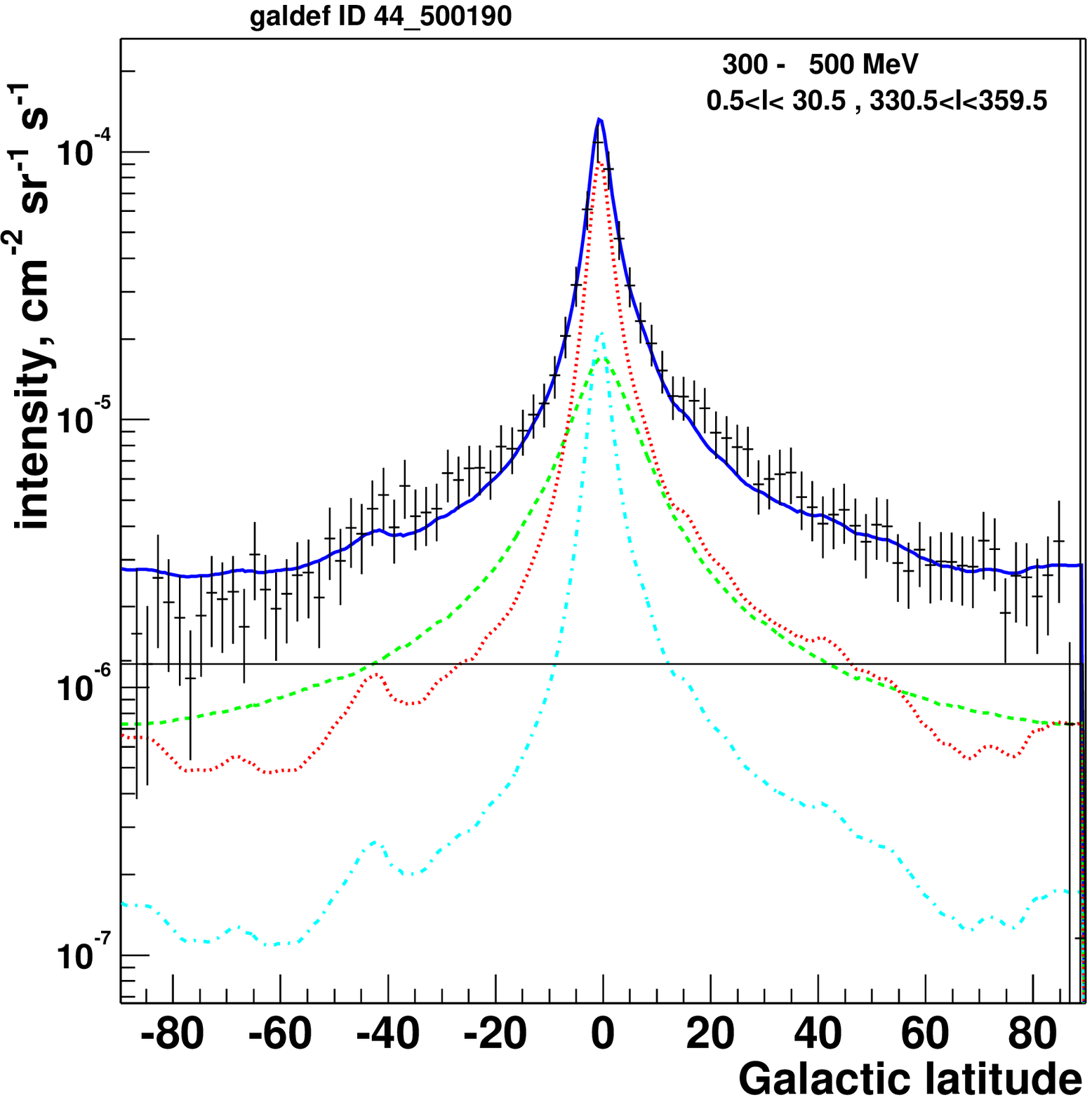}
\includegraphics[width=58mm]{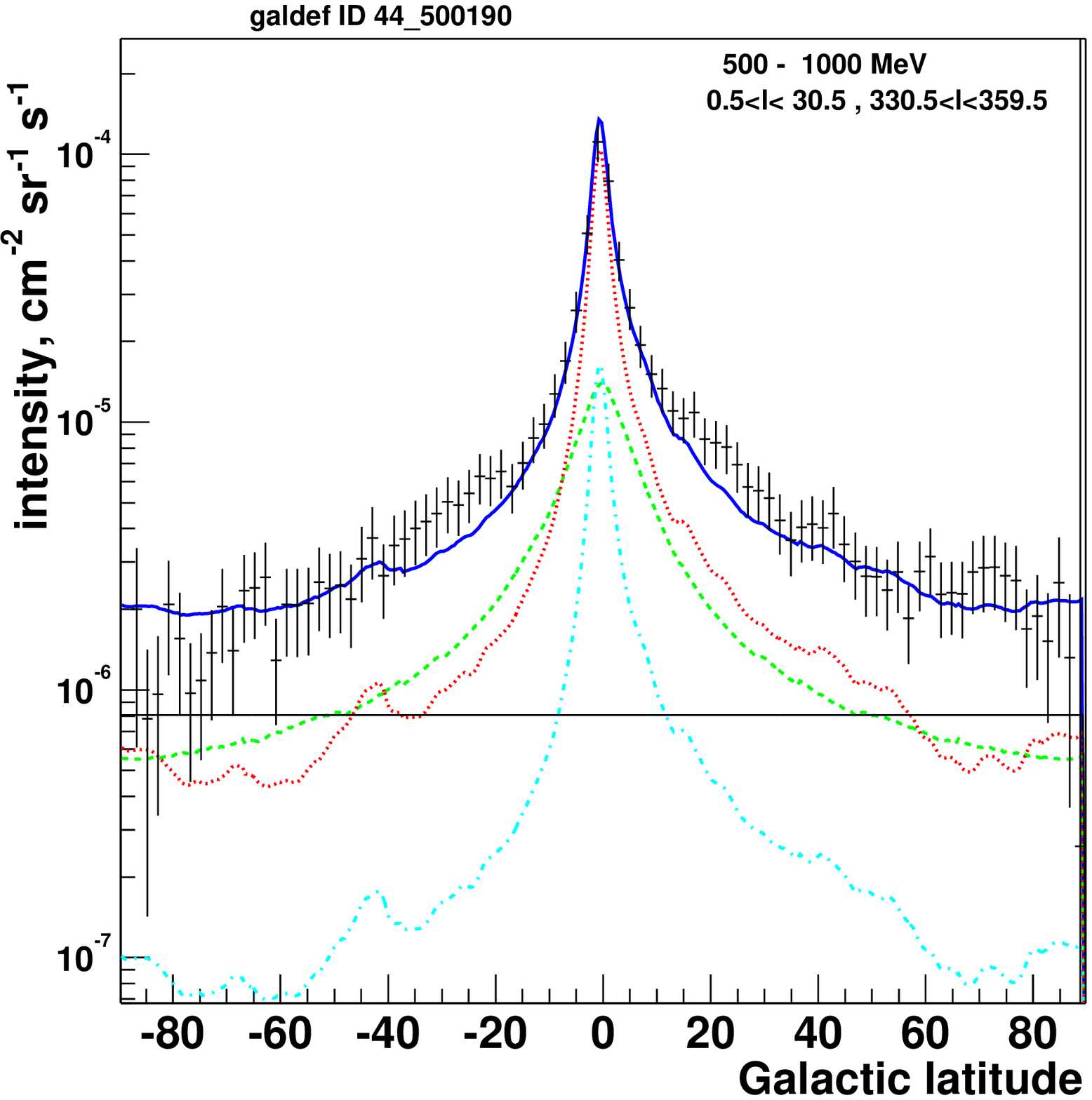}
\includegraphics[width=58mm]{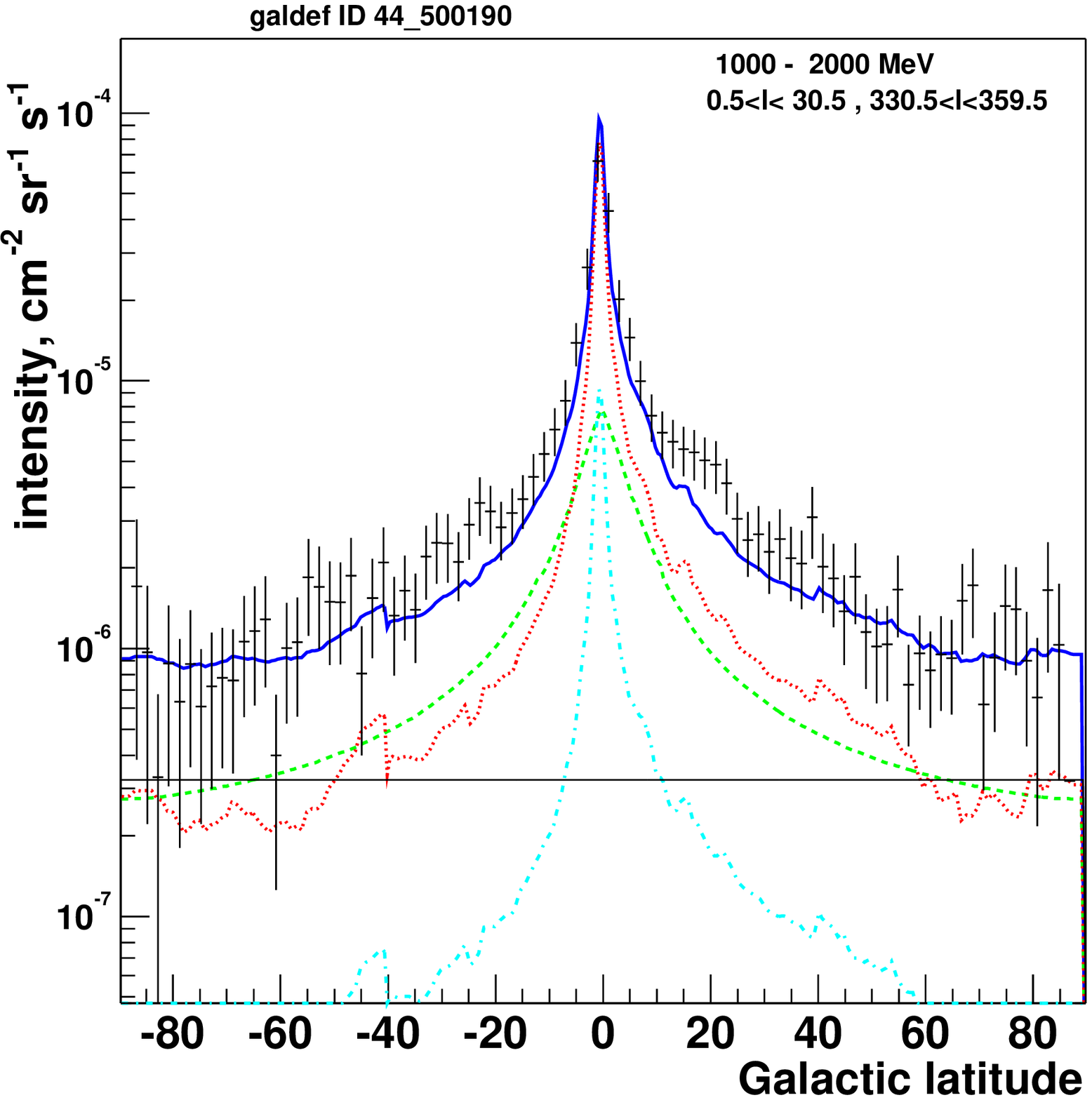}
\includegraphics[width=58mm]{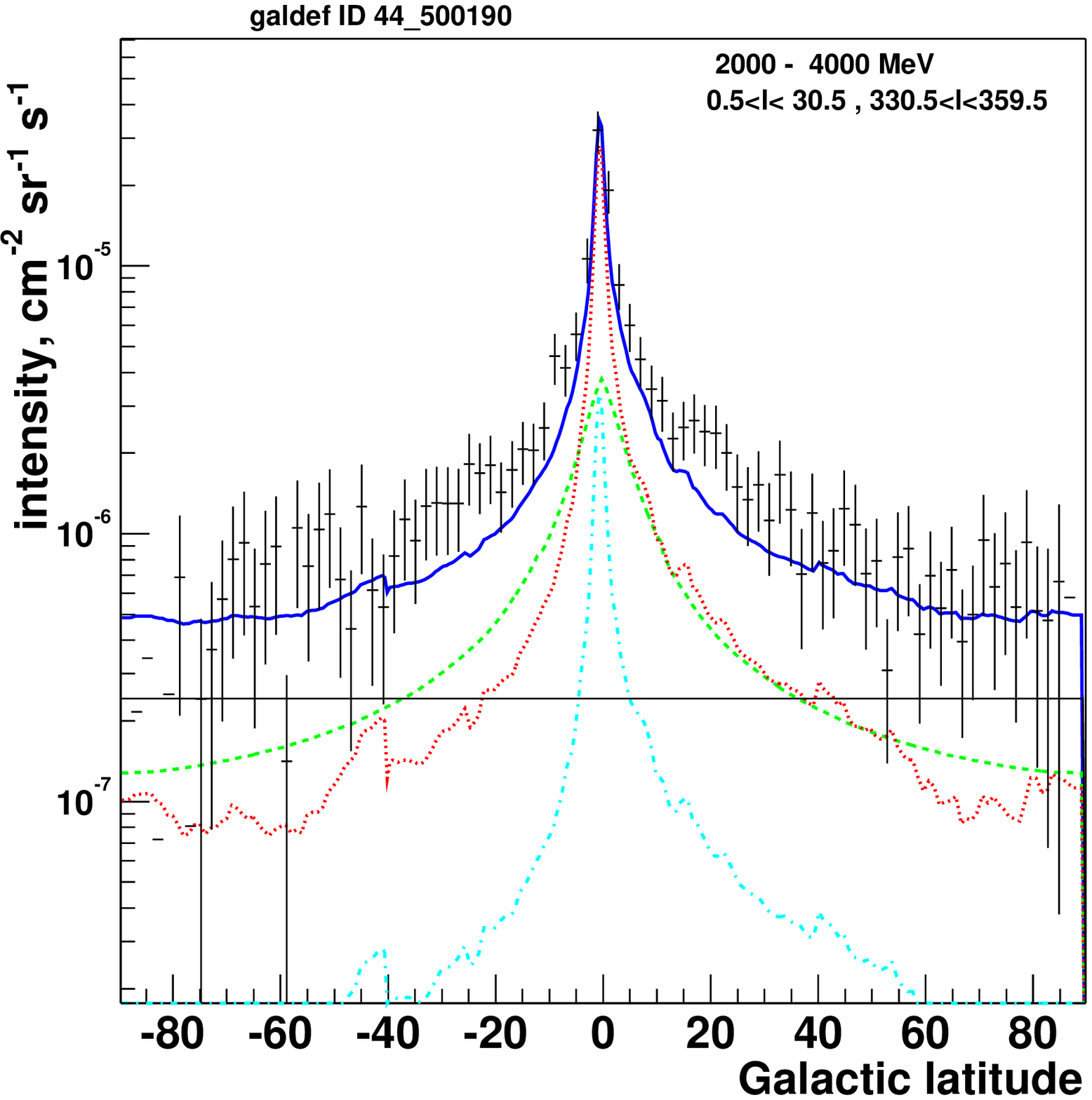}
\includegraphics[width=58mm]{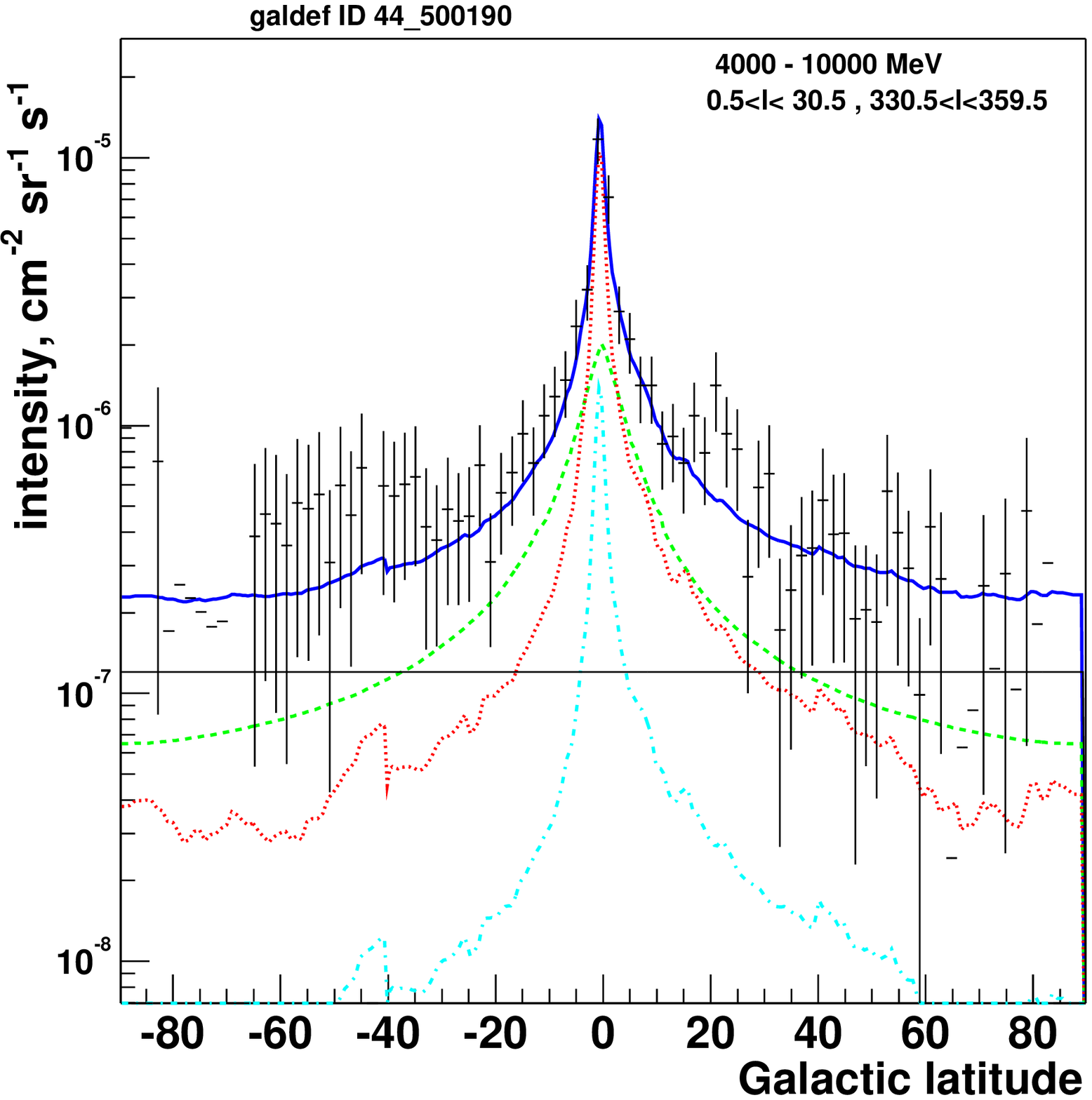}
\includegraphics[width=58mm]{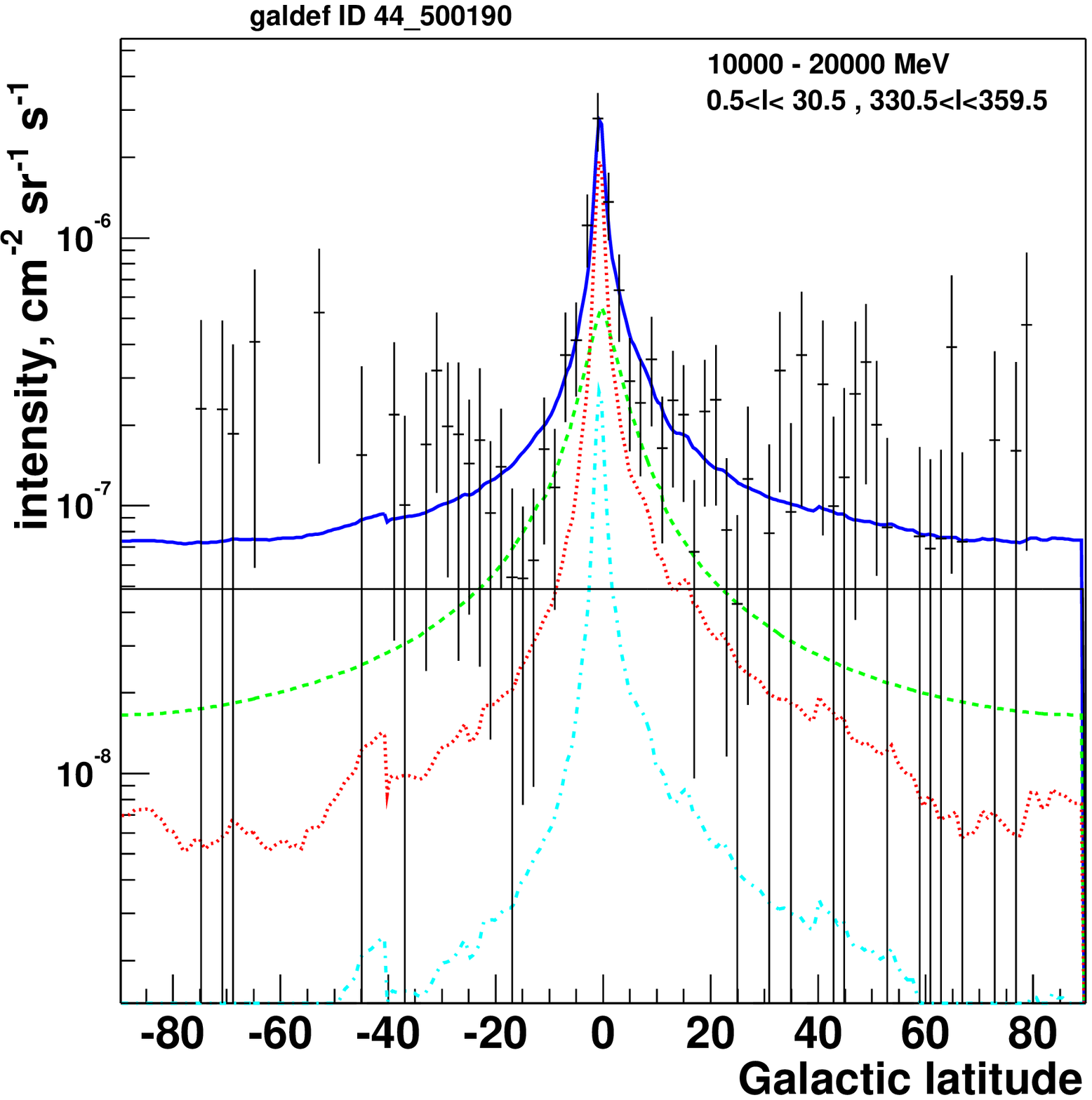}
\includegraphics[width=58mm]{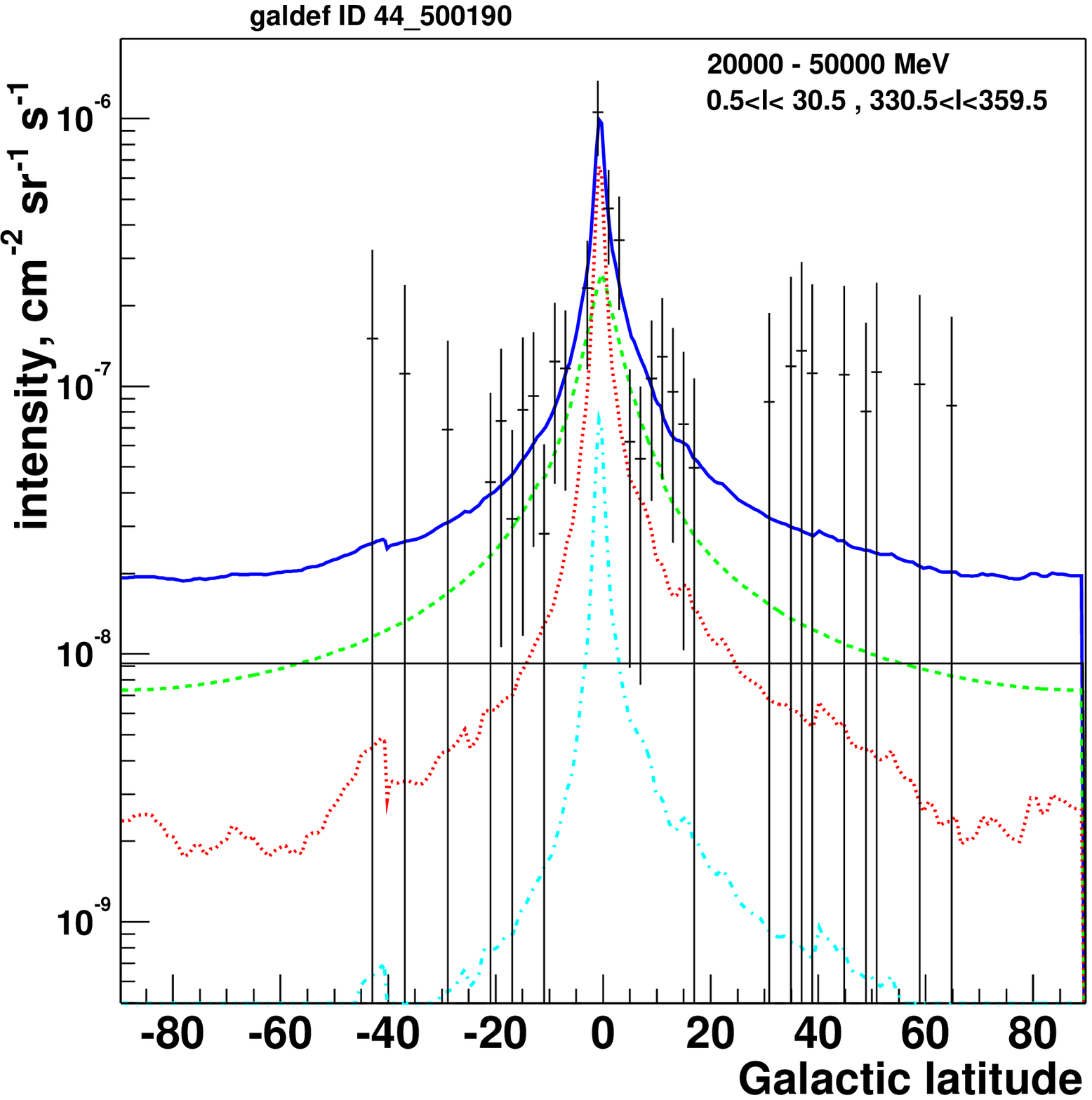}

\caption{Latitude profiles for optimized model (500190), 
inner Galaxy ($330^\circ<l<30^\circ$), compared with EGRET data
in 12 energy ranges 30 MeV -- 50 GeV.
Lines are coded as in Fig.~\ref{fig:spectrum_conventional}.}
    \label{fig:latitude_profiles_optimized}
\end{figure*}

\begin{figure*}
\centering
\includegraphics[width=58mm]{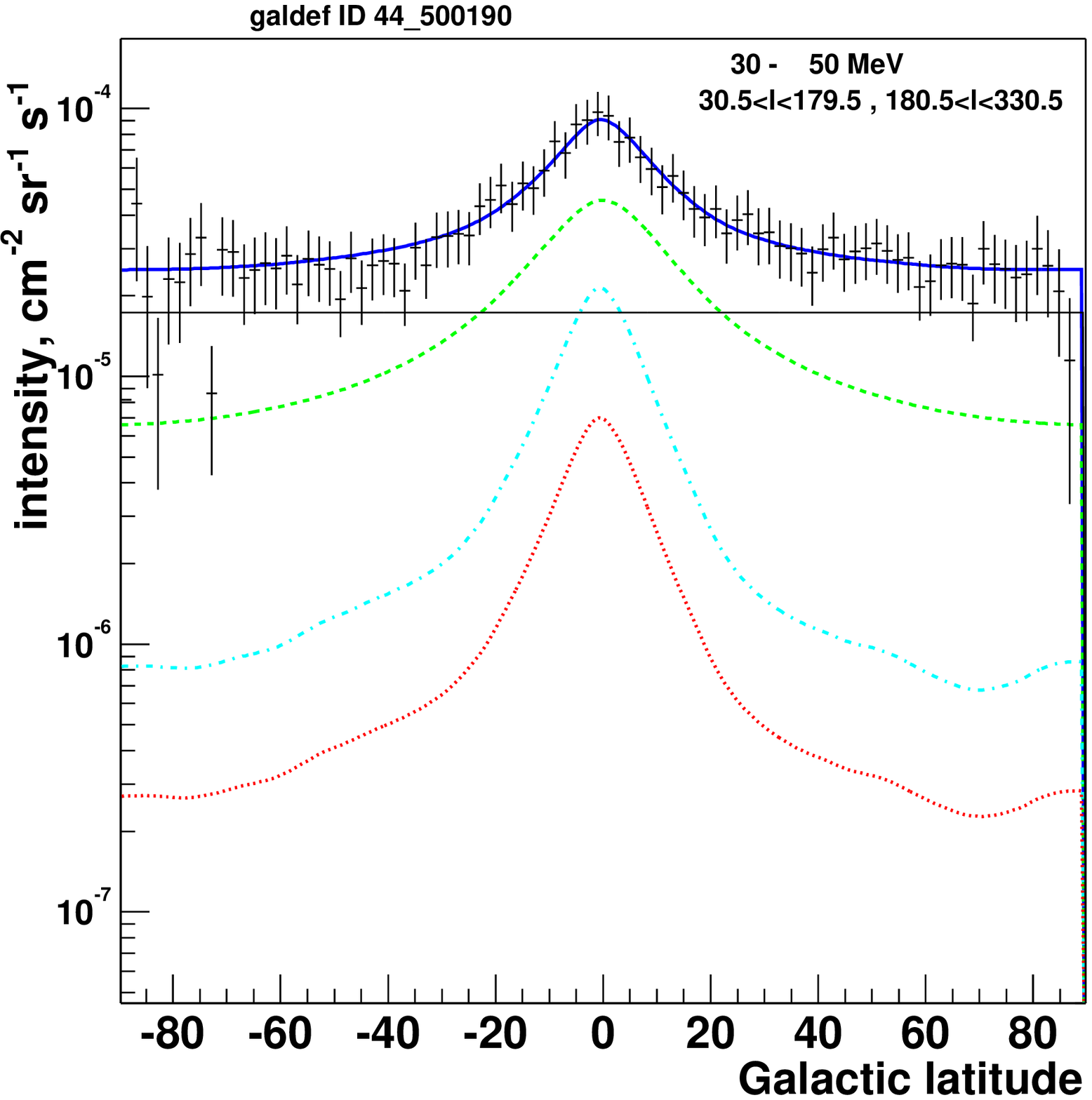}
\includegraphics[width=58mm]{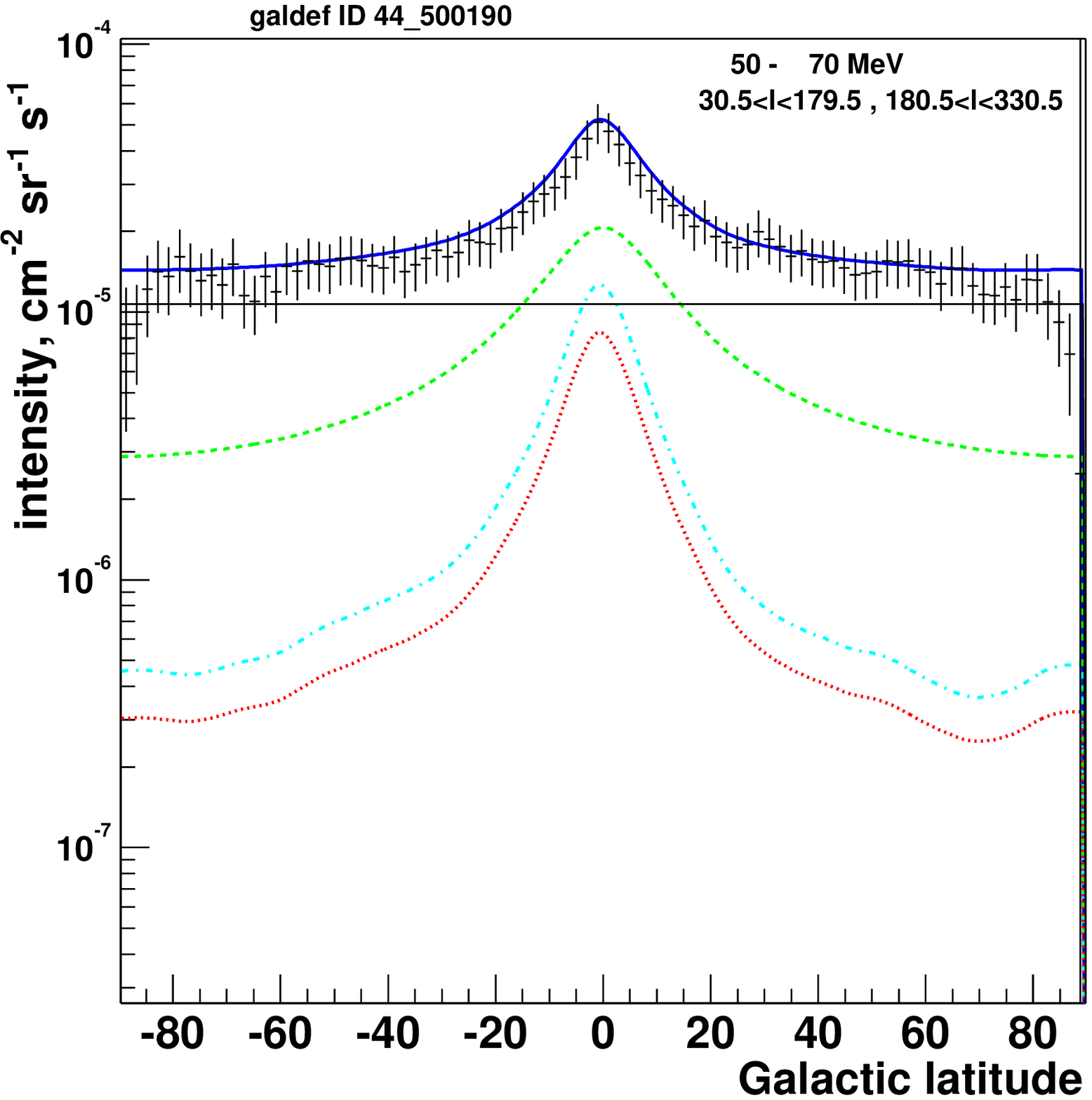}
\includegraphics[width=58mm]{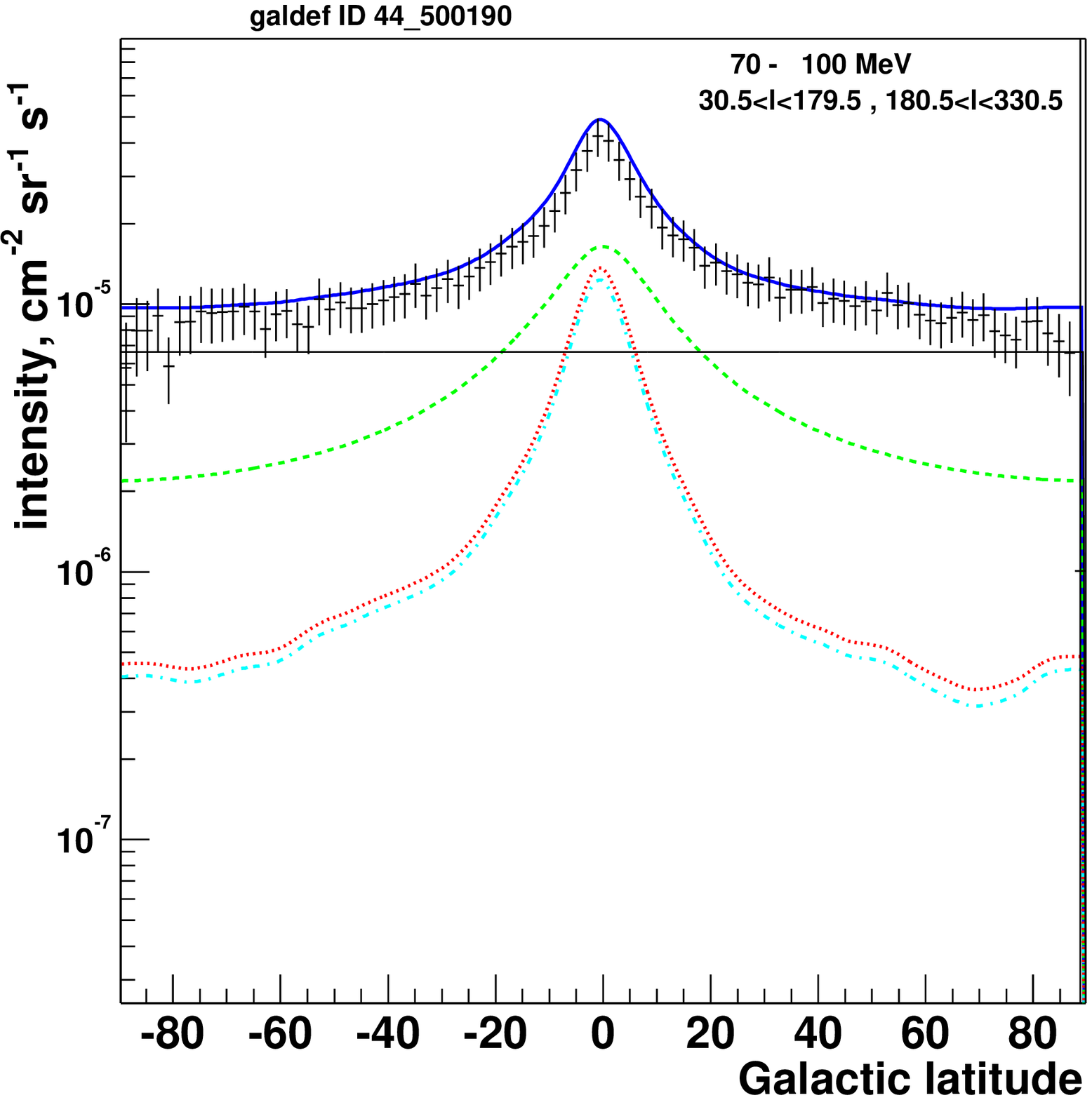}
\includegraphics[width=58mm]{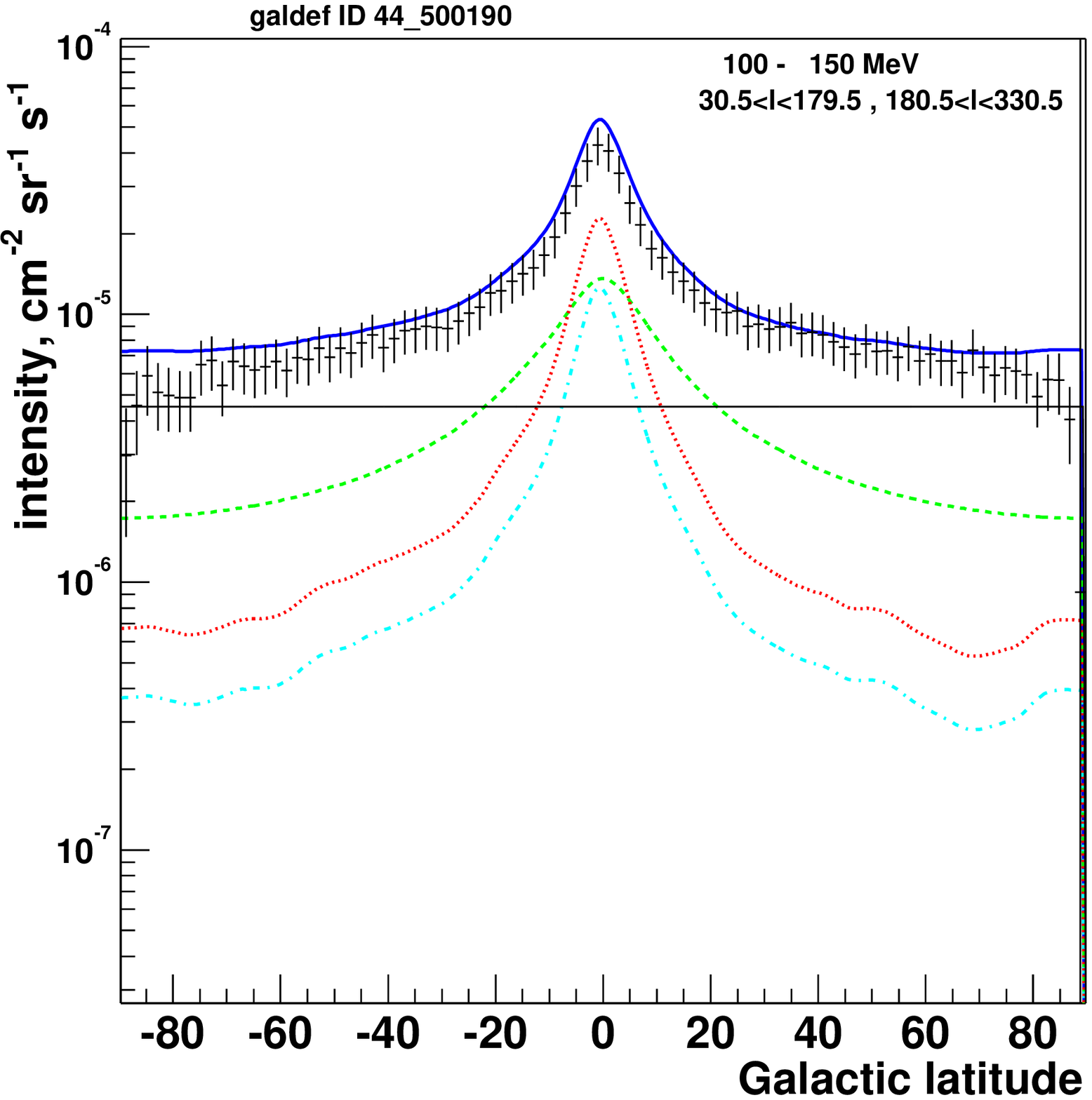}
\includegraphics[width=58mm]{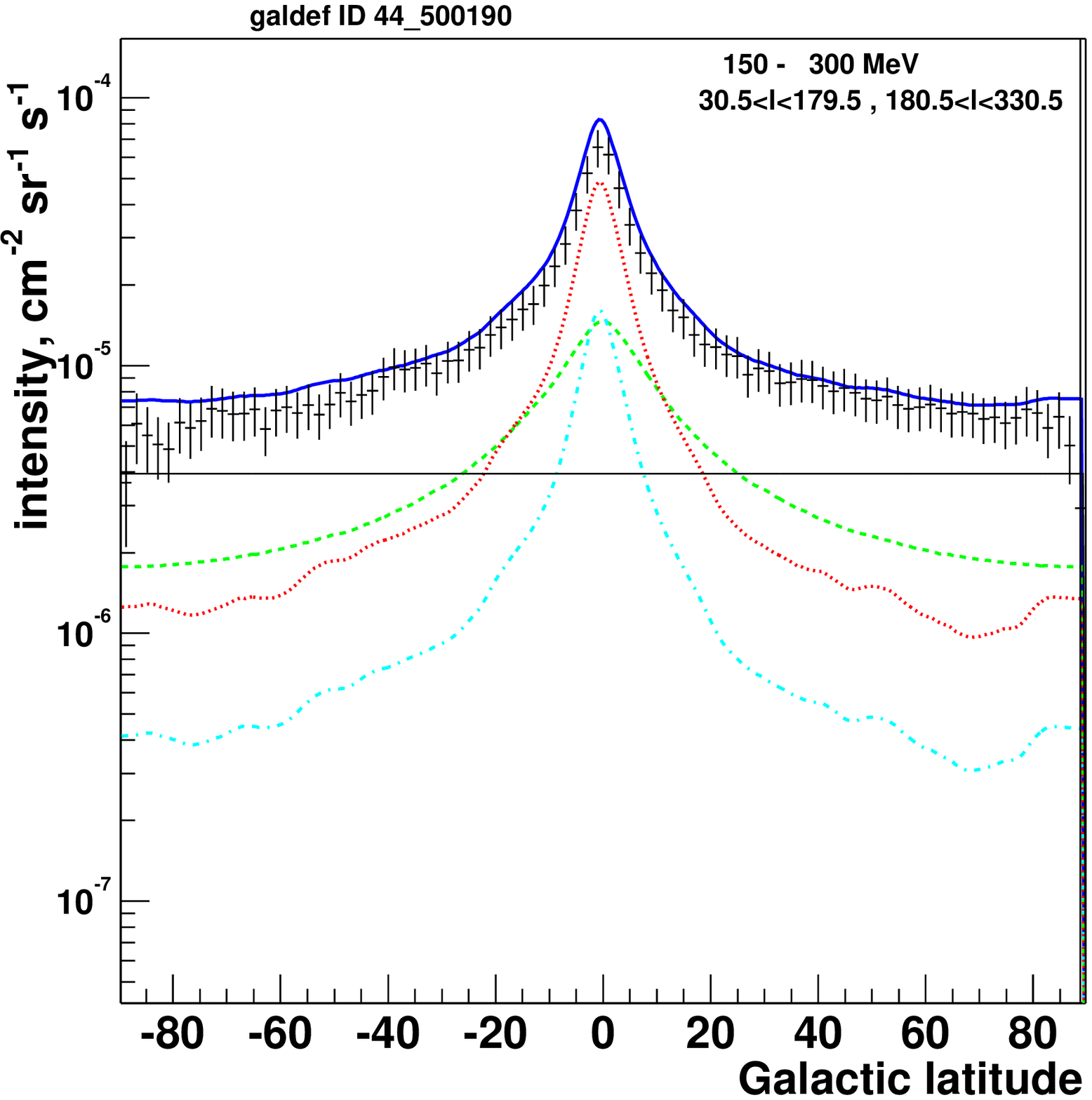}
\includegraphics[width=58mm]{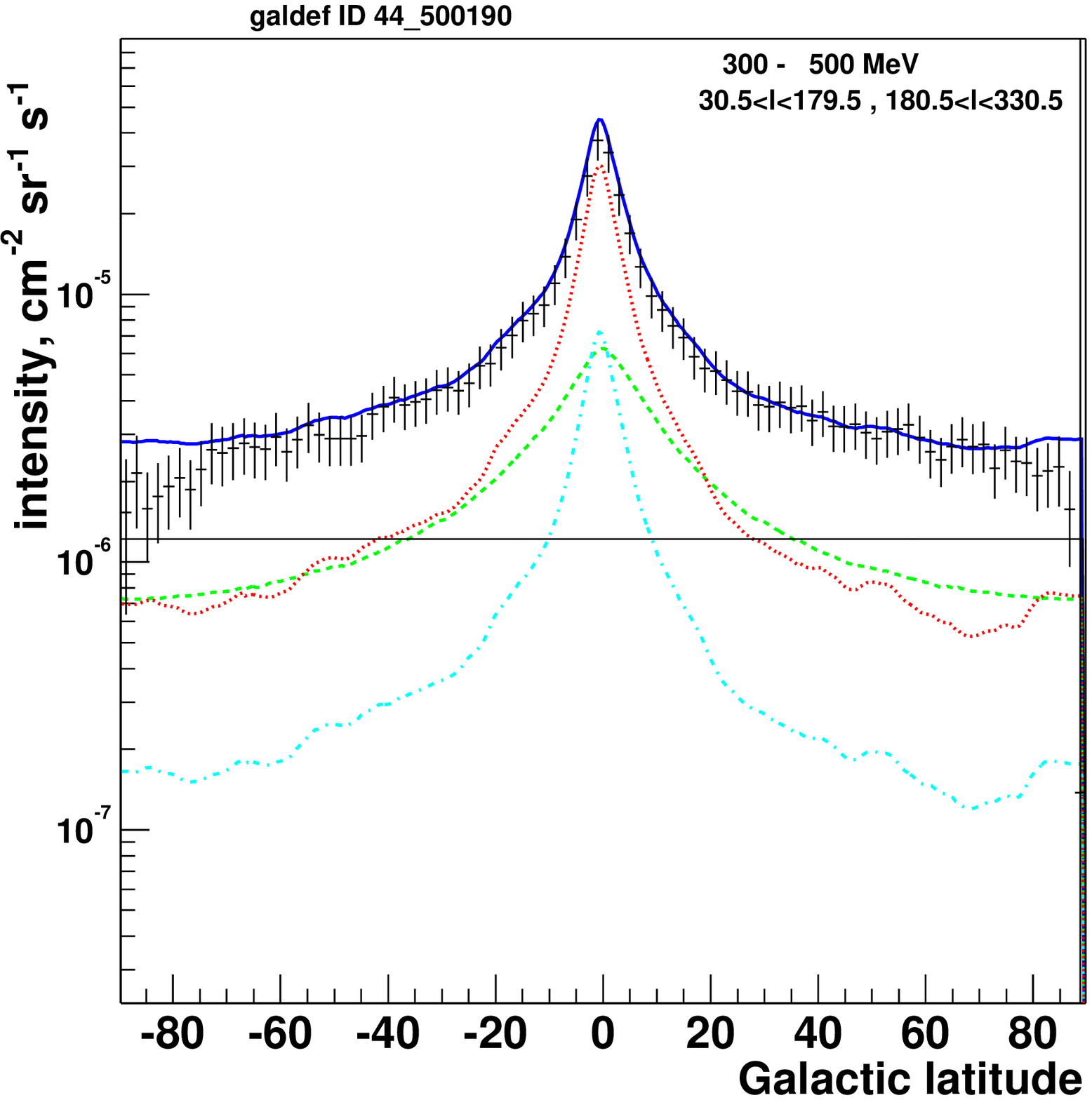}
\includegraphics[width=58mm]{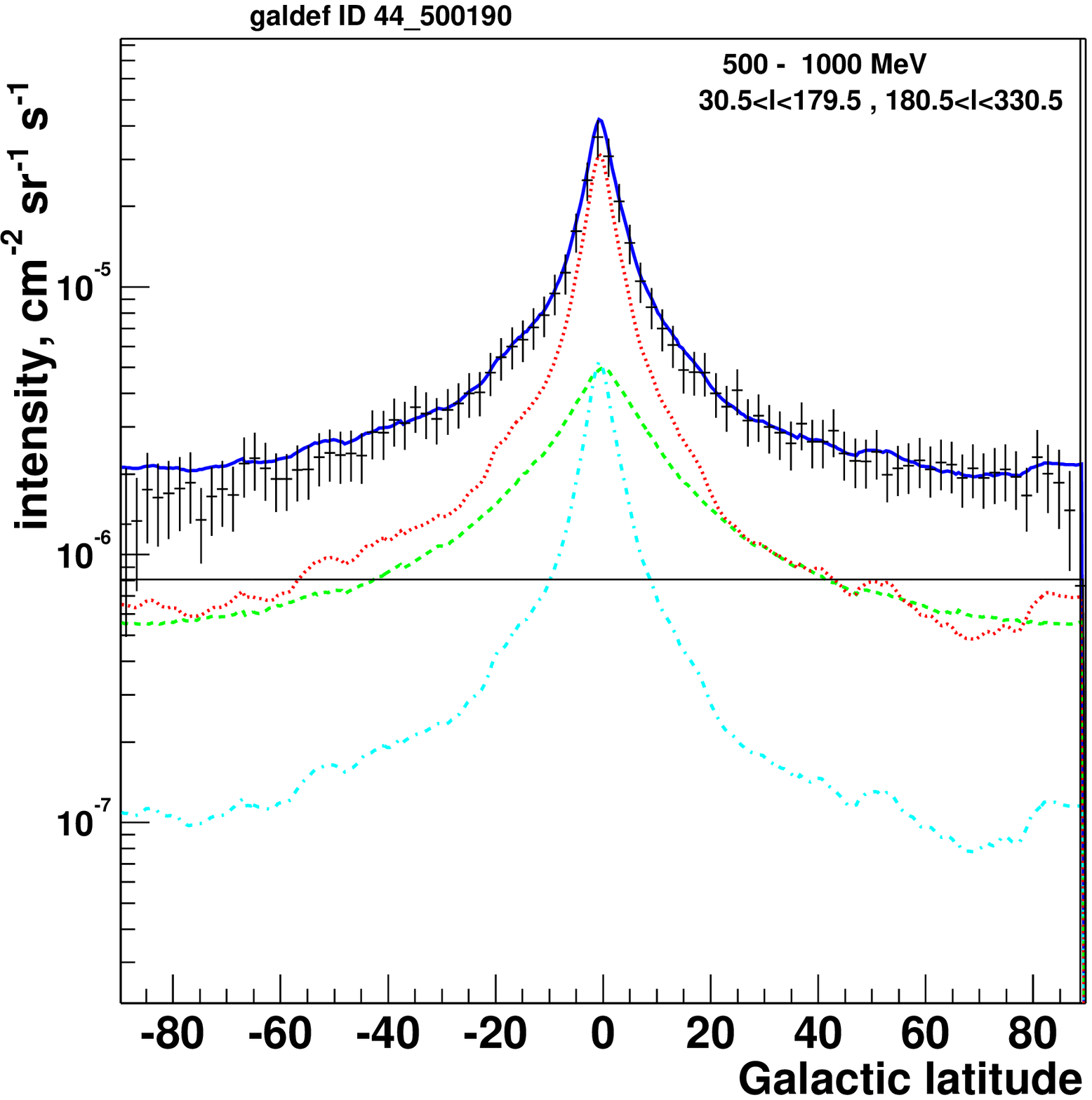}
\includegraphics[width=58mm]{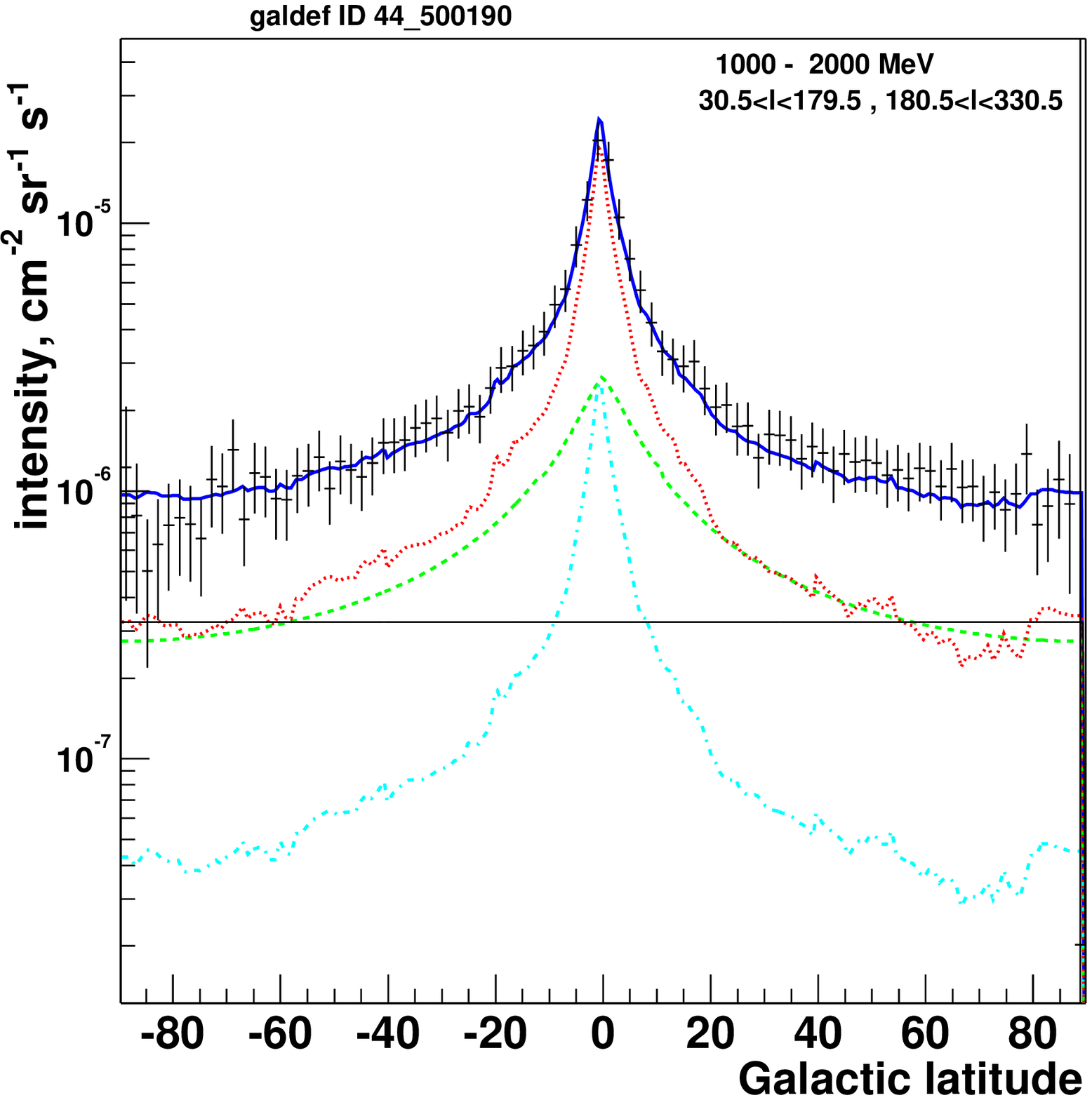}
\includegraphics[width=58mm]{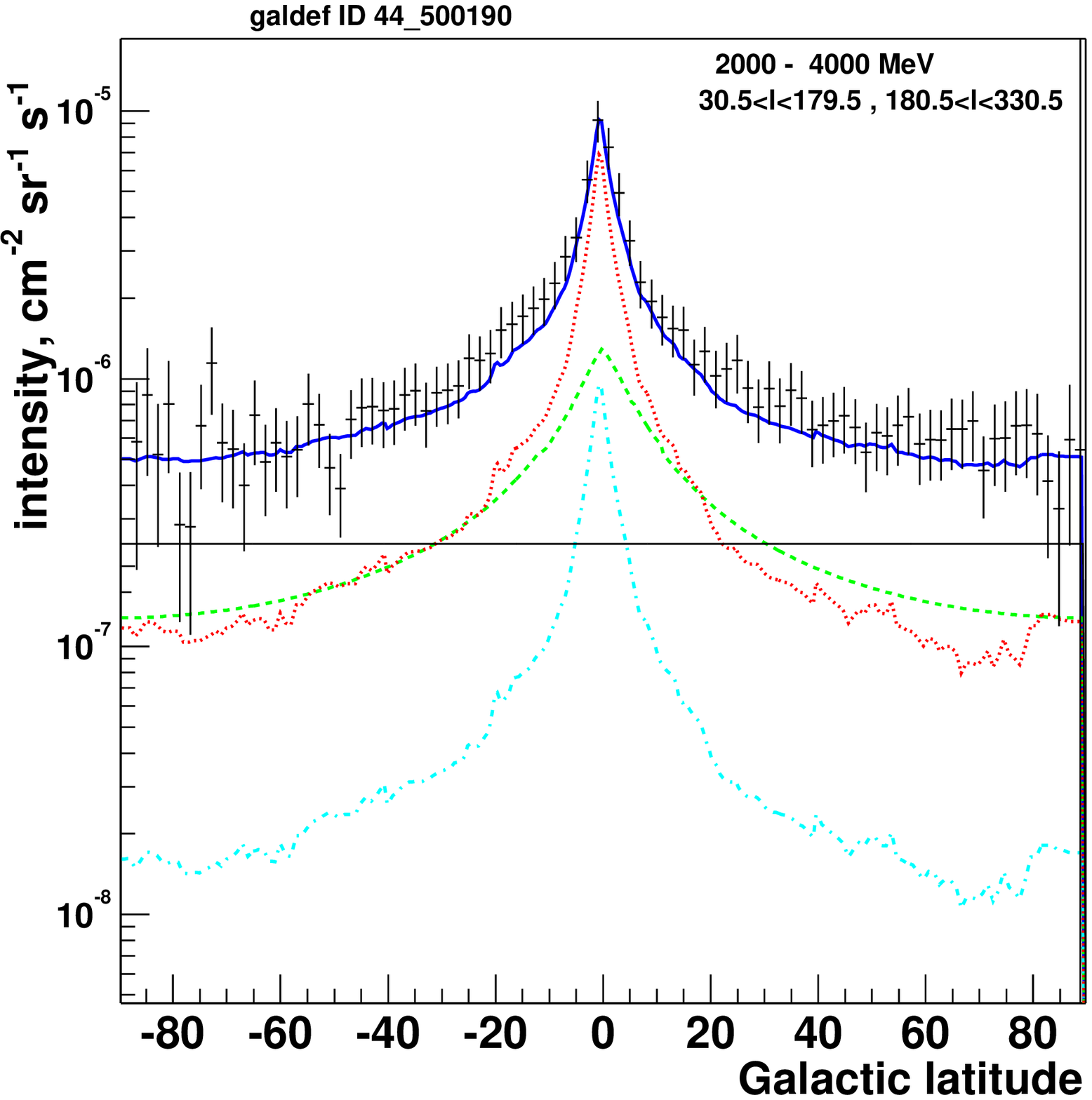}
\includegraphics[width=58mm]{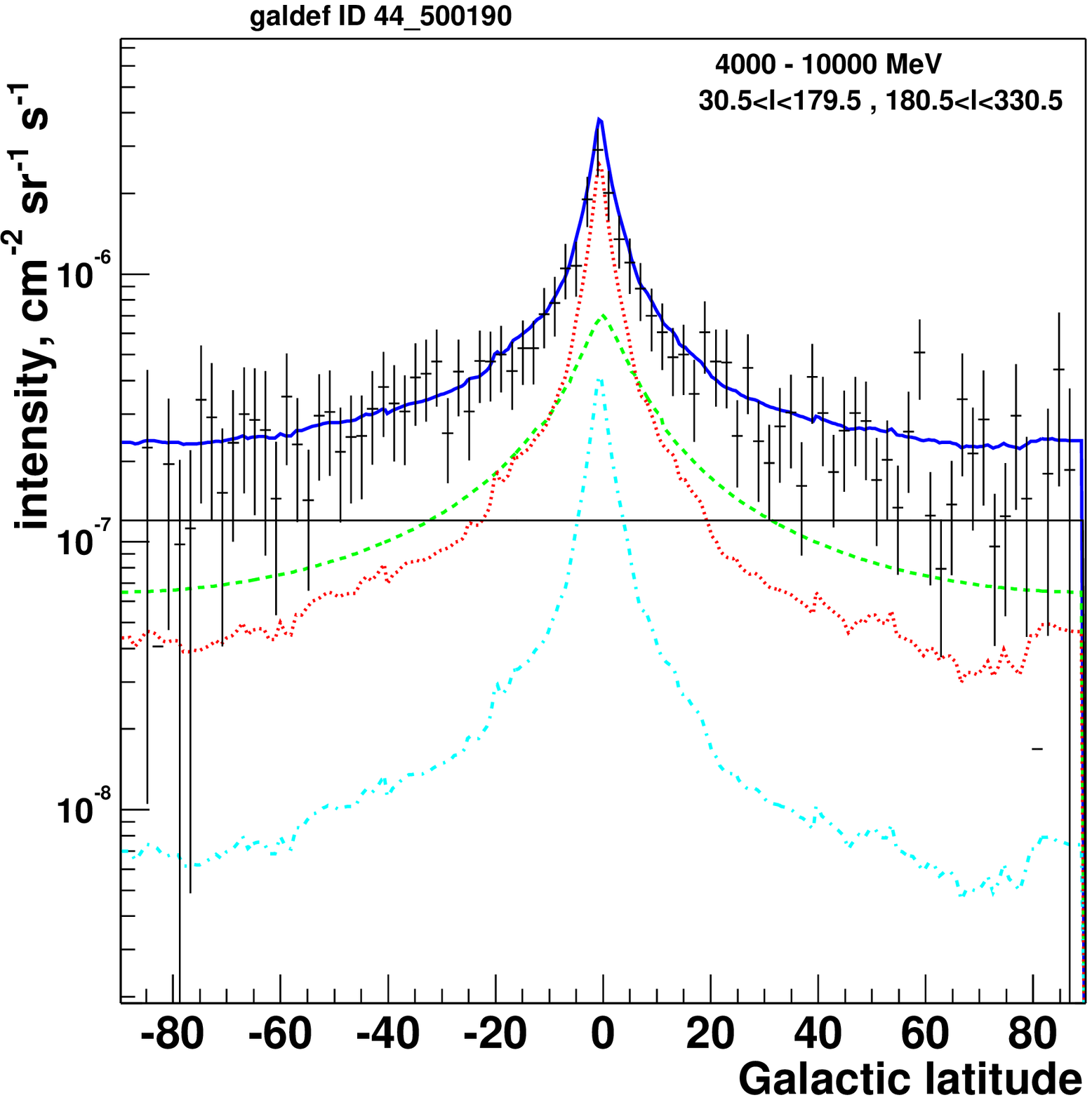}

\caption{Latitude profiles for optimized model (500190), 
except inner Galaxy ($30^\circ<l<330^\circ$), compared with EGRET data
in 10 energy ranges 30 MeV -- 10 GeV.
Lines are coded as in Fig.~\ref{fig:spectrum_conventional}.}
    \label{fig:latitude_profiles_optimized1}
\end{figure*}

Longitude profiles at low latitudes are shown in Fig.\
\ref{fig:longitude_profiles_optimized}. The agreement with the EGRET
data is generally good considering that the model does not attempt to
include details of Galactic structure (e.g., spiral arms), and the
systematic deviations reflect the lack of an exact fit to the spectra
in Fig.\ \ref{fig:spectrum_optimized}. The largest deviation ($\sim$20\%) is 
at 2--4 GeV, but this is still compatible with the systematic
errors of the EGRET data. Latitude profiles in the
longitude ranges $330^\circ<l<30^\circ, 30^\circ<l<330^\circ$ are
shown in Figs.\ \ref{fig:latitude_profiles_optimized},
\ref{fig:latitude_profiles_optimized1}, where the logarithmic scale
is chosen given the large dynamic range and
to facilitate the comparison at high Galactic latitudes. 
The agreement with EGRET is
again good, in particular the reproduction of the high-latitude
variation confirms the importance of the IC component which is much
broader than the gas-related \piodecay and bremsstrahlung emission.
In the inner Galaxy (Fig.\ \ref{fig:latitude_profiles_optimized})
there is evidence for an excess at intermediate latitudes, perhaps
related to an underestimate of the interstellar radiation field in the
Galactic halo, or special conditions in the Gould's Belt. The outer
Galaxy latitude profiles (Fig.\
\ref{fig:latitude_profiles_optimized1}) are in excellent agreement
with the data.

The $\chi^2$ values (Table~\ref{chisq}) confirm the visual conclusion
of the improvement of this model over the conventional and hard-electron
spectrum models.

The local electron spectrum (Fig.~\ref{fig:electrons}) {\it is}
compatible with the direct measurement considering fluctuations due to
energy losses and stochastic sources and propagation \citep{SM01}, and
in addition uncertainties in solar modulation at low energies. In
fact the agreement can be even better if we consider the uncertainty
in the ISRF, which can well be a factor 2 higher than our estimate.
The electron spectrum is consistent with the synchrotron spectral
index data \citep{SMR00}, since it differs from the conventional
model essentially only in the normalization, and this is in turn
consistent with synchrotron. The interstellar proton spectrum (Fig.\
\ref{fig:protons}) is also compatible with direct measurements; the
factor 1.8 may be attributed to fluctuations over the Galaxy relative
to the local value, and also to the uncertainty in the large-scale CR
gradient.

\begin{figure*}[!thb]
\centering
\includegraphics[width=58mm]{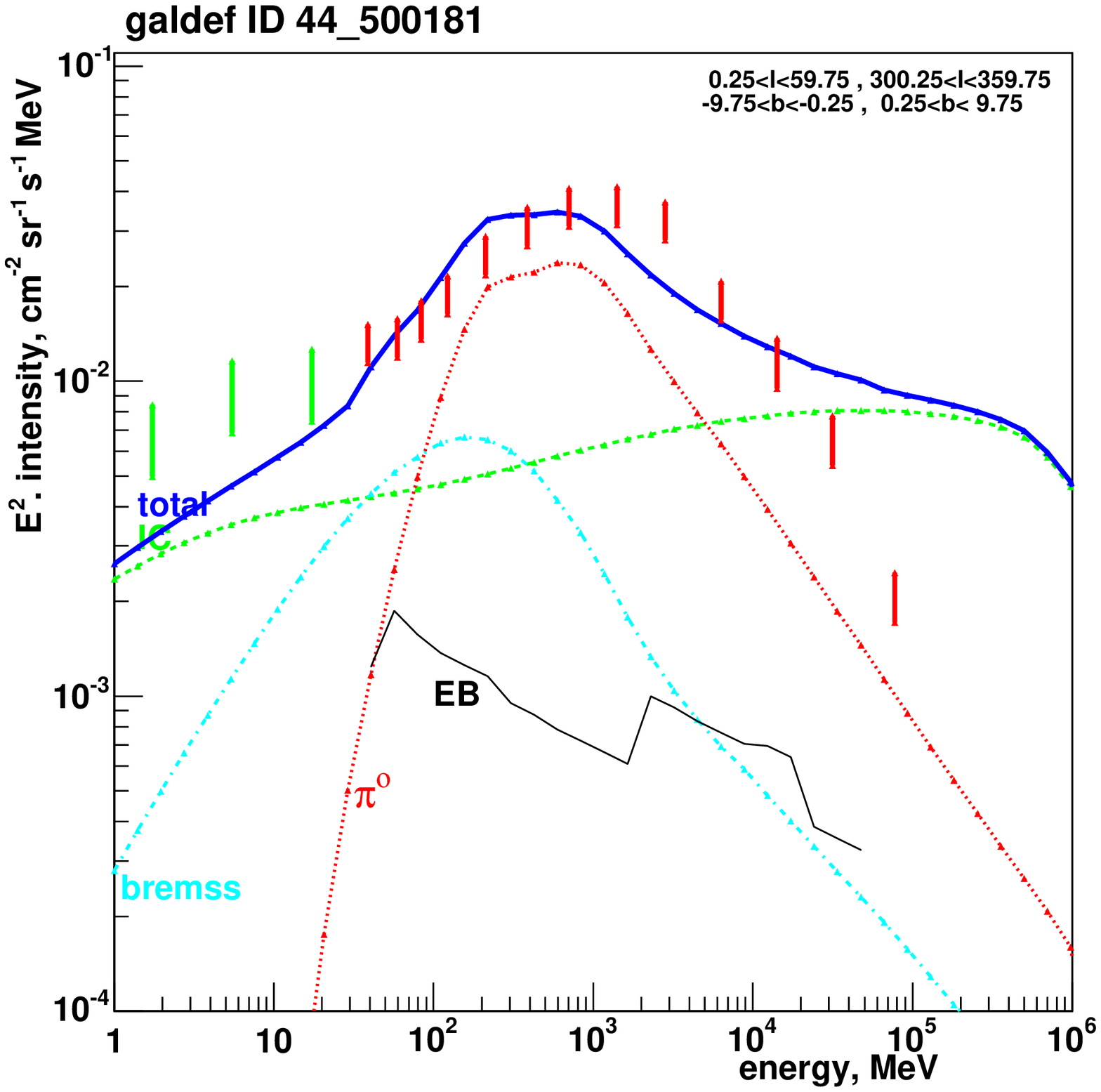}
\includegraphics[width=58mm]{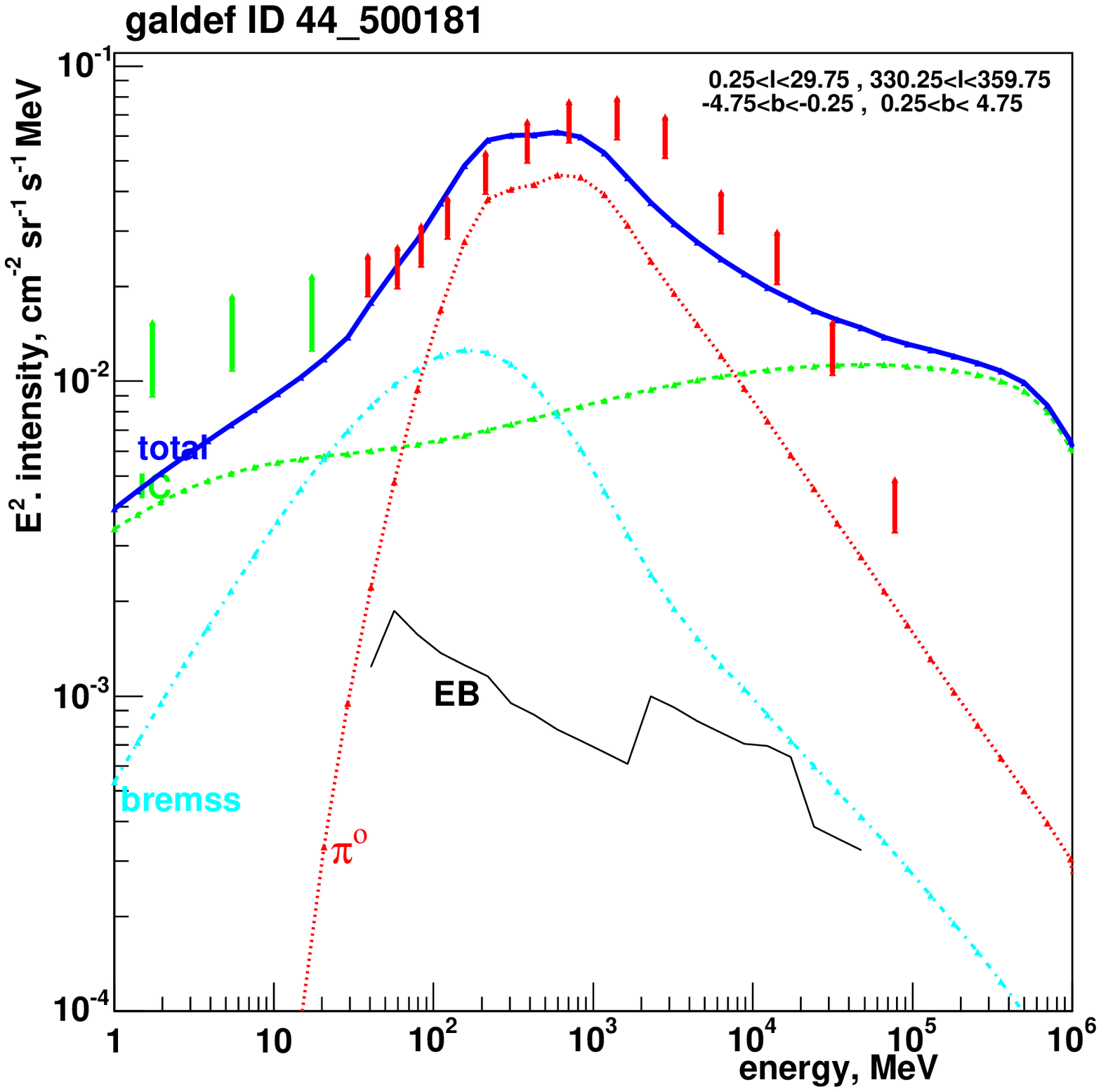}
\includegraphics[width=58mm]{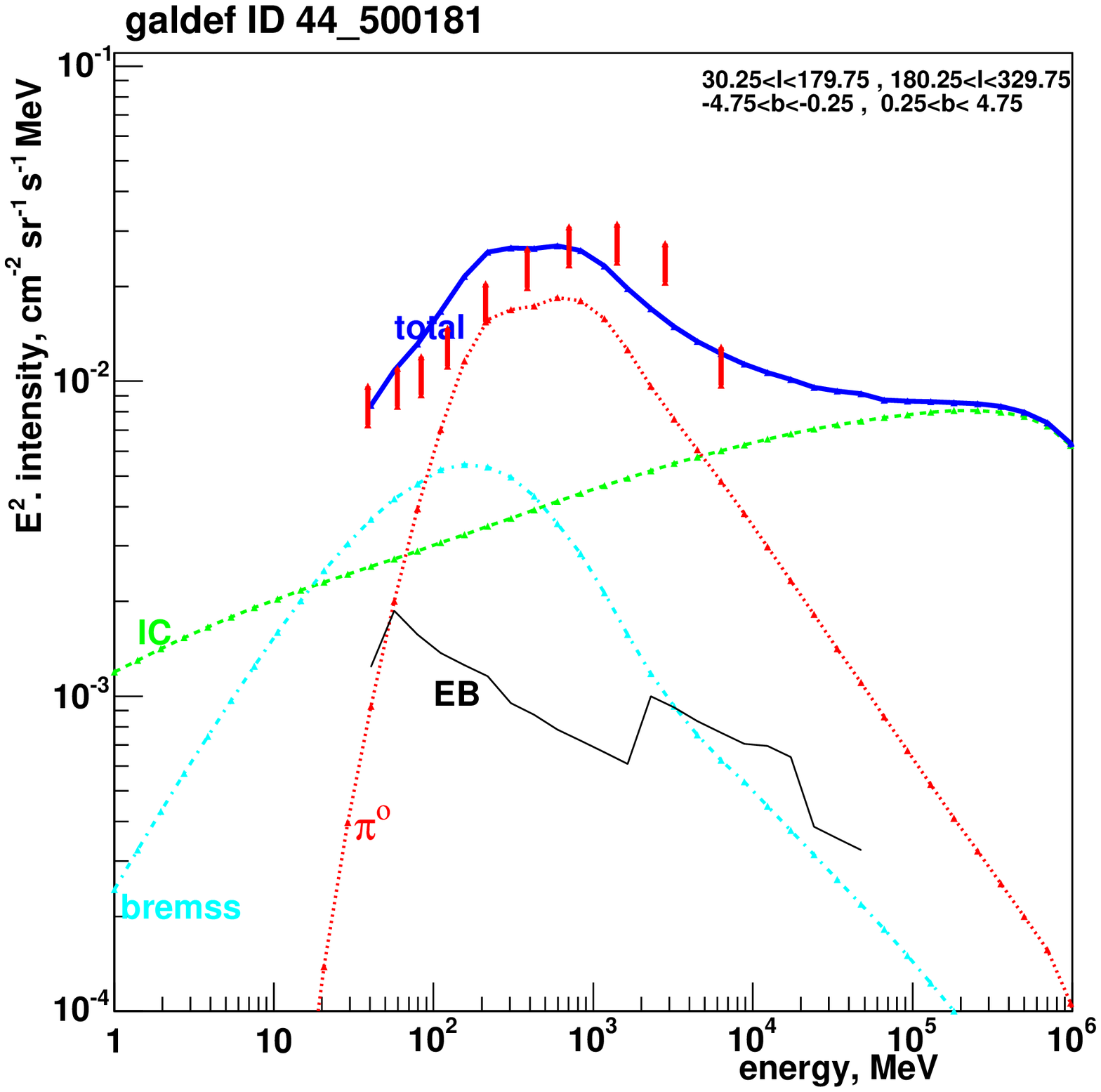}
\includegraphics[width=58mm]{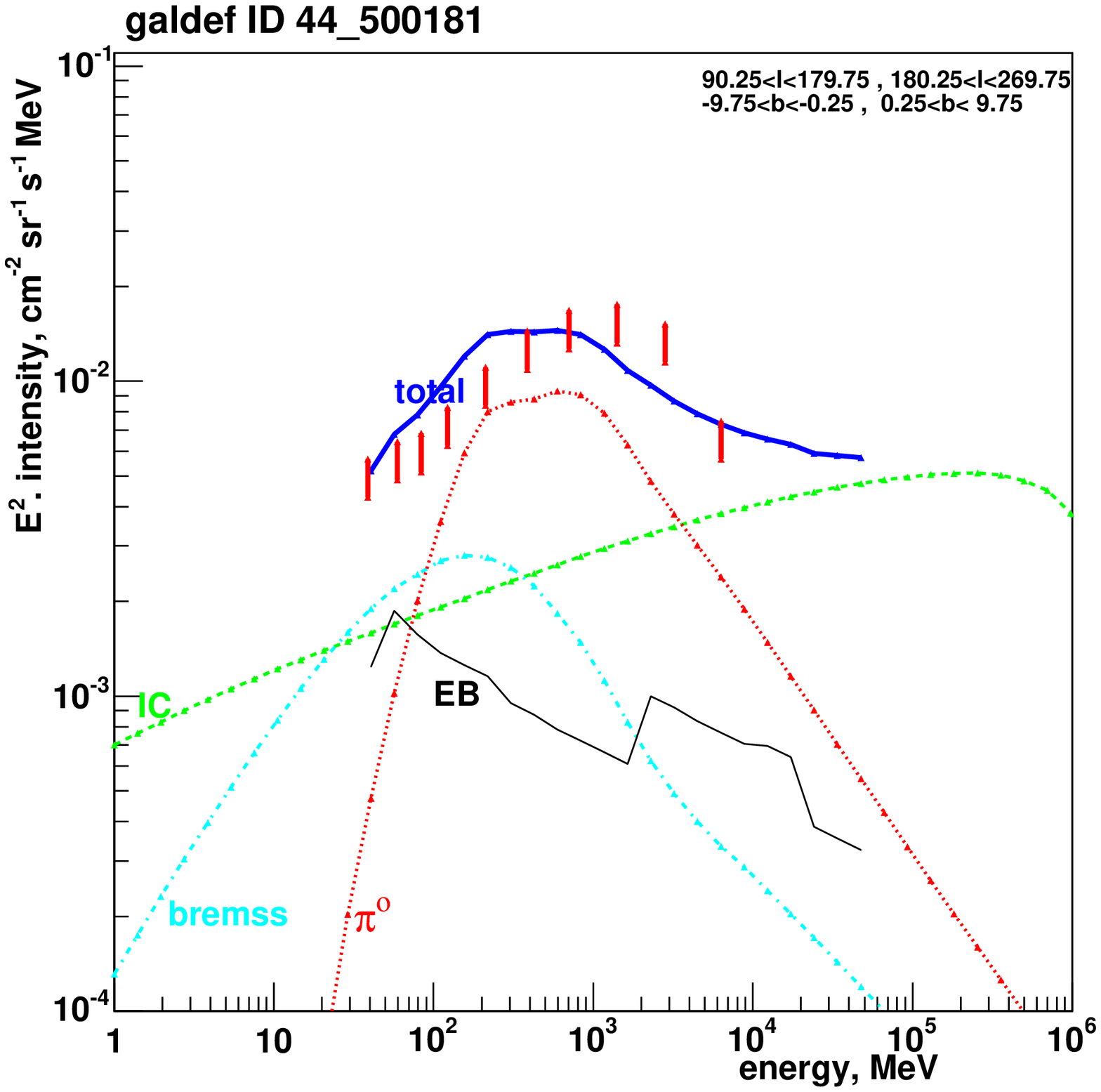}
\includegraphics[width=58mm]{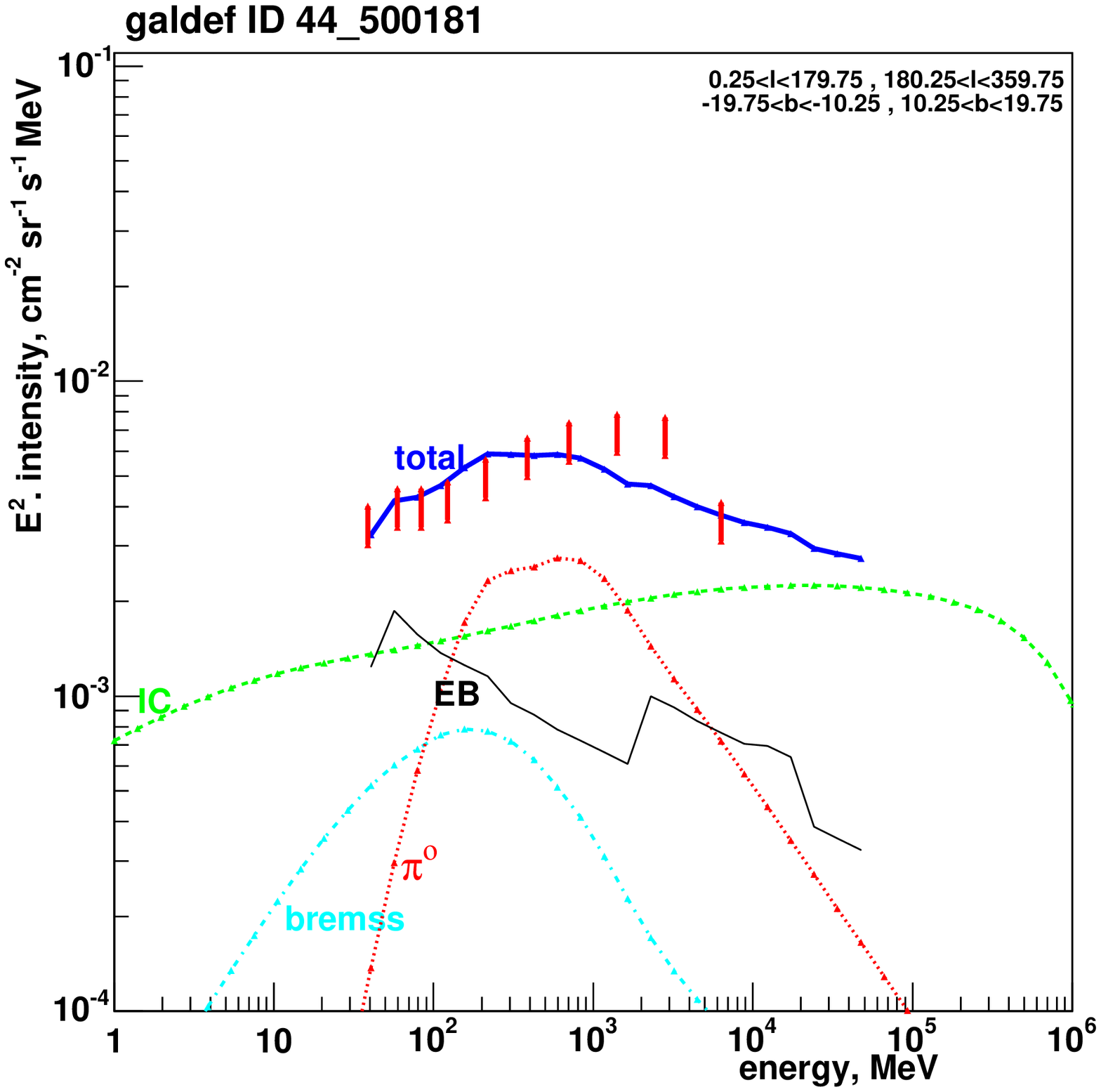}
\includegraphics[width=58mm]{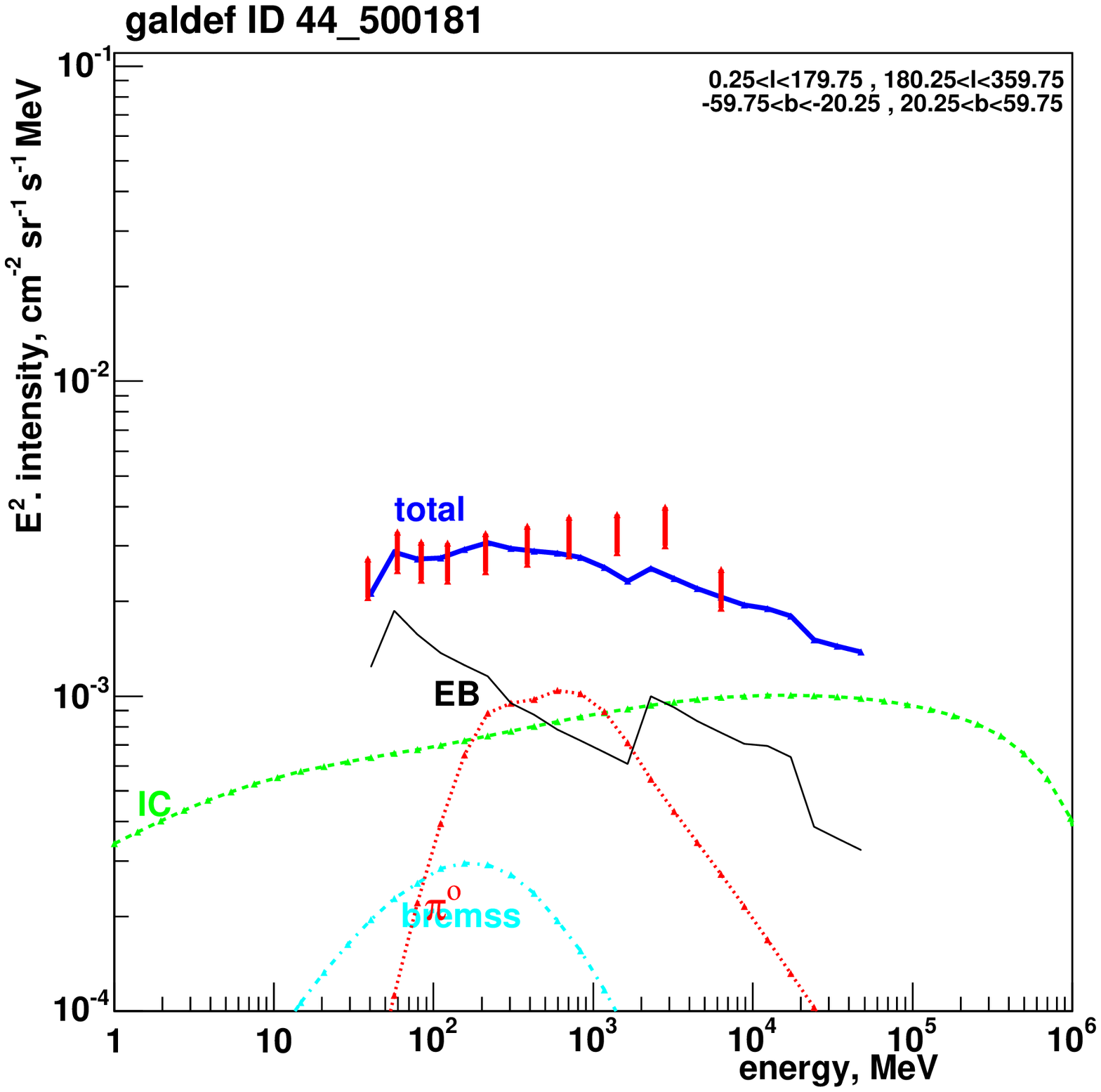}
\includegraphics[width=58mm]{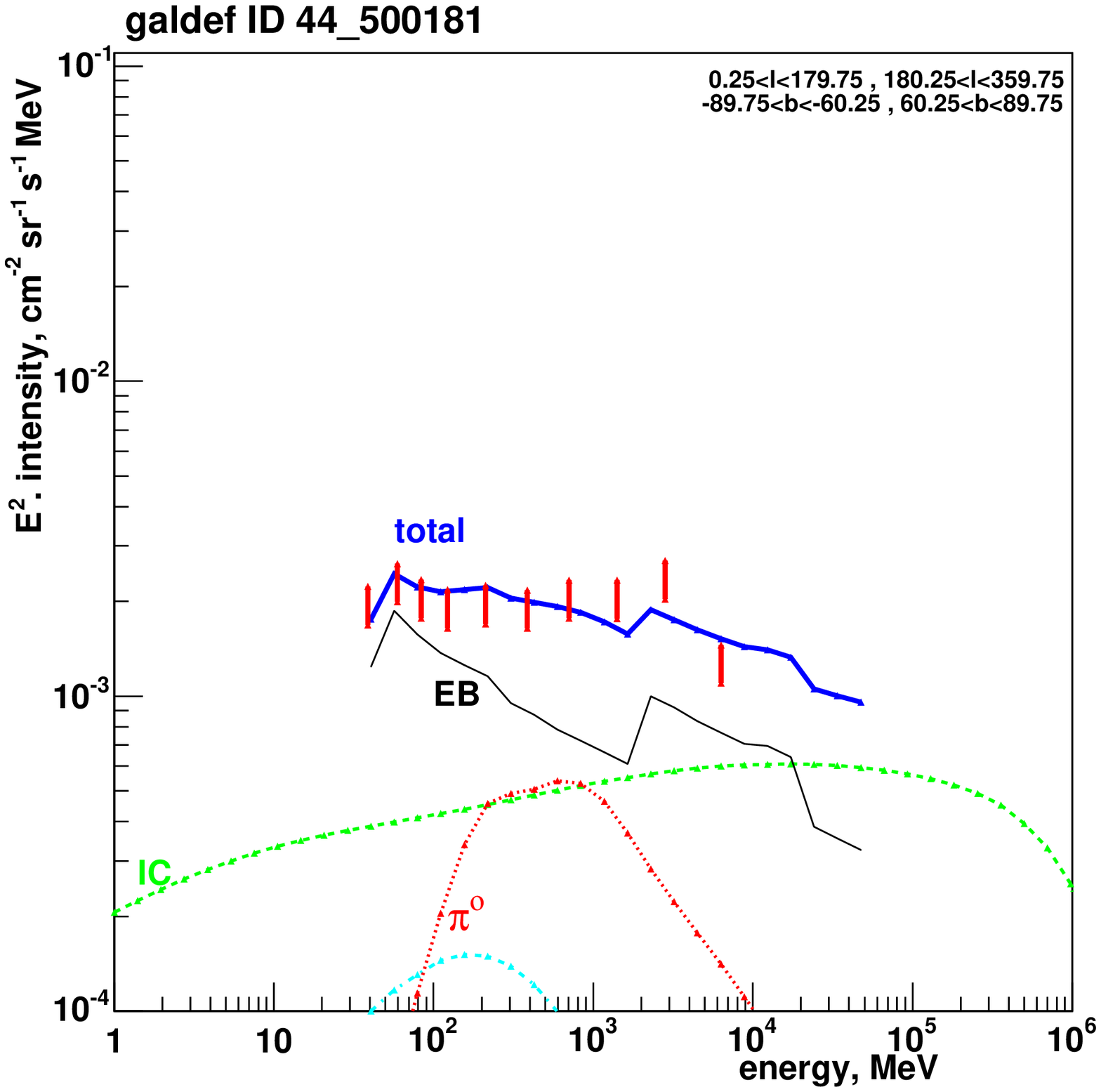}

\caption{\gray spectra of hard electron spectrum model (500181); regions and
coding as for Fig.~\ref{fig:spectrum_conventional}.
\label{fig:spectrum_hard} }
\end{figure*}

\placefigure{fig:spectrum_hard}

\begin{figure}[t]
\centering
\includegraphics[width=9cm]{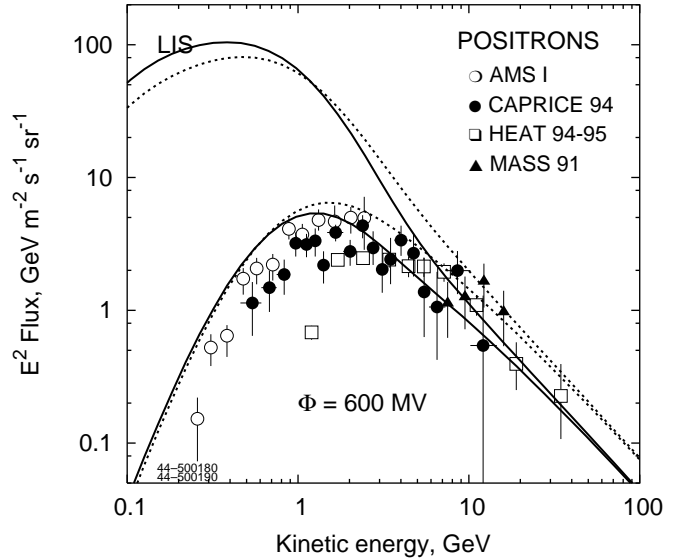}
 \caption{Positron spectra
for conventional and optimized models compared with data 
(upper curve -- LIS, lower -- modulated to 600 MV). The lines are coded as in 
Fig.~\ref{fig:protons}.
Data: AMS-I \citep{leptons_ams}, CAPRICE 94 \citep{Boez00}, HEAT 94-95 \citep{duvernois01}
MASS 91 \citep{grimani02}. 
\label{fig:positrons}}
\end{figure}

\placefigure{fig:spectrum_optimized}

\placefigure{fig:longitude_profiles_optimized}

\placefigure{fig:latitude_profiles_optimized}

\placefigure{fig:latitude_profiles_optimized1}

Below 30 MeV the predicted spectrum lies about a factor 2 below the
COMPTEL data, as found previously \citep{SMR00}. There we proposed
that a contribution from compact sources is the most likely
explanation. Recent results from INTEGRAL \citep{integral} indeed
indicate a large contribution from sources in the hard X-ray band, and
this would be consistent with the MeV region marking a transition from
source-dominated to diffuse-dominated ridge emission.

\section{Secondary antiprotons, positrons, electrons} \label{sec:secondary}
\subsection{Tests of the nucleon spectrum}

A sensitive test of the proton spectrum using the $\bar p/p$ ratio has
been proposed \citep{MSR98} based on the fact that secondary
antiprotons and \grays are produced in the same $pp$-interactions.
Positrons are also produced in $pp$-collisions and thus may be used to
support the conclusions made from the antiproton test. While some
deviation from the locally-observed spectrum of primary protons is
acceptable, secondary antiprotons (and partly positrons) trace the
primary proton spectrum on scales up to $\sim$10 kpc over the Galaxy,
and hence allow us to put limits on deviations from the local
measurements.

Antiprotons and positrons were originally used to exclude the possibility of
a hard proton spectrum as the origin of the \gray GeV excess \citep{SMR00}.
However, even for a conventional nucleon spectrum, a problem appears in the
reacceleration model in the simultaneous fitting of 
secondary/primary nuclei ratios and the antiproton spectrum; 
the former fixes the propagation parameters which can be
used to predict antiprotons, but using the measured proton spectrum
leads to too few antiprotons. 
To reproduce both the secondary/primary nuclei ratios and the antiproton
spectrum, \citet{M02} suggested a change in the propagation mode
at low energies.
\citet{M03} proposed a
contribution to the primary CR spectrum from the ``local bubble'' as a
possible solution. Our present model provides an alternative 
to these solutions.

As was noted in \cite{SMR00} and \citet[][and references
therein]{M03}, the ``GeV excess'' in \grays and underproduction of
antiprotons in the reacceleration model may indicate that the nucleon
spectrum typical of large regions of the Galaxy differs moderately
from the local measurements. The problem with secondary antiprotons
in the reacceleration model has been extensively discussed in
\citet{M02,M03}. It is apparent that if the solution of the \gray GeV
excess can not be found in modifications of the electron spectrum
alone, the required modifications in the nucleon spectrum must satisfy
the constraints from both antiprotons and positrons.

Figs.\ \ref{fig:pbars} and \ref{fig:positrons} show the antiproton and
positron fluxes as calculated in the conventional and optimized
models. The modifications of the nucleon spectrum introduced in the
optimized model 
appear to be exactly what is required to reproduce both, antiproton and
diffuse \gray data, and the positron spectrum also agrees
at high energies. At low energies the calculated positron spectrum is
rather high but the solar modulation is a factor of $\sim$1000 at
these energies, and besides the scatter in the positron data may
indicate large systematic errors.

\subsection{Gamma-rays from secondary positrons and electrons}

Secondary positrons in CR produced in interactions of energetic
nucleons with interstellar gas are usually considered a minor
component of CR. This is indeed so in the heliosphere where the
positron to all-lepton ratio is small at all energies,
$e^+/e_{tot}\sim 0.1$. However, the secondary positron flux in the
interstellar medium is comparable to the electron flux at relatively
low energies $\sim$1 GeV because of the steeper spectrum of secondary
positrons.

\begin{figure}[!tb]
\centering
\includegraphics[width=9cm]{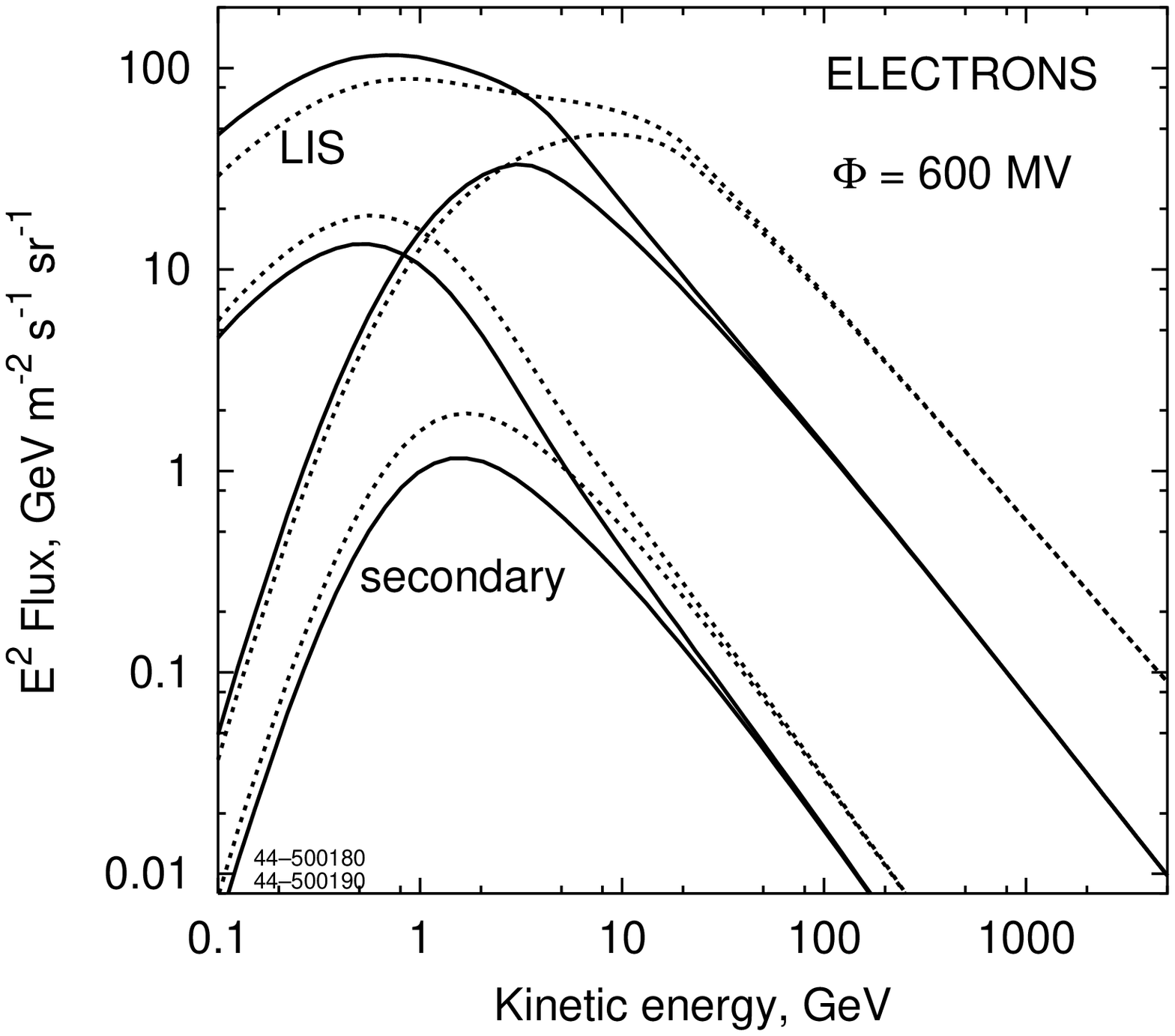}
\caption{Electron spectra for conventional (solid) and optimized models (dots).
Upper curve -- LIS, lower -- modulated to 600 MV.
Secondary electrons are shown separately for the same models.
\label{fig:sec_electrons}}
\end{figure}

The spectrum of \emph{secondary} positrons and electrons depends only
on the ambient spectrum of nucleons and the adopted propagation model.
Figs.~\ref{fig:electrons}, \ref{fig:positrons} show the spectra of
electrons and secondary positrons for the conventional and optimized
models. Secondary positrons contribute about half of the total
lepton flux at $\sim$1 GeV. Secondary electrons add up another 10\%
(Fig.\ \ref{fig:sec_electrons}). This leads to a considerable
contribution of secondary positrons and electrons to the diffuse \gray
flux via IC scattering and bremsstrahlung and significantly increases
the flux of diffuse Galactic \grays in MeV range. Therefore, secondary
positrons (and electrons) in CR can be directly traced in
$\gamma$-rays!

\placefigure{fig:sec_electrons}

Fig.~\ref{fig:spectrum_secondary} shows the contribution of secondary
positrons and electrons to the IC emission and
bremsstrahlung. Secondaries contribute more than 20\% of the total IC
in the 1--10 MeV energy range. More dramatic is the case of
bremsstrahlung, where secondaries contribute about 60\% of the total
below $\sim$200 MeV. It is the contribution of secondaries which
improves the agreement with data below some 100 MeV.

However the secondaries are not sufficient to explain the excess in
the 1--30 MeV range observed by COMPTEL, so that an additional
point-source contribution to the emission is still required here
\citep{SMR00}.  Evidence for such a point source contribution has
indeed recently been found  by INTEGRAL \citep{lebrun}.

\section{Discussion of uncertainties}\label{sec:discussion}

We do not discuss here possible calculation errors.
Derivation of such errors is a very complicated matter
given the many uncertainties in the
input. Those most probable are the uncertainties in the
$\pi^0$-production in $pp$-collisions at relatively \emph{low}
energies, nuclear cross sections, gas distribution in the Galaxy,
ISRF, systematic errors in the CR measurements, heliospheric modulation,
etc. Some possible errors and their effects
have been discussed in \citet{M02,M03}. Here we qualitatively mention
what we think may affect our conclusions and what may not.

\citet[][and references therein]{mori97} has re-evaluated the
$\pi^0$-production in $pp$-collisions using modern Monte Carlo event
generators HADRIN, PYTHIA, and FRITIOF. The HADRIN code, which is
designed to reproduce nuclear collisions at laboratory energies below
5 GeV and describes the threshold and resonance behavior of inelastic
hadron-nucleon interactions, shows good agreement with isobar model
calculations \citep{stecker70} at proton kinetic energy $T_p=0.97$
GeV. The isobar model is shown to reproduce the data on the secondary
$\pi^0$-production at low energies, in particular, at $T_p=0.56, 0.65,
0.97, 2.0$ GeV \citep[][and references therein]{dermer86}. However,
the data on $\pi^0$-production at GeV energies are now 40 years old,
they have large statistical errors and are very scattered indicating
possibly large systematic uncertainties. Given the lack of new data,
the deviations from the isobar model calculations by a factor of
$\sim$2 would be also consistent with the old data. However, the
comparison to the Galactic diffuse \gray emission is now rather
precise, and the uncertainty in the $\pi^0$-production at low energies
may be critical. Our required flattening of the proton spectrum below
10 GeV could thus be understood as a compensation for errors in the
$\pi^0$-production physics. At high energies, a comparison of
\citet*{badhwar} and \citet{stephens} scaling model with resuls of
PYTHIA and FRITIOF shows generally a good agreement, but all of them
overpredict the cross sections at high rapidities. Though it may
result in systematic uncertainties of the \gray flux above $\sim$100
GeV, this, however, is of less concern given the large error bars in
the EGRET data in this energy range.

\begin{figure}[!tb]
\centering 
\includegraphics[width=9cm]{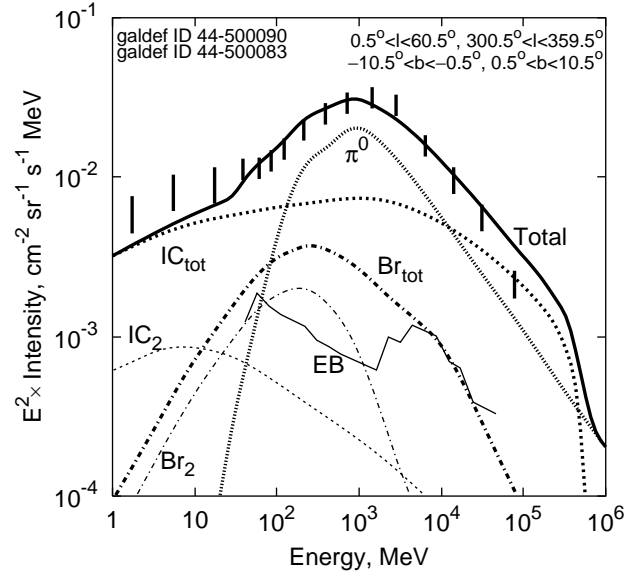}
\caption{\gray spectrum of optimized model with (thick lines) and without 
(thin lines) primary electrons, to show the contribution of secondary electrons and positrons.
Br$_{\rm tot}$ and Br$_2$ labels denote the total bremsstrahlung and the separate
contribution from secondary leptons, correspondingly. Similarly,
IC$_{\rm tot}$, IC$_2$ indicate the total IC and the contribution from secondaries.
\label{fig:spectrum_secondary}}
\end{figure}

\placefigure{fig:spectrum_secondary}

Possible errors in the cross
section of the CR nuclei affect the derived propagation
parameters such the diffusion coefficient, Alfv\'en speed etc.
While they may be important for calculation of CR isotopic 
abundances, they do not affect much the calculation of
the diffuse \grays as we normalize the particle spectra to the
given local values. 

Errors in the Galactic gas distribution are not so important in the
case of stable and long-lived nuclei. Such errors are compensated
simultaneously for all species
by the corresponding adjustment of the propagation parameters
(diffusion coefficient). 
In case of $\gamma$-rays, we compare with the large sky regions
so that the error in the column density in any particular direction
produces a minor effect.

For the calculation of the spectrum of \grays arising from IC
scattering and electron energy losses, the full ISRF as function of
position and wavelength is required, which was not available in the
literature. The ISRF was evaluated in \citet{SMR00} using
emissivities based on stellar populations and dust emission. Given
the complicated and uncertain input in this calculation a factor of
two error is quite possible. The inaccuracies in the ISRF are
compensated in our model by adjustment of the CR electron
flux. Therefore, if the ISRF energy density is in reality higher, it
will result in lower normalization of the electron injection spectrum
making it closer to the local one.

Our knowledge of the heliospheric modulation is still incomplete, and
it remains the source of a large uncertainty in the propagation
models. Over the last years Ulysses made its measurements at different
heliolatitudes so we know more about the solar magnetic field
configuration and the solar wind velocity distribution. However the
modulation parameters are usually still determined based on the
assumed {\it ad hoc} interstellar nucleon spectrum. Estimates of the
modulation made using the simplest force-field approximation show that
the modulation changes the proton intensity below 1 GeV by a factor
of 10 (Fig.\ \ref{fig:protons}), and by a factor of 100--1000 in case
of electrons and positrons (Figs.\ \ref{fig:electrons},
\ref{fig:positrons}). This makes it difficult to speculate about the
reasons for deviations of the calculated spectrum from the measured
one by a factor of a few at \emph{low} energies.
Modulation is important at energies below $\sim$10 GeV/nucleon, while 
it is negligible at higher energies. 
The \grays with energy $>$10 GeV are produced by protons of $>$100 GeV 
and by electrons of $>$10 GeV where modulation has (almost) no effect.
Lower energy \grays may be affected by the uncertainties
in the solar modulation, but this is compensated
by the adjustment of the interstellar spectra. 
In our optimized model, the diffuse \grays themselves are used to
constrain the interstellar particle spectra at low energies, while the 
constraints from the local measurements are relaxed.

Finally, due to the random nature of SN explosions the CR spectrum
fluctuates in space and time \citep[see simulations in][]{SM01a}.
Further, since more CR sources are concentrated in the spiral arms
\citep[e.g.,][]{CaseBhattacharya96}, the CR intensity in the arms
might be higher, while the sun is located in the interarm
region. These effects can cause the locally measured CR
intensity to differ from the large-scale average by the required
factor of 2--4.

\section{Conclusions}
We have revisited the compatibility of diffuse Galactic continuum
\gray models with the EGRET data. We confirm that the ``conventional
model'' based on the locally observed electron spectrum is
inadequate, for all sky regions. A conventional model plus hard
sources in the inner Galaxy is also inadequate, since this cannot
explain the excess outside of the Galactic plane. Models with a hard
electron injection spectrum, while reproducing the EGRET spectrum in
the few GeV region over much of the sky, are not compatible with the
locally observed electron spectrum (the expected fluctuations are not
sufficient) and are inconsistent with EGRET data above 10 GeV.

A new model, with relatively mild deviations of the electron and
proton spectra from local, is shown give a good reproduction of the
diffuse \gray sky. The agreement extends from 30 MeV to 100 GeV. It
also gives a very good representation of the latitude distribution of
the emission from the plane to the poles, and of the longitude
distribution. IC emission is a major component of the diffuse
emission and dominates outside the Galactic plane and at energies
below 100 MeV. The model reproduces simultaneously the $\gamma$-rays,
synchrotron, CR secondary/primary ratios, antiprotons and
positrons. In this sense it goes a long way towards realizing our
original goal, stated in \citet{SMR00}, to reproduce astronomical and
directly-measured data on cosmic rays in the context of a single model
of the high-energy Galaxy.

Obviously our optimized model is far from unique, both in the choice
of parameters like halo size, diffusion coefficient, source gradient, cosmic-ray spectra etc.
The purpose of this paper is to show that it is  possible to 
construct at least one  model which is consistent with all the relevant data
within understandable limits on CR fluctuations and solar modulation.

Based on the optimized model, a new \EB\ spectrum has been derived
\citep{SMR04}.

\begin{acknowledgements}
We would like to particularly thank David Bertsch for assistance and
discussions on the subject of the events and instrumental response of
the EGRET telescope above 10 GeV and Seth Digel for providing the
kinematically analysed \hi\ and CO data used in this work. A part of
this work has been done during a visit of Igor Moskalenko to the
Max-Planck-Institut f\"ur extraterrestrische Physik in Garching; the
warm hospitality and financial support of the Gamma Ray Group is
gratefully acknowledged. Igor Moskalenko acknowledges partial support
from a NASA Astrophysics Theory Program grant. Olaf Reimer
acknowledges support from the BMBF through DLR grant QV0002.
\end{acknowledgements}

\appendix
\section{Description of \hi\ and CO data \label{appendixHICO}}

The \hi\ and CO data used in this work are based on more recent
surveys than those used in \citet{strongmattox96} and \citet{SMR00}.
They were provided by S.~Digel (private communication) who provided
the following description.

The annular maps are generated for 8 ranges of Galactocentric distance
on the assumption of uniform circular rotation with the rotation
curve of \citet{clemens85} parameterized for $R_\odot$ = 8.5 kpc,
$V_\odot$ = 220 km/s. Emission beyond the terminal velocity is
assigned to the tangent point, and emission at slightly forbidden
velocities in the outer Galaxy is assigned to the local annulus
(7.5--9.5 kpc). The longitude ranges within $10^\circ$ of $l = 0$ and
$l = 180^\circ$ are excluded from the integrations in all annuli. The
boundaries of the of Galactocentric distance annuli are 1.5, 3.5,
5.5, 7.5, 9.5, 11.5, 13.5, 15.5, 50 kpc.

The CO data are from the 115 GHz line survey of \citet{dame87} and
the latitude range is $|b| < 25^\circ$. The coverage is not complete
within this latitude range but little or no significant CO emission
is believed to be missed.  The maps are of CO line intensity
integrated over the (longitude-dependent) velocity range of each
annulus and they have angular resolution $0.5^\circ\times0.5^\circ$
\citep[set by the sampling pattern of the constituent surveys;
see][]{dame87}. The units are velocity integrated radiation
temperature (W$_{CO}$), corrected to the intensity scale of
\citet{bronfman88}, in K km/s.

The \hi\ data are a composite of several 21-cm line surveys (Table
\ref{HI_surveys}), which were interpolated to a uniform grid.
Calibrations were checked against the Bell Labs \hi\ horn survey of
\citet{stark92}. Brightness temperatures T$_b$ were converted to
column densities of atomic hydrogen on the assumption $T_{spin}$ =
125 K uniformly. The few positions with $T_b >$ 110 K had $T_b$
truncated to 110 K. The maps have units of column density
$N($\hi$)/10^{20}$ atom cm$^{-2}$. Angular resolution is somewhat
better than 1$^\circ$ and the maps extend to $|b| = 40^\circ$.

\placetable{HI_surveys}


\begin{deluxetable}{lcl}
\tablecolumns{3}
\tablewidth{0mm}
\tabletypesize{\footnotesize}
\tablecaption{HI surveys used for annular ring maps.
\label{HI_surveys}}
\tablehead{
\colhead{} &
\colhead{Angular resolution} &
\colhead{} \\
\colhead{Survey} &
\colhead{(HPBW), deg} &
\colhead{Region of Sky}
}
\startdata
\citet{weaver73} &	36'&	$|b| < 10^\circ$, $l = 10-250^\circ  $     \\
\citet{heiles74} &	36'&	$|b| > 10^\circ$, $\delta > -30^\circ$     \\
\citet{kerr86}   &	48'&	$|b| < 10^\circ$, $l = 240-350^\circ $     \\
\citet*{cleary79} &	48'&	$|b| > 10^\circ$, $\delta < -30^\circ$     \\
\enddata
\end{deluxetable}

The $W_{CO}$ and $N($\hi$)$ maps described above were generated in
1996 using the then best available surveys of CO and \hi. Since that
time, surveys with greater coverage, angular resolution, sensitivity,
and improved calibration have been published, see \citet{dame01} and
\citet{burton94}. However these improvements would hardly affect the
results presented in this paper.

\end{document}